**Peer Review: Objectivity, Anonymity, Trust**

A Dissertation Presented

by

**Hans Daniel Ucko**

to

The Graduate School

in Partial Fulfillment of the

Requirements

for the degree of

**Doctor of Philosophy**

in

**Philosophy**

Stony Brook University

**May 2020**

**Stony Brook University**

The Graduate School

**Hans Daniel Ucko**

We, the dissertation committee for the above candidate for the
Doctor of Philosophy degree, hereby recommend
acceptance of this dissertation.

**Robert Crease,**
**Professor**
**Department of Philosophy**

**Anne O'Byrne**
**Associate Professor**
**Department of Philosophy**

**Harvey Cormier**
**Associate Professor**
**Department of Philosophy**

**George Sterman**
**Distinguished Professor**
**Department of Physics and Astronomy**

This dissertation is accepted by the Graduate School

Eric Wertheimer
Dean of the Graduate School

Abstract of the Dissertation

**Peer Review: Objectivity, Anonymity, Trust**

by

**Hans Daniel Ucko**

**Doctor of Philosophy**

in

**Philosophy**

Stony Brook University

**2020**

Objectivity is a concept that looms large over science. As a society we approve of objective approaches, methods, and attitudes. Objectivity can mean a lot of things: a correspondence with reality, a reliable epistemological process, or an attitude or stance of a scientist.

Peer review is an important part of scientific evaluation. Results do not count as *science* until they are made public, or published. The means by which scientific manuscripts are found to be suitable for publication is peer review, in which usually an editor consults one or more referees to ask for a judgment on the scientific work by an author or authors. These referees are asked to be representatives of a community, that is, they are asked for an objective evaluation not only from their standpoint but instead standing in for an entire community.

This dissertation is focused on the role of objectivity in peer review. Through an examination of aspects of peer review including anonymity, trust, expertise, and the question of who has standing to evaluate research, we find that objectivity in peer review differs significantly from other uses of the term objectivity in science. In peer review it is not required for this objectivity to have correspondence to an outside world, instead it is enough for it to operate inside the "rules" of the community. Neither is the objectivity here empirical in the sense of using data about the scientific problem in question. Rather, the objectivity is one of judgment, cleaving to the epistemological standards of a community that are formed by background assumptions and beliefs. As a consequence, we highlight the role of subjectivity in what is usually taken as a practice of objectivity, and arrive at the insight that objectivity is not defined by one core value, but a balance of transparency, confidentiality, trust, representation, and living up to community standards. As such, objectivity in peer review is a highly specific sense of the term that is not reducible to that used in other aspects of scientific practice.

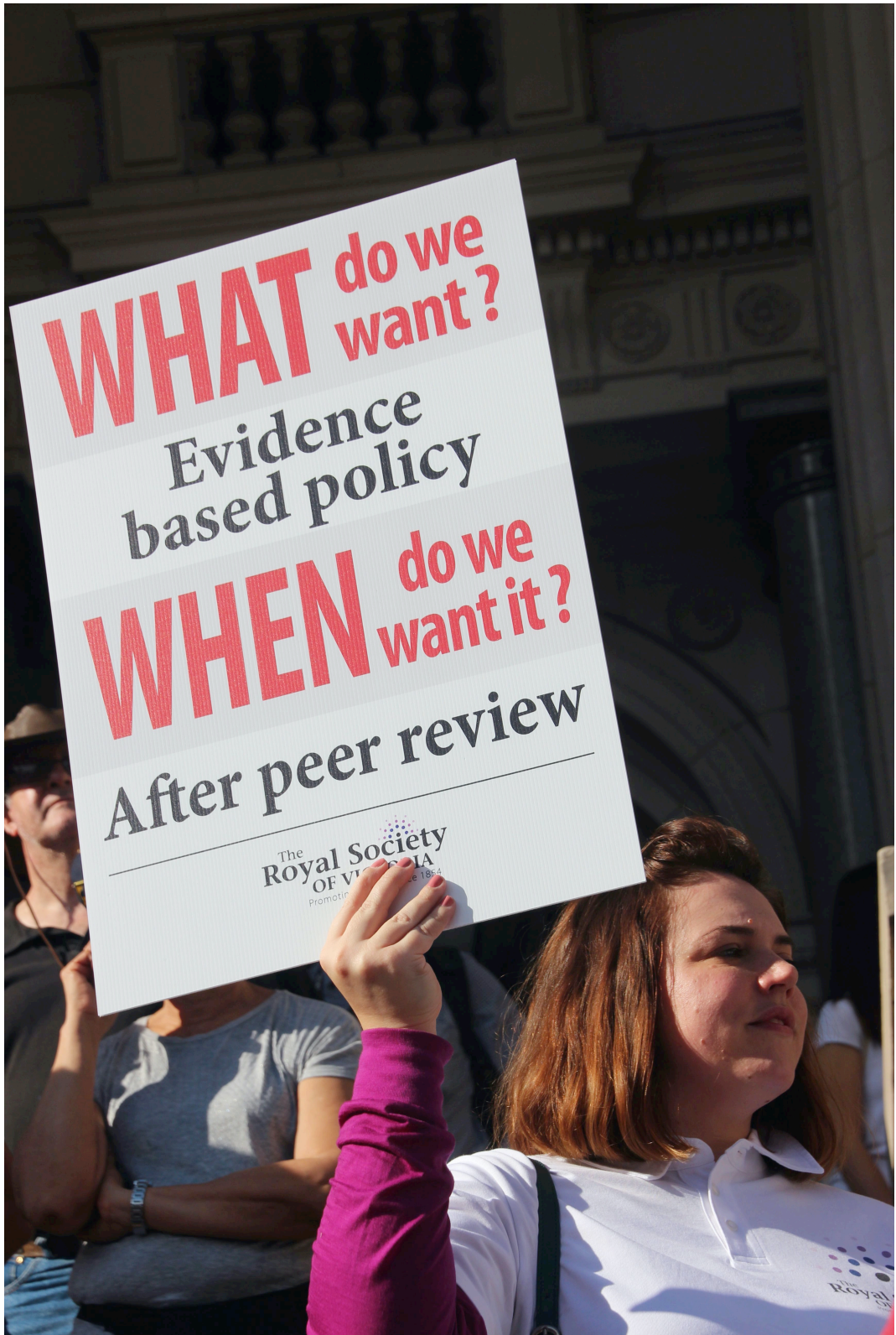

Photo by John Englart (Takver) 2017 at
https://flickr.com/photos/81043308@N00/34051555122

# Table of contents

# Introduction

## Peer review, an ancient new tradition

In 1936, Albert Einstein, together with collaborator Nathan Rosen, submitted a paper to the *Physical Review* titled "Do gravitational waves exist?". Einstein had already published several papers in the *Physical Review*, including the Einstein-Podolsky-Rosen paradox paper,[1] but with this paper he had a different experience, an experience that would cause him to stop submitting original research to the *Physical Review*. What so offended Einstein was that he received a critical referee report, together with an invitation to resubmit. Einstein wrote an angry letter to the editor explaining that the submission had been sent "for publication" and that he had not authorized it to be shown to other people, including the "so-called expert" the *Physical Review* had consulted.[2]

In the end, the paper was not resubmitted to the *Physical Review*, but was published in a revised version with different conclusions, suspiciously in line with the criticism of the *Physical Review* referee, in the *Journal of the Franklin Institute*.[3] Daniel Kennefick's detective work revealed that the referee from the *Physical Review* had befriended a collaborator of Einstein and managed to repackage his criticism in a way that was more palatable to Einstein. The revised paper had a new title: "On Gravitational Waves", answered the question on the existence of gravitational waves with a "yes" rather than the "no" of the first version, and gravitational waves

---

[1] Einstein, A., B. Podolsky, and N. Rosen. 1935. "Can Quantum-Mechanical Description of Physical Reality Be Considered Complete?" *Physical Review* 47 (10): 777–80. https://doi.org/10.1103/PhysRev.47.777.

[2] The story of Einstein's encounter with peer review is told in Kennefick, Daniel. 2005. "Einstein Versus the Physical Review." *Physics Today* 58 (9): 43–48. https://doi.org/10.1063/1.2117822.

[3] Einstein, A., and N. Rosen. 1937. "On Gravitational Waves." Journal of the Franklin Institute 223 (1): 43–54. https://doi.org/10.1016/S0016-0032(37)90583-0.

were in fact observed for the first time in 2015.[4]

Why was Einstein so offended by his treatment at the hands of the *Physical Review*? An explanation offered by Roberto Lalli, a historian of science, is that Einstein was German and spent his early career in a system where papers were not reviewed by external experts, but rather accepted on the say-so of the editor. "Many theoretical physicists coming from Germany or German-speaking countries, for instance, were unfamiliar with the practice of refereeing. The editors of authoritative German physics journals, including *Annalen der Physik*, did not usually seek external advice, and the publication of new papers by established scholars was almost automatic."[5] Einstein should however have been aware of refereeing, since he himself reviewed many papers for the *Proceedings of the Prussian Academy of Sciences*, which casts doubt on the cultural differences being the reason for the offense. A more likely explanation is that, as a member himself, Einstein's own papers were published without any questions or demands for revisions by the Prussian Academy.

The reaction to this story when it was first published in 2005 was one of incredulity. It is fascinating to consider that Einstein, probably the world's most readily recognized scientific personality, would be so thrown by simple criticism to result in such a temper tantrum. The incredulity in the reaction highlights just how much we take peer review for granted today. Not only scholars, but also the public, take peer review very seriously. A common sign at recent pro-science demonstrations in the United States and elsewhere read "What do we want? Evidence-based science! When do we want it? After peer review!" in the style of the protest chants of the labor movement. Peer review has become a recognized stamp of quality that lay people "know" is part of what makes science "official". Upon the discovery of the Higgs Boson, long sought after in physics, a science blog quipped "CERN's Higgs boson discovery passes peer review, becomes actual science."[6]

What is clear from the Einstein story, however, is that the notion of peer review by an external expert, which we today take very much for granted, was still not an established practice for many journals in the 1930s. This includes the British journal *Nature*, which did not start using referees consistently until the 1970s.[7] At the time of Einstein's submission, review by referees was not universally applied at the *Physical Review* either, and would not become so until the 1960s. Indeed, Einstein's previous submissions to *Physical Review* had been published without external review. It may well be that Einstein as an already established luminary was not used to having his submissions questioned in this fashion, but the editors of the journals he published in catered to his preferences, like they would for other established scholars.

---

[4] Abbott, B. P., R. Abbott, T. D. Abbott, M. R. Abernathy, F. Acernese, K. Ackley, C. Adams, et al. 2016. "Observation of Gravitational Waves from a Binary Black Hole Merger." *Physical Review Letters* 116 (6) 061102. https://doi.org/10.1103/PhysRevLett.116.061102.

[5] Lalli, Roberto. 2016. "'Dirty Work', but Someone Has to Do It: Howard P. Robertson and the Refereeing Practices of Physical Review in the 1930s." Notes Rec. 70 (2): 151–74. https://doi.org/10.1098/rsnr.2015.0022. p. 163.

[6] Sebastian Anthony, "CERN's Higgs Boson Discovery Passes Peer Review, Becomes Actual Science," ExtremeTech, Ziff Davis, Inc., 2012, http://www.extremetech.com/extreme/135756-cerns-higgs-boson-discovery-passes-peer-review-becomes-actual-science. My thanks to Melinda Baldwin for drawing my attention to this post.

[7] Baldwin, Melinda. 2015. "Credibility, Peer Review, and Nature, 1945–1990." *Notes Rec.* 69 (3): 337–52. https://doi.org/10.1098/rsnr.2015.0029.

Today journals that *do not* practice peer review with external referees are viewed with suspicion, and the absence of peer review (or very light perfunctory review) is a hallmark of predatory open access journals,[8] journals that charge authors to publish and perform minimal peer review, if any. So how come peer review, something that is now so valued that its absence is a cause for concern, caused Einstein such consternation?

I was motivated to start this project while working as a journal editor, noticing how invested people are in peer review. There is a great deal of faith in peer review. Anecdotal instances of peer review going wrong is explained as being down to individual bad actors. Most often, defenders of peer review not only refuse to admit anything might be systematically wrong, but display great anger at how the process can be corrupted by bad actors. People are very invested in peer review, but it remains poorly understood. As an activity by scientists it seems it should be scientific in procedure, but I will endeavor to show that it is not that simple to make a direct mapping between scientific method and peer review protocol.

---

[8] See for more details on predatory open access journals Moher, David, Larissa Shamseer, Kelly D. Cobey, Manoj M. Lalu, James Galipeau, Marc T. Avey, Nadera Ahmadzai, et al. 2017. "Stop This Waste of People, Animals and Money." *Nature News* 549 (7670): 23. https://doi.org/10.1038/549023a; Ana Bocanegra-Valle, "How Credible Are Open Access Emerging Journals?:: A Situational Analysis in the Humanities," in *Publishing Research in English as an Additional Language*, Practices, Pathways and Potentials (University of Adelaide Press, 2017), 121–50; Agnes Grudniewicz et al., "Predatory Journals: No Definition, No Defence," *Nature* 576, no. 7786 (December 2019): 210–12, https://doi.org/10.1038/d41586-019-03759-y.

**Introducing the project**

This is a study on the philosophical aspects and implications of peer review. The study will mostly focus on peer review as practiced in journals in the natural sciences, with occasional examples from other disciplines. One reason for focusing on natural sciences is that the themes of objectivity in peer review bear particularly strong relation to common conceptions of the scientific method. Natural science journals are also larger in number, with more articles published by year per journal, and are thus more likely to establish systematic peer review to deal with the submission influx. The epistemological standards of natural science have also affected those in other fields. The effects of scientism has led to a push to make otherwise unrelated fields like economics, history, and even theology "more scientific". Finally, and in the same vein, it is a historical fact that today's peer review methods, as adopted across academia, originated in the natural sciences and spread to other fields.[9]

Objectivity is the main theme of this dissertation. I will show that the objectivity at play in peer review differs qualitatively from objectivity in other parts of science. The objectivity in peer review is one of judgment and not empiricism in the moment of evaluation, cleaving to community standards developed socially inside of a scientific community, even as those standards are rooted in and concern themselves with empirical science. I will further compare this to other accounts of objectivity and highlight the contrasts and their consequences. In what follows, I will introduce the mechanics and history of peer review in Chapters One and Two, before setting out what I mean by objectivity in peer review in Chapter Three. One of the frequent strategies for bolstering objectivity and lessening the effect of subjectivity is by strategies of anonymizing some or all of the actors, and I will be discussing the philosophical implications of this in Chapter Four. Matters of trust and expertise play in to the dynamics of peer review, which will be discussed in Chapter Five. Finally, in Chapter Six, I will discuss some emerging trends in peer review, and the philosophical consequences of these changes.

---

[9] Mario Biagioli, in making a similar narrowing of scope, notes that "First, peer review was applied to scientific publications before it became standard in the humanities and social sciences, thus providing substantial evidence about the genealogy of peer review and its early direct relation to early book censorship." Biagioli, Mario. 2002. "From Book Censorship to Academic Peer Review." *Emergences: Journal for the Study of Media & Composite Cultures* 12 (1): 11–45. https://doi.org/10.1080/1045722022000003435, p. 12.

# Chapter One: Description of Peer Review

Today, the principal method of communication of new scientific results is in papers in academic journals. A paper is an account of original research performed, and is an example of primary scientific communication. "At a mundane level, every scientific paper must contribute *something* new to the archive. It must suggest a new scientific problem, propose a new type of investigation, present new data, argue for a new theory, or offer a new explanation. Ideally, it should combine several such virtues."[10] The norm in science is to codify results through a communication system (journal publications). "Research results do not count as scientific unless they are reported, disseminated, shared, and eventually transformed into communal property, by being formally *published*."[11] The publication process raises the following philosophical issues: because science is meant to be speaking a universal language, consideration of the process brushes up against notions of *objectivity*. In terms of the evaluation of submitted articles and the process by which they are deemed suitable for publication, the *judgment* of several actors is called upon. These actors originate from *communities of practice*, whose members are identified from their *expertise*. And through all of this, as a social process, *trust* between actors is a prerequisite for entering into the publication process as a valid means of communicating results.

Ludwig Fleck treated the various means by which knowledge is communicated into four separate categories: 1) popular science, 2) expert science that is divided up into a) journal science and b) vademecum science, and finally 3) textbook science as an additional socio-intellectual form.[12] This last category is a more pedagogical form that aids the induction of people into the expert thought collective, "introductory science."[13] Fleck makes use of his schema of thought collectives, which might be treated as equivalent to a scientific community, to draw the boundaries of these categories,[14] making the distinction between an esoteric circle of experts and an exoteric circle of the wider society.[15] A recognized expert in radiation physics, say, would be at the center of an esoteric circle of "general experts" which Fleck describes as "scientists working on related problems" into which he includes all physicists. The exoteric circle, which is larger and concentric, comprises the more or less educated amateurs who take an interest in physics, readers of popular science. While popular science is strictly speaking for non-experts, it also is created in the aim of creating a world-view [Weltanschauung], "a special structure [Gebilde] emerging from an emotive selection of popular knowledge from various fields."[16] This form of knowledge is not enough for those matters that demand expert knowledge, however it is based on the esoteric knowledge of the expert. It is a characteristic trait of popular science to be vivid and compelling, expressed with great certainty, focusing less on process, details, and caveats, and more on the results, which are themselves presented as simple facts. "Popular exoteric knowledge stems from specialized esoteric knowledge. Owing to simplification, vividness, and absolute certainty it appears secure, more rounded, and more firmly joined together. It shapes specific public opinion as well as the Weltanschauung and in this form reacts in turn upon the expert."[17] We will return to the question of this distinction and who science communication is intended for in Chapter Six, which contains a discussion on Open Access.

---

[10] Ziman, J. M. *Real Science: What It Is, and What It Means*. Cambridge: Cambridge University Press, 2002, p. 41.

[11] *Ibid.*, p. 110.

[12] Fleck, Ludwik, Thaddeus J Trenn, Frederick Bradley, and F Bradley. 1981. *Genesis and Development of a Scientific Fact*. Chicago: University of Chicago Press. http://public.eblib.com/choice/publicfullrecord.aspx? p=3563087, p. 172-191.

[13] *Ibid.*, p. 174.

[14] *Ibid.,* p. 156.

Expert science, on the other hand, is split in two by Fleck, into *journal science* and *vademecum science*.[18] While vademecum science represents an established, if esoteric, systematic synopsis of established knowledge in a field, journal science is provisional, at the very edge of scientific research. Journal science however makes every effort to link up with vademecum science, because the aim is to incorporate the journal science in a paper with the vademecum science. "Any paper published in a scientific journal contains in the introduction or the conclusion just such a connection with vademecum science as proof that the author aims at incorporating his paper in the vademecum and regards its present state as provisional."[19] As Fleck points out, the very way in which scientific papers are written, with caveats and carefully modest expressions, highlights the contrast with the impersonal and definitive vademecum science. It is important to realize that the vademecum is not just a compendium of papers. Such papers often contradict each other in an active field of research, and "does not yield a closed system, which is the goal of vademecum science."[20] Vademecum science represents an established corpus of knowledge, not the frontier of active research. The content of journals is however this frontier, and the eventual transition of this content into the vademecum is aided by peer review. Peer review is only a feature in what Fleck calls journal science, not in any of the other categories he identifies. This indicates that peer review is in some way part of the transformation from journal science into vademecum science.

---

[15] *Ibid.,* p. 164.

[16] *Ibid.*, p. 174-175.

[17] *Ibid.*, p. 176.

[18] *Ibid.*, p. 173.

[19] *Ibid.*, p. 182.

[20] *Ibid.*, p. 184.

At this stage, a *thick description* of peer review is required. This term was first coined by Gilbert Ryle,[21] but the concept was developed further by Clifford Gertz, who used it to describe his method of ethnography. To him, ethnography is not simply a matter of recording what is seen, but also to find the context of what is seen. As Gertz puts it, "[t]he ethnographer `inscribes' social discourse; he writes it down. In so doing, he turns it from a passing event, which exists only in its own moment of occurrence, into an account, which exists in its inscriptions and can be reconsulted."[22] In the spirit of this, we will attempt to thus describe the process by which scientific results are made public, or are published as it were.

A paper is not a full account of what happened in the laboratory in the style of lab notes, but rather an extended argument around original experimental or theoretical results. It is written in a style as to be convincing to the reader and to make a case for the addition of the research and its result to corpus of knowledge. A paper has a case to make, and does not pretend to make a disinterested account of the research performed, rather its aim is to persuade readers and advocate for the veracity of the results and the conclusions. It is up to the referees to evaluate those claims skeptically. This skepticism is very much part of scientific norms.[23]

A journal is primarily composed of a selection of such papers, though many also contain non-research items such as editorials, and secondary pieces such as comments, corrections, errata, most of which are tied to previously published work. Nevertheless, the primary purpose of a journal is to present original research. Journals vary significantly in selectivity, spanning the range from the highly selective journals at the top (for instance *Nature, Science, Cell*) to ones that only demand that a paper be free from detectable error (for instance *Scientific Reports, PLOS One*). It used to be the case that journals limited the numbers of papers to what would fit inside an issue of a journal, but in the Internet age, such limitations now seem arbitrary, and any journals that do any filtering now increasingly do so with reference to their criteria of importance or interest. The description of criteria are intended as guidelines for authors as to help them ascertain whether their research matches the journal in question, and for referees as a guideline for how to evaluate a study on questions beyond mere validity.

A measure of the importance of an article can be through considering the work it inspires. Scientific articles are identified by listing the authors, the title of the paper, and the journal in which it appeared, together with a volume, page or article number, and publication year, all to locate it exactly. Of late, a universal digital object identifier, or DOI (https://www.doi.org/), has grown in popularity as a means of referring to articles. In writing a paper, authors cite relevant prior work in order to back up their claims, or make reference to the origins of prior work on which the present paper builds. This can be reference to a technique, an established value for a constant or parameter, or a theoretical underpinning for the interpretation of experimental results. Credit of this kind is the currency of academic publishing. One should however note that not all citations are there to give credit. Sometimes authors cite works and results they wish to refute. Citing such works are an essential part of a paper's narrative, putting the work in context for the reader.

---

[21] The concept was developed in two articles, "The Thinking of Thoughts: What is `Le Penseur' doing?" and "Thinking and reflecting", Ryle, Gilbert. 2009. Collected papers. Vol. 2. London: Routledge.

[22] Geertz, Clifford (1973), "Thick Description: Toward an Interpretive Theory of Culture", in The Interpretation of Cultures: Selected Essays, New York: Basic Books, p. 18.

[23] "If originality is the motor of scientific progress, scepticism is the brake." Ziman, 2002, p. 42.

There are several ways of using references. One is of course used in the evaluation of papers before publication, during peer review. Another, post-publication, is for readers who can follow the references to find articles that may inform their research. The growing discipline of bibliometrics uses citation data to draw various conclusions about science and its practice. One very popular way of doing this is by using the number of citations to obtain a quantitative measure of a journal's profile in terms of the impact on a community by articles published in it, the Impact Factor. This is a measure of the influence of a journal, and is calculated by taking the number of citations to that journal in a given year of articles published in the two preceding years, divided by the number of articles published in that journal in that year. The Impact Factor was introduced in 1975 by Eugene Garfield[24] at the Institute for Scientific Information, and is published in the annual Journal Citation Reports by Clarivate Analytics. The JCR was previously the property of Thomson-Reuters. Because publication in reputable journals is tied to the advancement and reputation of a researcher, there is a race to publish in so-called "high impact journals" that perform well on this index. Given the nature of the formula for calculating the Impact Factor, smaller journals that manage to "select well" enjoy higher impact factors than larger journals with more inclusive criteria.[25]

One major issue with the Impact Factor is that, as an average quantity, it contains no information about the distribution of citations. Universally across journals, this distribution is heavily skewed, with most of the citations contributing to the Impact Factor of any journal coming from a small set of comparatively highly cited papers, and the rest describing a long tail of research cited less frequently and sometimes not at all inside the two-year window. It has been suggested to make public the citation distributions of journals in order to address this issue,[26] but it is not clear that such an approach will solve the problems of the misuse[27] of the impact factor, invoking the observation that histograms of journal citations look very similar at a glance. Another concern about the Impact Factor is the relative brevity of the two-year window. It takes time to read research and incorporate it into your own research, do the study, write the paper citing the original research, and to publish the research in order to register a citation for the original research. Such a lag time can vary significantly between fields. In addition, since the absolute number of citations scales with the number of publications in a given field, it can be misleading to compare Impact Factors of journals in different disciplines. Even inside of a given discipline, such as physics, the number of publications and practitioners can vary significantly across fields.[28]

---

[24] Garfield, Eugene (2007). "The evolution of the Science Citation Index" (PDF). *International Microbiology*. **10** (1): 65–69.

[25] Antonoyiannakis, Manolis, and Samindranath Mitra. 2009. "Editorial: Is PRL Too Large to Have an 'Impact'?" *Physical Review Letters* 102 (6): 060001. https://doi.org/10.1103/PhysRevLett.102.060001.

[26] Lariviere, Vincent, Veronique Kiermer, Catriona J. MacCallum, Marcia McNutt, Mark Patterson, Bernd Pulverer, Sowmya Swaminathan, Stuart Taylor, and Stephen Curry. "A Simple Proposal for the Publication of Journal Citation Distributions." *bioRxiv*, September 11, 2016, 62109. https://doi.org/10.1101/062109.

[27] See for instance Alberts, B. 2013. "Impact Factor Distortions." *Science* 340 (6134): 787–787. https://doi.org/10.1126/science.1240319, Pendlebury, David A. 2009. "The Use and Misuse of Journal Metrics and Other Citation Indicators." *Archivum Immunologiae et Therapiae Experimentalis* 57 (1): 1–11. https://doi.org/10.1007/s00005-009-0008-y, Shubert, Elliot. 2012. "Use and Misuse of the Impact Factor." *Systematics and Biodiversity* 10 (4): 391–94. https://doi.org/10.1080/14772000.2012.753716, Simons, K. 2008. "The Misused Impact Factor." *Science* 322 (5899): 165–165. https://doi.org/10.1126/science.1165316.

Other measures have been proposed to address concerns about the Impact Factor, such as the h-index[29], in its many variations,[30] the Eigenfactor,[31] SCImago journal ranking,[32] Source Normalized Impact per Paper (SNIP),[33] to name but some. What they have in common is that they rely on citations in one form or another. None have had the take-up of the Impact Factor, and inside of science there is plenty of skepticism[34] about citation metrics, however, when research has to be evaluated by non-experts, for instance in hiring, promotion, and tenure committees, in funding decisions by funding agencies, and in the decision making about major projects such as whether to fund the construction of new experimental facilities, citation metrics become a widely used,[35] if flawed, sorting mechanism. However, as online repositories such as arXiv[36] (an online repository of preprints from physics, which will be discussed more in the chapter on Open Access) and other open access repositories become more prevalent, the Impact Factor is becoming decoupled from citations, in some fields.[37] In addition, funding agencies are not unaware of the limitations of the Impact Factor, and are taking steps to limit the use of the Impact Factor in their decision-making.[38]

---

[28] Batista, Pablo D., Mônica G. Campiteli, and Osame Kinouchi. 2006. "Is It Possible to Compare Researchers with Different Scientific Interests?" *Scientometrics* 68 (1): 179–89. https://doi.org/10.1007/s11192-006-0090-4.

[29] Hirsch, J. E. "An Index to Quantify an Individual's Scientific Research Output." *Proceedings of the National Academy of Sciences of the United States of America* 102, no. 46 (November 15, 2005): 16569–72. https://doi.org/10.1073/pnas.0507655102.

[30] https://scholar.google.com/intl/en/scholar/metrics.html#metrics

[31] http://www.eigenfactor.org/

[32] http://www.scimagojr.com/

[33] http://journalinsights.elsevier.com/journals/0969-806X/snip

[34] Finch, Adam. "Can We Do Better than Existing Author Citation Metrics?" *Bioessays BioEssays* 32, no. 9 (2010): 744–47.

[35] Muller, Jerry Z. "The Tyranny of Metrics." *The Chronicle of Higher Education*, January 21, 2018. https://www.chronicle.com/article/The-Tyranny-of-Metrics/242269.

[36] http://arxiv.org/

[37] Lozano, George A., Vincent Larivière, and Yves Gingras. "The Weakening Relationship between the Impact Factor and Papers' Citations in the Digital Age." *Journal of the American Society for Information Science and Technology* 63, no. 11 (November 1, 2012): 2140–45. https://doi.org/10.1002/asi.22731.

[38] "The San Francisco Declaration on Research Assessment (DORA) was initiated by The American Society for Cell Biology (ASCB) together with a group of editors and publishers of scholarly journals and published in 2013. As of August 2016 DORA was signed by 825 research organisations and over 12,000 individual researchers. A general

## Agents and their roles

The specifics of journal peer review can vary quite widely, but the most common form involves a triad of agents, an author or several authors, a referee or several referees, and an editor. John Ziman described this triad relationship in the following manner, "The peer review process is evidently a highly reflexive and convoluted social activity, where a delicate balance has to be achieved between three distinct interests – those of the author, of the editor, and of the referee. […] any professional scientist of some standing may be called on from time to time to play any of these roles: it is as of every citizen must sometimes be the accused, sometimes the judge, and sometimes in the jury in a succession of criminal trials!."[39] There can be variations on all these roles. For authors, in some fields monographs are very common, in others, for instance high-energy particle physics experiments, a collaboration of hundreds of people is the norm. Author lists can be alphabetical, or ordered in respect to the norms of a field. However, the order of names, when not alphabetic, is not necessarily a guide to levels of contribution. For instance, one norm is to have the primary investigator — the senior researcher who is behind the project — as last author, and the junior researcher — who most probably did most of the day-to-day work — as the first author. Practices vary quite widely, with some fields even specifying individual contributions, though even this is sometimes insufficient even to advocates.[40]

Likewise, the referee or referees used in the review of a paper can range all the way from one or more editor-appointed referee or referees, to a group of hundreds of referees who all send impressions post-publication. In principle all peer review really means is that a submission is reviewed by peers of the author, but it was only relatively recently that the role of the referee was formalized.[41]

Editors can also be either full-time professionals, or active researchers who spend some part of their working life editing journals. The advantage of full-time editors is a certain detachment, a separation from the politics of a field, though this may mean less exposure to current research.

---

recommendation made by DORA was to stop the use of journal-based metrics, such as Journal Impact Factors, as a surrogate measure of the quality of individual research articles. It promotes evaluation of publications by research content rather than the journal in which it is published when assessing an individual scientist's contributions, or in hiring, promotion, or funding decisions. http://www.ascb.org/dora/" http://www.rcuk.ac.uk/media/news/180207/

[39] Ziman, J. M. 1984. *An Introduction to Science Studies: The Philosophical and Social Aspects of Science and Technology*. Cambridge [Cambridgeshire]; New York: Cambridge University Press. p. 64.

[40] "A general problem of using authorship positions as a means of estimating relative contributions is that, regardless of what system for valuing authorship positions is used, it will be misleading and unfair in most applications because relative contributions vary in ways that are not captured by fixed value assignments to authorship positions. " Helgesson, Gert, and Stefan Eriksson. 2018. "Authorship Order." *Learned Publishing*, August. 2018 https://doi.org/10.1002/leap.1191.

[41] "It was a geologist, George Greenough, who introduced the term 'referee' in 1817, importing into science a term he knew from his days as a law student." Csiszar, Alex. "Peer Review: Troubled from the Start." Nature 532, no. 7599 (April 19, 2016): 306–8. https://doi.org/10.1038/532306a., p. 307.

The advantage on the other hand of part-time adjunct editors is a closer relation to the field, though this can come with conflicts of interest.

Everyone involved in the review process seeks to achieve different things in it. Authors publish to establish priority, but also choose their publications venues with a view to legitimation. If it were only a matter of publishing first, authors would self-publish, especially in the age of the Internet. Authors wish to have their ideas accepted by their peers, which leads them to publish in journals that are respected in their community.

A reader, for which the referee can be said to be a proxy, wants peer review to identify relevant content, select said content in a trustworthy and accurate fashion. In an article, they want a paper that is part of the permanent record, and that is a permanent record that does not change in time.

For the editors, and the journals that employ them, the needs are to attract high-quality submissions, and to perform a filtering process, through the help of referees. Referees might be less keen to review for a journal that is not keeping to its own standards, that has become disreputable. It is therefore in the interest of editors to maintain the standards of the journal. In order to serve authors, referees, and readers, journals have to provide four services:[42]

1. Date stamping, or registration, of published ideas, to establish priority.

2. The certification of quality, such that one can be confident that research published in a journal meets the criteria of the journal, and that everything published in the journal has undergone the same peer review process.

3. Recording and archiving definitive versions of papers.

4. Dissemination to an audience of scholars.

These are the reasons why authors seek to publish, and the quality of these services informs the choice of journal. Referees, who are also authors, are similarly motivated to review for journals that provide these services. We will now expand on the roles of all parties in the peer review process.

## The author, or, why publish?

Why do scientific researchers need to publish their results? In order for science to become public knowledge, rather than private knowledge, it must be communicated. "Whatever scientists

---

think or say individually, their discoveries cannot be regarded as belonging to scientific knowledge until they have been reported to the world and put on permanent record. The fundamental social institution of science is thus its system of communication."[43]

There is a difference between communicating results, however, and having them verified. It is no coincidence that the Royal Society of London, established in 1660, has as it motto, *Nullius In Verba*, "take no one's word for it"![44] Publication can of course take different forms, and it would be wrong to claim no scientific publications appeared before the founding of the Royal Society. These were however usually in the form of books, a form which continued after journals were established, for instance Charles Darwin's *Origin of Species* (1859). However, as the pace of scientific research increase readers demanded faster access to the latest results, and the fashion began to shift towards journals as the preferred means of communicating results, since the turnover time for books is much longer than for a scientific article.

Whether evaluated by an editor of by a reviewer or referee, the idea was that for something to be considered as a scientific result, someone else had to approve it. Robert Boyle, one of the founders of the Royal Society, advocated a legalistic approach to experimental work, making a distinction between "witnessed experiments" and "thought experiments", and furthermore insisted that no scientific result was real until it had been witnessed. "In Boyle's view the capacity  of experiments to yield matters of fact depended not only on upon their actual performance but essentially upon the assurance of the relevant community that they had been so performed."[45] Boyle was not even satisfied with one witness, but required "a multiplication of witnesses" much like in a murder trial. Multiplication of witnesses was practically achieved through several means: public demonstration of experiments in the Royal Society assembly rooms; replication of experiments, and also what he came to call "virtual witnessing."[46] Virtual witnessing had as its goal to produce, in a reader's mind, an image of an experimental scene that would obviate the need for direct witnessing or for actual replication. This would be achieved through images and through a scientific report that was written in order to provide an "experimental essay" that would provide this third way of witnessing. A sufficiently explicit account could provide a form of witnessing. Indeed, a paper must be written "as if it were addressed to a hypothetical, very sceptical reader, who is already very well informed on the subject, and might therefore form the spearhead of critical opposition."[47]

The motivation for witnessing and a multiplication of witnesses might well have been empirical, but has other philosophical consequences. For instance, a connection can be made with Husserl's discussion on perspectives in perception. Noting that our external perception is by

---

necessity limited in that (for instance) we can only view an object from one side at a time, Husserl states that "no matter how completely we may perceive a thing, it is never given in perception with the characteristics that qualify it and make it up as a sensuous thing from all sides at once."[48] This complicates that notion that anyone might actually fully witness anything, since we necessarily have an incomplete view of whatever it is we are observing. We can extrapolate, "[v]iewing the front side of the table we can, whenever we like, orchestrate an intuitive presentational course, a reproductive course of aspects through which the non-visible side of the thing would be presented to us."[49] Having once seen the back of the table the "empty premonition now has a determinate prefiguring that it did not have previously."[50] We can further extrapolate this to other observers. If I were to take up the position of another observer, albeit one with their own egocentric viewpoint, if the object in question exists in a shared objective reality, this second observer can "fill in the blanks" of what is not accessible to the first observer. This would be the phenomenological rationale for a multiplication of witnesses.

It would be useful to ask oneself how deeply the witnessing has to go in order to qualify. Science has become much more complex since Boyle's days, and as this complexity increases, actually witnessing an experiment in person is a difficult request, never mind more theoretical studies such as computer simulations. As a result, the nature of witnessing and also the nature of reviewing must change. One cannot practically repeat studies to verify them before publication. "The referee cannot be expected to validate a discovery claim according to strict philosophical standards of proof. A referee's report can be no more than a `first reading', certifying that the material is original and cited, that the argument is clearly expressed and not implausible that the experimental procedures are technically competent, and that the conclusions are not contrary to established fact."[51] We not only need a multiplication of perspectives, but the limitations of the perspectives is more and more significant as our science becomes more complex.

Due to several constraints, the verification aspect of peer review is not as epistemologically rigid as one might hope. The constraints include that of time, since authors (and editors) want to publish exciting research quickly. Most journals ask for reports inside a few weeks, sometimes sooner for especially exciting results, so there is no time for replication studies as a part of the peer review process. The referee is also unlikely to have access to original or equivalent equipment for a replication study, even if time permitted. It is also the case that no referee can be a total match for the author(s) of a paper in terms of a match of expertise. The nature of the referee consultation process is such that referees with the most relevant expertise are often ruled out as being too connected to the authors, e.g. past collaborators, or competitors who would have a conflict of interest in acting as referees for a paper where they stood to benefit from its delayed appearance or nonappearance in the scientific literature.

This means that trust has become essential to the peer review dynamic, as referees have to assume authors have done what they describe, obtaining the results they describe, since referees cannot witness in person or repeat the study. The origins of this sense of trust is not simple goodwill and belief in the best of intentions, but is rather underpinned by a sense of togetherness in a community of practice. This will be explored later in Chapter Five, on trust in peer review. In addition, the question of who is included in this community of practice is a question of expertise and its role in determining who counts as a peer. This will also be discussed in more detail in Chapter Five.

---

[48] Husserl, Edmund. *The Essential Husserl: Basic Writings in Transcendental Phenomenology*. Edited by Donn Welton. Studies in Continental Thought. Bloomington, IN: Indiana University Press, 1999, p.221.

[49] *Ibid.*, p. 222.

[50] *Ibid.,* p. 225.

[51] Ziman 1984, p. 65.

The publication of research bolsters the reputation and stature of a researcher, particularly if that research is influential. We previously discussed how the number of citations can be a measure, if a crude one, of a scientific journal's impact. The credit from this also extends to the author or authors of the article. Sociologically, citations have the function of recognizing researchers for their contribution to knowledge.[52] The number of citations to an author's work thus has a correlation to the reputation of that author. The consequence of this is that to be a reputable researcher one must publish, and publish in reputable journals[53] at that. It has also been shown that the reputation of an author often has positive effects on that author's future success in publishing their research,[54] though journals are keen to downplay the idea that there is a favored group of authors.[55] Robert K. Merton has referred to this as the Matthew effect,[56] from the bible verse in the gospel of the same name.[57] The system of referencing we have distributes its gifts unevenly.

## The scientific referee persona

> "In effect, a referee acts as a representative of the scientific community.
> This involves a temporary reversal of social roles, from making the case

---

[52] "Citations operate sociologically to `recognize' researchers individually for their contributions to knowledge. Indeed, one of the features of post-academic science is the notion that the standing of a researcher amongst her peers can be assessed by counting `hits' in the *ScienceCitation Index* — an annual catalogue of all the *new* papers that have cited each *previous* paper stored in the archives." Ziman 2002, p. 260.

[53] "Although the communication network of science is not a formally closed system, the number of `reputable' journals — i.e. the journals scientists normally attend to in their research — is limited, and there is strong competition amongst scientific authors for access to this finite resource." Ziman 1984, p. 63.

[54] Bravo, Giangiacomo, Mike Farjam, Francisco Grimaldo Moreno, Aliaksandr Birukou, and Flaminio Squazzoni. "Hidden Connections: Network Effects on Editorial Decisions in Four Computer Science Journals." Journal of Informetrics 12, no. 1 (February 1, 2018): 101–12. https://doi.org/10.1016/j.joi.2017.12.002.

[55] "Fair Success Rates." *Nature Materials* 14, no. 10 (October 2015): 961. https://doi.org/10.1038/nmat4445.

[56] Merton, Robert K. "The Matthew Effect in Science." Science 159, no. 3810 (1968): 56–63.

[57] For to every one who has will more be given, and he will have abundance; but from him who has not, even what he has will be taken away. — Matthew 25:29, RSV.

for a research claim to trying to pick holes in one."[58]

Who can be a referee? Historians have different ideas about when peer review proper started. The definition of what constitutes a referee depends on one's definition of peer review. Depending on what one counts as peer review, one can peg its starting date at different stages. For instance Alex Csiszar claims that the start of peer review occurs in the 19th Century,[59] because he claims that this is when the systematic practice of sending referral requests to external experts for their reviews originated. By contrast, David Kronick has a much more inclusive view of what constitutes peer review, writing that "In the broadest sense of the term, peer review can be said to have existed ever since people began to identify and communicate what they thought was new knowledge."[60] In the same article he describes the practice at the Royal Society of Edinburgh in a fashion that seems very systematic indeed, "Memoirs sent by correspondence are distributed according to the subject matter to those members who are most versed in these matters. The report of their identity is not known to the author. Nothing is printed in this review which is not stamped with the mark of utility."[61] Though this practice does not rely on external referees, it could certainly be described as review by peers, and hence peer review. We will deal with these distinctions in more detail in Chapter Two.

For either of these accounts, Kronick's or Csiszar's, or others that we will detail later, there is an entry test to become a referee. Ziman writes, "At the very lowest level, an academic scientist scarcely exists until his or her work has been published in a reputable scientific journal."[62] People do not become referees by virtue of their academic qualifications alone; the stress in peer review is on *peer*, in peer review an author's submission is evaluated by colleagues in the same specialization.[63] Another definition of the word "peer" is social equal. The interaction between the agents is very much informed by the norms and practices of the research community in question.[64] Referees are chosen because they are part of the same community and culture as the authors. The training of competent new referees is important for authors as well as editors. Senior scholars frequently help junior ones negotiate the peer review process, both as authors and as referees. Journal editors do frequent outreach and publish materials on how to review for their journals.[65] In addition to this, services like Publons[66] and Enago[67] have sprung up that provide training services for referees outside of any specific journal.[68]

---

[58] Ziman 2002, p. 42.

[59] Csiszar, Alex. "Peer Review: Troubled from the Start." Nature 532, no. 7599 (April 19, 2016): 306–8. https://doi.org/10.1038/532306a.

[60] Kronick, D. A. "Peer Review in 18th-Century Scientific Journalism." *JAMA* 263, no. 10 (March 9, 1990): 1321–22, p. 1321.

[61] *Ibid.,* p. 1321.

[62] Ziman 1984, p. 70.

[63] *Ibid.,* p. 64.

[64] "In order for criticism to be relevant to a position it must appeal to something accepted by those who hold the position criticized. […] This cannot occur at the whim of individuals but must be a function of public standards or criteria to which members of the scientific community are or feel themselves bound. Longino, Helen E. *Science as Social Knowledge: Values and Objectivity in Scientific Inquiry*. Princeton, N.J.: Princeton University Press, 1990. p. 77.

It is possible (in fact probable) that there are differences in the research focus between the author and referee, since one of the virtues valued by science is originality, as already mentioned. Journals seek to publish new results; the publication of established facts is left for textbooks, so in the pursuit of unchartered territory, it comes as no surprise that not all scientists in a discipline are working on the exact same problem. In addition, choosing a referee that is too close to the subject of the submission raises the possibility of conflicts of interest. For small fields the number of perfectly matched peers might be very small, such that competition for finite resources and attention may become an issue. The issue of proper disciplinarity is however not straight-forward. Robert Frodeman and Jonathan Parker, in writing about the Broader Impacts Criterion (BIC) standard that is relevant for evaluating National Science Foundation proposals, have highlighted that "there are no clear and unequivocal disciplinary boundaries to be drawn simply on the basis of epistemology."[69] In a later article with Adam Briggle, Frodeman observes a "dedisciplining of peer review" at funding agencies, as manifested by BIC and Institutional Review Boards, which seeks to find a balance between scientific autonomy and public accountability.[70]

In deciding who counts as a peer, the entry into this category is by the demonstrated adherence to commonly held norms.[71] Referees whose reports indicate that they do not adhere to the norms are not consulted again by editors. This also happens if referees are viewed as being insufficiently critical, not backing up their arguments sufficiently well, or are particularly slow about returning reports. Referees are not consulted evenly, and a small group of referees tends to do most of the reviewing for any field, journal, or region.[72] There are many reasons behind this, reputation of these referees as scholars, personal relationships with editors, a sliding scale of willingness to review. Editors will naturally return to referees who they have formed a good working relationship with, but as a result the reviewing pool has very deep and shallow ends on each extreme, and the slope of the floor is far from linear.

---

[65] See for instance http://www.nature.com/authors/policies/peer_review.html, http://pubs.acs.org/page/review/index.html, https://publishingsupport.iopscience.iop.org/questions/general-review-procedure-on-iop-journals/

[66] https://publons.com/home/

[67] https://www.enago.com/academy/

[68] "The Publons Academy is a practical peer review training course for early career researchers developed together with expert academics and editors to teach you the core competencies and skills needed of a peer reviewer." https://publons.com/community/academy/

[69] Robert Frodeman and Jonathan Parker, "Intellectual Merit and Broader Impact: The National Science Foundation's Broader Impacts Criterion and the Question of Peer Review," *Social Epistemology* 23, no. 3–4 (2009): 337–45, https://doi.org/10.1080/02691720903438144, p. 343.

[70] Robert Frodeman and Adam Briggle, "The Dedisciplining of Peer Review," *Minerva* 50, no. 1 (March 2012): 3–19, https://doi.org/10.1007/s11024-012-9192-8.

[71] "The shift from "refereeing" to "peer review" is not merely a linguistic curiosity. Calling external refereeing at journals or funding bodies "peer review" established that the evaluation of a paper or grant proposal could only be done by experts—the peers of the person who submitted the work." Baldwin, Melinda. 2018. "Scientific Autonomy, Public Accountability, and the Rise of 'Peer Review' in the Cold War United States." Isis 109 (3): 538–58. https://doi.org/10.1086/700070 p. 548.

In any form of peer review that involved the triad structure of author-referee-editor, the referee's role is that of an advisor to the editor.[73] The referee does not accept or reject submissions; only the editor does that, albeit usually in concert with the recommendations of the referees. Researchers are very invested in quality reviewing, especially of their own papers, and many researchers have taken it upon themselves to weigh in on what constitutes a good report,[74] and (to authors) how to respond to criticism.[75] Journals also have an interest in good referee reports, and put out materials[76] and statements[77] in order to help referees review papers for them. The job of the referee[78] is not to help write the paper as a quasi-coauthor; to make an otherwise unpublishable paper publishable by directing the research performed; or to ensure that the paper cites the referee's own work. Rather the referee is advised to be honest with the authors,[79] and only recommend that they revise and resubmit if the paper can reasonably be made publishable in a limited number of review cycles.

---

[72] "In each country we measured, a reasonably small proportion of reviewers (10%-20%) were responsible for half of the reviews done in their country." https://publons.com/blog/spread-of-peer-review-workload/

[73] "The job of the referee is to provide expert and unambiguous advice to the editor about whether or not a paper is publishable. The referee advises, the editor decides." Berk, Jonathan, Campbell R. Harvey, and David A. Hirshleifer. "Preparing a Referee Report: Guidelines and Perspectives." SSRN Scholarly Paper. Rochester, NY: Social Science Research Network, November 21, 2016. https://papers.ssrn.com/abstract=2547191.

[74] Berk, Jonathan B., Campbell R. Harvey, and David Hirshleifer. "How to Write an Effective Referee Report and Improve the Scientific Review Process." *Journal of Economic Perspectives* 31, no. 1 (February 2017): 231–44. https://doi.org/10.1257/jep.31.1.231.

[75] Martin, Brian. "Writing a Helpful Referee's Report." *Journal of Scholarly Publishing* 39, no. 3 (May 11, 2008): 301–6. https://doi.org/10.1353/scp.0.0008.

[76] https://publishingsupport.iopscience.iop.org/becoming-a-journal-reviewer/

[77] https://www.maa.org/press/periodicals/guide-for-referees; https://journals.aps.org/prl/referees/advice-referees-physical-review-letters

[78] Berk et al., 2016, p. 3.

[79] "A revise recommendation to the editor is a serious commitment given that A-level publications are rare. You should not make a revise recommendation to simply defer judgment. It is not helpful for editors to hear that 'this paper seems ok, but I am not sure, let's see what the authors can do.' Most of the work you put into the refereeing process should be at the initial stage. That said, it is crucial that you stick with the paper and make sure the paper attains the journal's high standard when it is resubmitted." Berk et al., 2016, p. 3.

## The editor

The nature of editors varies widely across journals, even journals in the same field. Some editors are adjunct or sometimes known as associate editors. This group of editors work part-time, primarily working in positions in academia, and editing journals a fraction of that time. The exact title of these editors varies between journals, and some associate editors work full time. For the purposes of this description these editors will be referred to as adjunct editors. These editors are likely to be active in the field of the journal, or journal section, that they edit. As such, such editors are vulnerable to accusations of being biased in favor of some research over other research. The advantage of such editors for journals is however that they have such a close connection, and, because they are usually senior figures in their field of specialization, have a degree of authority.

Another mode of editor is the professional, full-time editor. This editor is usually not personally engaged in research, or if they are, they will only be doing it in their spare time and not as an official part of their position. Such editors might have the benefit of a less embedded perspective, but are vulnerable to accusations of an ivory-tower mentality. These editors are often not expert in whatever subfield they handle, though they should ideally have some related expertise. [80]Professional editors are more expensive than adjuncts, and are usually thus only found at journals at the top rank of the field in question.

Many journals operate with a mixture of professional and adjunct editors, in an attempt to achieve the best of both worlds. The motivation for this can vary, for instance, a journal might be so small that it can make do with one or more part-time editors, or, in larger journals, some smaller subfields might not warrant a dedicated full-time editor.

The relationship with editors and any editorial board also varies significantly. Some journals have one or more professional editors who mostly manage the influx of submissions, assigning them to a board of editors who are themselves active researchers in their field. Other journals use their boards primarily as consultants in a more or less formal capacity, or for adjudication for tricky cases. Whatever their formal functions, editorial boards are usually populated with senior figures in their respective fields, all whom lend the journal authority.

The range of editorial setups is very wide. One might think that top journals would gravitate towards one approach, but it is not even true that all top journals even use professional editors. For instance, the journals of the American Chemical Society (ACS) operate with an editor-in-chief who is also a working scientist and a group of editors, also working scientists, who handle manuscripts in their own subfields.[81] By contrast, the journals of the American Physical Society[82] (APS) rely primarily on professional editors, though many also have adjunct editors appointed as appropriate. The editors of *Nature* journals, a group of commercial journals published by Springer-Nature, are usually professional editors,[83] whereas *Science*, published by the American Association for the Advancement of Science (AAAS) has a smaller in-house staff[84] and an extensive editorial board of reviewing editors.[85] Often these editorial setups have occurred contingently, without an initial plan, but sometimes an approach can be the result of a deliberate desire to shape a journal's identity. Nature journals, as mentioned, are staffed by professional editors, but this is a quite deliberate decision made in order to safeguard their independence, "A principal distinction between a Nature journal and its societal counterparts is that we have no editorial board made up of practicing academics. […] [which] means that we are independent from the direct influence of any university, society or funding agency…"[86] Decisions about the relationship with an editorial board, including whether to have one in the first place, are at least in part informed by a journal or publisher's relationship with the community it is supposed to serve, much like the decision of whether or not to have full-time or part-time editors.

---

[80] "Editors of a journal that tries to cover a whole discipline can only have special competence for a limited number of manuscripts." Hirschauer, Stefan. "Editorial Judgments: A Praxeology of 'Voting' in Peer Review." *Social Studies of Science* 40, no. 1 (February 1, 2010): 71–103. https://doi.org/10.1177/0306312709335405, p. 84.

It might be tempting to consider the editor as primarily a manager of the correspondence between authors and referees. Put as a question, why do we not just allow authors and referees to communicate freely? Nowadays, a technological solution that allows for anonymized communication could even do the job of safeguarding the confidentiality of the process. However, what the editor brings to the process, ideally, is judgment. It is true that journals have different resources and that editors thus work in a more or less interventionist manner, however, the ideal is that the editor is engaged in the process.

Turning to my own experience as an editor for *Physical Review Letters*, my job frequently involves extraordinary communications directly with referees to further discuss their reports, especially where the recommendation of the referee does not adhere to the standards of the journal. This is not phrased as an attempt to alter the recommendation, but to obtain better justification for a decision in either direction. Not all journal editors engage in this exact practice of engaging in discourse with the referee beyond the submitted report, but a uniting description of editors that puts the focus on expertise is that while referees are experts in the subject matter of the article, the editor is an expert on the criteria of a journal.[87] Phenomenologically, this expertise often manifests itself in a feeling that the manuscript "just doesn't look like" the kind of contribution that suits the journal, in a way that is sometimes hard to express in rigorous objective criteria. As a praxis, this is reminiscent of Dreyfus and Dreyfus's account of expertise, in which a person passes through five stages from "beginner" to "expert", where they describe the expert as being completely engaged in skillful performance by the end of the transition to expert status.[88] An editorial model that was more mechanical, with the editor as simply a correspondence manager, would not employ this kind of expertise in decision-making.

---

[81] An example of this set-up is the one for the Journal of the American Chemical Society (JACS), the ACS flagship journal. In this journal, there is a senior editor, a managing editor and their staff, and a editorial board of associate editors, who handle manuscript in their subfields. https://pubs.acs.org/page/jacsat/editors.html. The exact makeup of the editorial staff varies across ACS journals, but no ACS journal relies primarily on professional editors.

[82] https://journals.aps.org/

[83] See for instance https://www.nature.com/nature/about/editors, https://www.nature.com/nphys/about/editors, https://www.nature.com/ncomms/about/editors.

[84] http://www.sciencemag.org/about/team-members

[85] http://www.sciencemag.org/about/editors-and-editorial-boards

[86] "Meet the Editors." *Nature Physics* 13, no. 5 (May 2017): 415. https://doi.org/10.1038/nphys4137. See also https://www.nature.com/nature/for-authors/editorial-criteria-and-processes, "Nature does not employ an editorial board of senior scientists, nor is it affiliated to a scientific society or institution, thus its decisions are independent, unbiased by scientific or national prejudices of particular individuals. Decisions are quicker, and editorial criteria can be made uniform across disciplines. The judgement about which papers will interest a broad readership is made by Nature's editors, not its referees. One reason is because each referee sees only a tiny fraction of the papers submitted and is deeply knowledgeable about one field, whereas the editors, who see all the papers submitted, can have a broader perspective and a wider context from which to view the paper."

## Schematic of the peer review process

There is of course a whole sequence of events that lead to the writing of a manuscript, from an idea, the request for funding, the painstaking research, the interpretation of results, and the writing of a scientific account. All of these steps are worthy of consideration on their own merit, but from the perspective of publication, the review process starts with the submission of a manuscript. The first task of the editors is deciding an article is fit for review by referees. Submissions are can be turned away by journals for reasons of scope, failure to meet certain levels of scholarship, or that the result is not interesting enough. These rejections are variously called desk rejections,[89] editorial rejections,[90] or rejections without external review,[91] to name some examples, and the way these many names for the same procedure are used can give some hints about a journal's attitude towards this practice. A desk rejection gives a bureaucratic impressions, whereas editorial rejection, or rejection without external review, has the air of something more deliberate.

---

[87] "Peer reviewing evolved from the need of editors to choose among a surplus of submitted manuscripts and the growing inability of an editor to possess enough expertise to judge quality in all specialized fields that a journal might cover." Mayland, H.F., R.E. Sojka, and John C. Burnham. "How Journal Editors Came to Develop and Critique Peer Review Procedures." In *ACSESS Publications*. American Society of Agronomy, Crop Science Society of America, Soil Science Society of America, 1992. https://doi.org/10.2134/1992.researchethics.c5. Abstract.

[88] Dreyfus, Hubert L., Stuart E. Dreyfus, and Tom Athanasiou. *Mind over Machine: The Power of Human Intuition and Expertise in the Era of the Computer*. New York: Free Press, 1986, p. 31.

[89] Elsevier. "5 Ways You Can Ensure Your Manuscript Avoids the Desk Reject Pile." Authors' Update. Accessed February 11, 2018. https://www.elsevier.com/authors-update/story/publishing-tips/5-ways-you-can-ensure-your-manuscript-avoids-the-desk-reject-pile.

[90] Picciotto, Marina. "Why Editorial Rejection?" *Journal of Neuroscience* 38, no. 1 (January 3, 2018): 1–2. https://doi.org/10.1523/JNEUROSCI.3465-17.2017.

[91] "Editorial: Thank a DAE!" Physical Review Journals, January 7, 2014. https://journals.aps.org/edannounce/PhysRevLett.112.010001.

In some journals such rejections are reserved for articles that are obviously wrong, or evidently do not fit the scope of the journal, but in some more selective journals, in principle perfectly publishable manuscripts are turned away because the results are not sufficiently important or interesting. Such rejection rates can vary quite widely,[92] though because of confidentiality such numbers are usually not available in great detail. If a journal filters on criteria beyond apparent validity, and receives many submissions, a calculation is made that rejecting more manuscripts before regular peer review is imperative to conserve resources for the sake of manuscripts that seem like they may be publishable in the journal. Indeed, in some highly selective journals like Nature and Science, rejection before peer review is overwhelmingly the norm. Some society journals review a larger fraction of their submissions than their actual acceptance rate warrants, because peer review is viewed as a service to the community and as something that researchers in the community have a right to.

The practice of editorial rejection puts the question of judgment at the forefront in a way that is more controversial than then the rejection is done on the recommendation of a referee. Of course it is the charge of the editor to interpret the criticism of the referees correctly and to make appropriate decisions, but when an editor who is not necessarily a practicing researcher, in the case of professional editors, or one who is and who might be accused of being too entrenched in the current debate to assume a disinterested perspective, the question of judgment becomes important. Such judgments are made from the perspective of socially communicated expertise of a field in the case of professional editors, or drawing on one's direct expertise in the case of academic editors who are also active researchers. We will return to this in more detail in Chapter Five.

Some journals practice post-publication peer review, in which the paper is published immediately and is then reviewed after it has appeared. Usually such journals practice open peer review, such that the reports are also publicly available. An example of this is the review process practiced by *Atmospheric Chemistry and Physics* (ACP), introduced in 2001.[93] After pre-screening, the article appears immediately on the journal's website, and classified as a "discussion paper." The paper is then assigned to peer reviewers much in the same way as for pre-publication review, but the discussion between the referees and authors is public, and other interested readers can also participate. For ACP, a second phase of review then starts, with revisions and review much more like the model of traditional journals. The article, if at this point acceptable, is then promoted to the main journal, together with referee reports and the authors' response to said reports. Another approach towards open review is that practiced by SciPost. In this model, the journal exists as an overlay on a preprint server, in this case the physics

---

repository arXiv.org, and authors can ask to have their papers considered by this overlay journal. An open review process occurs, in accordance with their principle of peer-witnessed reviewing[94] in full view of the community. If acceptable, the article is then published together with the review correspondence. However, the vast majority of journals practice pre-publication peer review. Some do publish referee reports[95] to accompany some or all of the published papers, though here as well practices vary.

Whichever model or peer review the journal practices, while out to review, referees compare the article to the standards of the journals. These standards vary, and can range from journals accepting articles simply being free from detectable error, to an acceptable article demonstrating a very important result or results. In some cases articles of a more methodological bent are acceptable, and go on to be highly useful resources for the community in question. These techniques become so prevalent that they are usually known primarily by their acronyms or initialisms, like Density Functional Theory (DFT), Scanning Tunneling Microscopy (STM), Magnetic Resonance Imaging (MRI), etc.

The editor sends a decision to the authors, usually with reports and an editorial decision letter. The recommendations can be to accept as is, accept after minor revisions, revise and resubmit, or reject. There are many reasons to not publish an article. In a paper on peer review, Jerome P. Kassier and Edward W. Campion, taking medicine as their example, list, comprehensively reasons for rejection for an article.[96] There can for instance be deficiencies in (study) design, such that the validity of the results are in question due to bad procedures, bad samples, systematic errors, insufficient amount of data collected, insufficient breadth of study, and a number of biases. A paper might also be rejected due to bad presentation, which includes failing to properly motivate the study; a failure to cite relevant prior work in the field; an unclear explication of the experiment; inadequate data presentation; omission or manipulation of data; poor writing, including excessive jargon, over-selling or hyping results, or by the study just being boring. There can also be problems with interpretation: the data can be preliminary, inconclusive, not lend themselves to the authors' conclusions. In addition, the results can be too specific, not generalizable, not properly contextualized, or further development of the results of the study can be unrealistic for economic reasons. A final reason to reject concerns the importance of the research. The results in a published paper should not be trivial, unoriginal, predictable, nor should the results be of narrow interest and too specialized. The implications of the study should be clear and the issue at hand should not be outdated.

It should not be concluded, however, that success in avoiding falling into one of these deficiencies means that an article is universally suitable for publication. The judgment of editors can be based on other criteria and is sometimes limited by space concerns[97]. For funding agencies, an *ex ante* attempt to gauge societal impact is also often attempted.[98] Many journals, particularly selective ones, prefer to frame their policies from a standpoint that publication is not the default, and that authors (and positive-leaning referees) have to affirmatively show that a paper is suitable for the journal, not that it is unsuitable.

---

[94] "Peer-witnessed refereeing: Scientific publications should undergo the strictest possible peer refereeing process, witnessed by the community instead of hidden behind closed doors." https://scipost.org/about

[95] "Nature Communications uses a transparent peer review system for manuscripts submitted from January 2016, where in order to improve the openness of our peer review system we are publishing the reviewer comments to the authors and author rebuttal letters of our research articles online as a supplementary peer review file. Authors are given the opportunity to opt out of this scheme at the completion of the peer review process, before the paper is accepted. In agreeing to review a manuscript, reviewers give their consent to the potential publication of the reviewer comments made to authors." https://www.nature.com/ncomms/journal-policies/guide-to-referees

[96] Kassirer J.P., and Campion E.W., "Peer Review: Crude and Understudied, but Indispensable." *JAMA* 272, no. 2 (1994). p. 96–97. doi:10.1001/jama.1994.03520020022005,, p. 97.

If the judgment of the editor after considering the referee evaluation is sufficiently positive, the authors may resubmit to the journal. Most journals would then go back to the previous referees, if available, or consult new ones for cause. Since reviewing papers is not the main job of referees, who usually work as researchers in their own right, this part of the review process can be very time-consuming, so many journals limit the number of times a manuscript can be resubmitted. Improving a manuscript iteratively until it reaches a minimal threshold of publishability is not the goal of the review process.

Responding to criticism from referees can be challenging. The message the authors receive is relayed by the editor, who may not even pass on all of the reviewer criticism to the authors — some may be remarks only intended for the editor. The authors need to figure out how to respond to the decision letter, which can sometimes involve quite a bit of guesswork as to the editor's intentions. In many cases, journals rely on form letters[99] that do not contain a lot of information, putting the focus instead on referee reports, which may be more or less detailed in their criticism.

It is at this stage of the process that the triad of editor-author-referee really becomes important, because all parties have made their initial statement. In the case of the authors that was the submission, the reviewers have provided their reports, and the editor has provided a decision letter. If peer review were a game, all players have now made one move. The ball is now in the authors' court.

Assuming the authors can respond to the criticism, and that the referees have not identified some fatal flaw in the work, reviewers are consulted again and the submission is (hopefully) accepted for publication. In some cases, the reviewers are not convinced, or the authors have resubmitted in the face of overwhelming negative criticism, and the submission is rejected with finality. Depending on the criticism, the authors can at this point consider resubmitting to another journal, possibly one with less stringent standards than the first.

In some journals, rejected authors can appeal to an editorial board for a final degree of reconsideration of their case. Policies on this extra layer of consideration vary from journal to journal, for instance, the Nature journals, as mentioned previously, do not have an editorial board, and all appeals are evaluated solely by the in-house editors. In general journals have more or less formalized appeals procedures, for some an appeal simply consists of having the editor consider the manuscript again,[100] possibly without consulting anyone else for additional input, and in some the appeals procedure is enshrined in official policy.[101] For many journals, the number of manuscripts published after appeals is much smaller than the regular acceptance rate of the journal, however, considering that journals' definition of the word "appeal" differs widely, so it is hard to compare post-appeal acceptance rates across different policies and practices.

---

[97] "Nature has space to publish only 8% or so of the 200 papers submitted each week, hence its selection criteria are rigorous. Many submissions are declined without being sent for review." https://www.nature.com/nature/for-authors/editorial-criteria-and-processes

[98] Holbrook, J. B., and R. Frodeman, "Peer Review and the Ex Ante Assessment of Societal Impacts," *Research Evaluation* 20, no. 3 (September 1, 2011): 239–46, https://doi.org/10.3152/095820211X12941371876788.

[99] "As an editor you likely go through a weekly (or perhaps daily!) circuit of emailing out review requests, manuscript acceptance decisions, and rejection letters. With the frequency that you're sending near duplicate emails to authors and reviewers, have you considered creating email templates to save time?" https://blog.scholasticahq.com/post/peer-review-email-templates/

## Conclusion

Given the variations in approach, one could be tempted in concluding that there is no one thing called peer review, but rather individual instances of peer review. However, though details and implementations may vary, general themes are constant across the multitude of examples of peer review. At essence what is at stake is a community that participates in a process of knowledge production. The exact way or whether this involves the familiar triad of author-referee-editor is a detail that is not at significant qualitative with this main point. We therefore can feel justified in talking about peer review as if it is a thing that can be subjected to philosophical analysis.

Variations do exist: editors can be professional full-time editors or part-time academic adjuncts, and referees can review the paper before or after publication, in a more or less transparent process. Journal policies usually reflect the particulars of the community they seek to serve, and are as such often contingent, arrived at in a pragmatic fashion, rather than motivated by some overarching philosophy. While the roles of editors and referees may vary quite widely, the triad structure of author-referee-editor is reproduced in one form or another, which highlight one of the functions of peer review, verification. Science only counts as science if it is communicated, and the decisions structure for what to communicate is peer review.[102] In the age of the Internet the function of peer review as a filter is particularly important to allow for readers to discriminate between more or less important research results. The function and form of peer review has evolved throughout history, which we will explore in the next chapter.

---

[100] "An appeal letter is only read by the editors, so sensitive information not meant to be seen by the referees can be included." "How to Write an Appeal Letter : Methagora." http://blogs.nature.com/methagora/2013/09/how-to-write-an-appeal-letter.html.

[101] "If your manuscript is rejected, and if you believe a pertinent point was overlooked or misunderstood by the reviewers, you may appeal the editorial decision by contacting the editor. If you appeal to the editor and are not satisfied with the editor's response, the next step in the APA editorial appeal procedure is to contact the APA Chief Editorial Advisor. If a satisfactory resolution is still not achieved, and you still believe that key factors have been overlooked in the review process, you may appeal to the Publications and Communications (P&C) Board." http://www.apa.org/pubs/authors/appeals.aspx

[102] "The results of original research are customarily published in the *primary literature* of science […] This consists mainly of `papers' […] A scientific paper is the contribution of a a named individual (or group of individuals) to the communal stock of knowledge." Ziman 2000, p. 34.

# Chapter Two: History

As we saw in the last chapter, the practice of peer review can vary quite widely between disciplines, but also between journals and publishers inside of those disciplines. The historical development of the *de jure* protocols and the *de facto* practices of peer review are of interest in this project inasmuch as they were adopted for reasons that bear philosophical analysis. One should not ignore practical contingencies as a motivator for developing procedures, but that is not our focus here. Rather we wish to examine the philosophical motivation, implications, and consequences of variations on the theme of peer review. For instance, the adoption of anonymity for referees, and sometimes authors, was done in service of making the peer review process more objective. This will be discussed more in Chapter Four. Additional concerns brought out by considering the history of peer review includes the changing nature of expertise, with the attendant boundary-creation of who counts as a peer, to be discussed further in Chapters Five and Six. And, understanding the history of journals is vital when one seeks to grapple with the Open Access question, which in itself addresses notions of expert communities and who science, as expressed in publications, is really for.

In attempting to discuss the evolution of peer review from its beginnings to what we would recognize as such today, it is almost as if we are back to the era of whig history of science. Many accounts of peer review draw a straight line back to the establishment of the *Philosophical Transactions* in 1665 and confidently declare Henry Oldenburg as the originator of peer review. However, as historians of peer review have shown, the project of fixing the "start" of peer review depends a lot on what markers one adopts. Unlike for history of science, there has been nothing equivalent to Science and Technology Studies (STS) with its various schools of thought in the consideration of the history of peer review.

To give an example of what the shift in approach in STS represented: in his most famous book, *The Structure of Scientific Revolutions*, Thomas Kuhn described how physics had historically been presented as an iterative enterprise of adding knowledge to an even-expanding pile. Textbooks on Newton focus on his work on the laws of mechanics, and not his corpuscular theory of light, never mind his extensive theological writings. All figures are drafted in as signposts on the way to the current state of enlightenment, erasing the inconvenient bits that do not fit the narrative. The emergence of Science and Technology Studies (STS) complicated this image of steady progress, and Kuhn popularized the term "paradigm shift"[103] to describe a fundamental change in the practice of a scientific discipline, such as the change from Newtonian to Einsteinian dynamics, the discovery of oxygen, or the germ theory of disease. The study of history of science is further complicated by the invention of new terms, and repurposing of old ones. Though people can certainly have been said to have done investigations of the natural world from quite early on, we cannot base a history of science based on additions of knowledge alone. What is required to have something that we can call science is something more akin to a process than individual achievements. The start time of what can really be called a scientific process is however up for debate. Looking rather to the emergence of social systems, David Wootton,[104] taking astronomy as an example of a scientific discipline, fixes the beginning of a modern science between 1572 (Tycho Brahe's observation of a nova) and 1704 (the publication date of Newton's *Optics*). Wootton fixes this span by paying attention to a number of criteria, such as having a research program, a community of experts, and a willingness to "question every long-established certainty"[105] What is also needed is to have a sense of priority and how to claim it, which is now through publications, which are subject to peer review.

---

[103] Kuhn, Thomas S. 1962. *The Structure of Scientific Revolutions, 3rd Edition by Kuhn, Thomas S. Published by The University of Chicago Press 3rd (Third) Edition (1996) Paperback*. New ed of 3 Revised edition. University of Chicago Press.

To be fair, questions of "is this/is this not peer review" are common in the study of the history of peer review, but there is nothing in history of peer review that compares to the methodological shift in STS. While STS has shaken up the traditional view of the "beginnings of science", historical accounts of peer review seem committed to a linear model of progress towards the present day. Many works on peer review seem to start from the present day, a varied picture for sure, but mostly variations on the tripartite scheme of authors, referees, and editors that we encountered in chapter one.

A seminal text that has influenced the study of peer review was written not by historians, but by the sociologists Harriet Zuckerman and Robert K. Merton, "Patterns of Evaluation in Science: Institutionalisation, Structure and Functions of the Referee System."[106] This treatment has proven very influential in the study of how science progresses, but it has also inadvertently succeeded in instilling in people's minds the idea that peer review starts with Henry Oldenburg at the Royal Society in 1665, with the foundation of the *Philosophical Transactions*. Henry Oldenburg was indeed the person responsible for *Philosophical Transactions*, as the Royal Society's Secretary, however, he ran it very much as a personal project,[107] not consulting experts or even engaging in dialogue with the authors about their submissions.[108] Oldenburg's protocol

---

[104] Wootton, David. 2015. *The Invention of Science: A New History of the Scientific Revolution*. London: Allen Lane an imprint of Penguin Books.

[105] Wootton p. 1.

[106] Zuckerman, Harriet, and Robert K. Merton. 1971. "Patterns of Evaluation in Science: Institutionalisation, Structure and Functions of the Referee System." *Minerva* 9 (1): 66–100. https://doi.org/10.1007/BF01553188.

[107] "*Transactions* under Oldenburg looked very different from a modern science journal, and also from the formal learned society publication it would become in the eighteenth and nineteenth centuries. There was no formal submission process, and Oldenburg was the publisher, compiler, and even—as he occasionally called himself—the author." Fyfe, Aileen, Julie McDougall-Waters, and Noah Moxham. 2015. "350 Years of Scientific Periodicals." *Notes and Records of the Royal Society of London* 69 (3): 227–39. https://doi.org/10.1098/rsnr.2015.0036, p.230.

[108] "Close examination of Oldenburg's practices as editor of the *Transactions*, from 1665 until his death in 1677, reveals how different his role was from that of the modern scholarly journal editor. He did not receive submissions from authors and choose among them on the basis of intellectual merit, let alone engage in systematic

bears very little resemblance to what we think of as peer review these days, and developments of submissions committees and refereeing systems followed in the later history of scholarly publication.

As seen in chapter one, Steven Shapin and Simon Schaffer did their part in propagating the myth of Oldenburg in *Leviathan and the Air Pump*, where they create an account of the witnessing of scientific results as being essential to their acceptance into the scientific corpus, with particular attention paid to "virtual witnessing", through publications, and trace it back to the same period as Zuckerman and Merton did.[109] Charles Bazerman names Henry Oldenburg as the founder of the first scientific journal in English, and credits him with bringing together "a previously dispersed scientific community, which had communicated primarily through books."[110] In support of the notion of 17th century peer review, Mario Biagioli describes an instance of a form of peer review of this time, "For instance, in 1671 Nehemiah Grew's manuscript of Anatomy of Plants was given to Henri (sic) Oldenburg (the secretary of the Society), who read it and passed it to another member, John Wilkins, who after reading it gave a very positive report to the whole Society urging them to read it too."[111] Michael Mabe in 2009 repeats this assertion, "*Philosophical Transactions* from its outset did not publish all the material it received; the Council of the Royal Society reviewed the contributions sent to Oldenburg before approving a selection of them for publication. Albeit primitive, this is the first recorded instance of 'peer review'."[112] There are many more examples of this assertion.[113] Authors mostly acknowledge that the practical details of peer review have changed substantially since the 17th century, but the claim of lineage is nevertheless present.

On the other hand, it can also reasonably be said that Oldenburg's was not the first instance of peer review. Mario Biagioli claims that while several accounts point to quality control as the motivator for the foundation of peer review, for early academic publications in the 17th century, the focus was not really on quality control as much as to avoid publishing anything that would offend the church or the state. If one restricts one's definition of "peer review" to a work being in some way reviewed by one's peers, then Mario Biagioli's argument, that the birth of peer review coincides with the foundation of the Royal Academies of England and France in the 17th century,[114] is on strong footing. This privilege, granting  the Royal Academies the right to publish

---

their own works, was "an extraordinary exception from the licensing and censorship systems" that had been established since the popularization of the modular-type printing press. These systems, established by the political and religious authorities, were put in place in response to perceived political and religious dangers from printing, which somewhat contradicts the origin story with Oldenburg the gentleman scholar collating information he found useful for the Transactions for no motivation other than knowledge. The novelty was not that the academies were exempt from government control, but rather that they were allowed to administer such controls themselves, which can be said to be an early instance of peer review. Even then, Biagioli points out that "[t]he logic of peer review, however, predates royal academies and can be seen at play in earlier private academies and religious institutions — institutions that had not received special royal dispensations about publishing."[115] In addition, Ray Spier cites an instance of a peer review procedure from 9th Century Syria, in which doctors would write up notes of their treatments for the judgment of their peers.[116] Depending on one's definition of peer review, one can find instances dating even earlier. Casting the net wide and searching for examples of "scholarly review" that nevertheless match the criteria laid down for peer review, Mark Hopper identifies examples as early as 4th Century BCE.[117]

However one would classify these early attempts at judgment through peer review, it seems unlikely they would pass muster as a systematic approach of evaluation, sufficiently durable to assume the status of an Institution, as it seems peer review is today. In their account of the publication practices of the *Philosophical Transactions*, Noah Moxham and Aileen Fyfe refer to the period they cover as the "pre-history" of peer review. While not a stand-in for all scientific journals, *Philosophical Transactions* is a good case study for historians because of its long history as the world's oldest running scientific journal, and its rich archives. Many other journals, though well established in years, have not maintained correspondence archives to the same degree, so a narrative of evolving editorial practice is hard to establish.

Moxham and Fyfe point to distinct turning points in peer review, drawn from the history of the *Philosophical Transactions*. While critical of the picture of the origins of peer review with Henry Oldenburg, they acknowledge that Zuckerman and Merton's account recognizes that the review system evolved, and that today's peer review did not appear fully formed. However, this nuance did not apparently propagate as much as the Myth of Oldenburg, possibly because the Zuckerman and Merton account was thin on details on the intervening years[118] between the 17th Century and present day. In parallel, Melinda Baldwin[119] observes that this account has become so ubiquitous that it is now commonplace to see pieces on peer review starting with a confident declaration that peer review as we think of it today owes its origins to Henry Oldenburg. Moxham and Fyfe, Baldwin, and others draw a useful distinction between peers commenting on each others work and the much more systematic forms of peer review that would come later. This is philosophically significant because peers commenting on each others work is not the same as the systematic consultation of independent referees. What an independent referee does is to put themselves in a position of being a skeptic about a paper, and as we will show later this role of the referee only emerged later, in the 19th Century.

---

[115] "For instance, when the Roman Accademia dei Lincei (a private academy) decided to publish and endorse Galileo's Assayer in 1623, it circulated the manuscript among several of its top members to read and edit it before it was submitted to the censor for the imprimatur." Biagioli, note 10, p. 37.

[116] "Perhaps the first documented description of a peer-review process is in a book called *Ethics of the Physician* by Ishap bin Ali Al Rahwi (CE 854–931) of Al Raha, Syria. This work, and its later variants or manuals, states that it is the duty of a visiting physician to make duplicate notes of the condition of the patient on each visit. When the patient had been cured or had died, the notes of the physician were examined by a local council of physicians, who would adjudicate as to whether the physician had performed according to the standards that then prevailed. On the basis of their rulings, the practising physician could be sued for damages by a maltreated patient", Spier, Ray. 2002. "The History of the Peer-Review Process." *Trends in Biotechnology* 20 (8): 357–58. https://doi.org/10.1016/S0167-7799(02)01985-6., citing Al Kawi, M.Z (1997) History of medical records and peer review. *Ann. Saudi. Med.* 17, 277–278 and Ajlouni,K.M.andAl-Khalidi,U.(1997), Medical records, patient outcome, and peer review in eleventh-century Arab medicine. *Ann. Saudi Med.* 17, 326–327.

[117] "My argument is not merely that the general spirit of scholarly review is old. Argument, feedback, dialogue, and criticism are obviously ancient. Plato is sufficient proof of the long history of those practices. I make the stronger claim that there existed: 1) organized systems for facilitating review by peers; 2) in the context of publishing practices; 3) to improve academic works; and 4) to provide quality control for academic works long before the first scientific journals of the seventeenth century." Mark Hooper, "Scholarly Review, Old and New," *Journal of Scholarly Publishing* 51, no. 1 (October 2019): 63–75, https://doi.org/10.3138/jsp.51.1.04.

One development of interest in peer review is its transformation into a collective endeavor. While judgment on what to publish and not was always present, for labor and financial concerns if nothing else, the idea of accountability of process is one that emerged later. Moxham and Fyfe describe the formation of a "committee of papers" in the Royal Society, who would shoulder the editorial responsibility of selecting contributions for the *Philosophical Transactions* instead of the Secretary-Editor, a role inaugurated by Oldenburg. David Kronick, also writing about this development, argues that peer review proper started when the Royal Society, having taken financial responsibility for the Philosophical Transactions nearly a hundred years after its foundation, formed this "Committee on Papers" that had the role of reviewing all submissions.[120] The new statutes, "laid down that the committee should consider all papers communicated to the Society, in the order in which they had been read at meetings. Committee members met roughly every six weeks, were furnished with abstracts of the papers on which to base their judgements, and were supposed to reach their decision by secret ballot without discussion."[121] Certainly this was systematic, though the review was still internal and did not involve a third party. Kronick also has an even earlier 18th Century data point, The Royal Society of Edinburgh, which introduced a system of review in 1731, with the stated policy that "Memoirs sent by correspondence are distributed according to the subject matter to those members who are most versed in these matters. The report of their identity is not known to the author. Nothing is printed in this review which is not stamped with the mark of utility.[122]" In addition, Kronick discusses the practice of peer review emerging at societies such as the Académie Royale de Médecine, which adopted similar practices in 1782, with responsibility for review falling on the society officers of the society and four members elected at large. An additional example offered is The Literary and Philosophical Society of Manchester, which had its own Committee on Papers established in 1785, consisting of "the president, vice president, secretary, treasurer, and librarian along with six other members elected at large."[123] These publishers, societies all, adopted these procedures to safeguard their reputation, also frequently publishing disclaimers disavowing responsibility of the contents, much like the Royal Society did[124]. A similar, and earlier, scheme of peer review is described by Alex Csiszar as emerging in the Royal Academy of Sciences in Paris in 1699 with the introduction of early reviewers, called rapporteurs, in the publishing system of the Academy. "Decisions about the value of new inventions, as well as about the suitability of scientific and technical memoirs, came to involve the writing of reports by commissions consisting of two or three academicians. These would be read out at meetings and recorded in the minutes, and some would even find their way into print. The collective authority of the Academy was increasingly concentrated in these judgments."[125] Unlike the Royal Society's Committee of Papers, these reports were public. The Royal Society of Edinburgh also had statutes that "would explicitly allow committee members to discuss the merits of the papers."[126] Bazerman, for his part, credits Oldenburg as an innovator, but also makes a claim about "real" peer review starting with the Committee on Papers, which Bazerman equates to a form of refereeing.[127] However, as Bazerman also points out, this innovation did not have referees as entirely separate persons from the journal's editors, which problematizes the assignment of the shift from unilateral editor to committee as the historical start of peer review. A committee of papers or of readers is more similar to the thought collective discussed by Fleck (see chapter one), but it cannot be described an entirely separate actor in the process.

---

[118] "They recognized that 'the referee system did not appear all at once' but 'evolved'; however, their discussion of the early Royal Society (reflecting Merton's earlier work on science in seventeenth-century England) was followed by a leap to the twentieth century, thus resulting in their paper being widely cited to support the invention of peer review in 1665." Moxham and Fyfe, p. 864.

[119] "This version of peer review's history has been widely circulated and repeated in other scholarly papers, creating the pervasive impression that refereeing has been a part of science ever since the first scientific journal was created." Baldwin 2018, p. 540.

[120] "The Royal Society of London is frequently assigned the credit for having introduced the concept of refereeing or reviewing scientific manuscripts for publication in 1752. At that time the society finally, after almost 100 years of existence, took over the fiscal responsibility for the Philosophical Transactions. The society established what they called a "Committee on Papers," whose function it was to review all articles that were published in the Transactions. Before this the selection of articles had been nominally the responsibility of the respective secretaries of the society, starting with its first editor, Henry Oldenburg." Kronick, p. 1321.

[121] Moxham and Fyfe, p. 871.

---

[122] Kronick, p. 1321.

[123] Kronick, p.1322

[124] Moxham and Fyfe, p. 869.

[125] Csiszar, Alex. 2018. *The Scientific Journal*. University of Chicago Press, p. 29.

[126] Moxham and Fyfe, p. 871.

[127] "Through this innovation, the Royal Society established the role of editorial board cum referee. The editorial function was maintained and strengthened by removal of the responsibilities from any one individual's hands. In order to maintain authority, the editor cannot be perceived as exercising it, but rather must take a distanced stance on all decisions which might be likely to be perceived as injurious to others, The invention of editorial boards to handle issues of general policy and of referees to handle issues concerning individual contribution not only helps the editor maintain authority and trust by assigning responsibility to other individuals, but it further allows the journal to establish a corporate identity, representing the field as a whole. Perceived scientific eminence of editorial board members and referees, as well as distribution among the various subcommunities of the larger scientific community, help maintain the authority of the journal as an institution through the appearance of fairness and generalized competence." Bazerman, p. 137

What motivated the turn towards these kinds of committees of papers? In the case of the Royal Society, Moxham and Fyfe reveal that this was motivated by criticism of the Society. After the death of Oldenburg in 1677, the *Transactions* had been edited by a succession of Secretary-Editors, many bankrolling the journal from their own personal funds.[128] Criticism of the manner in which articles was selected was such that the Society enacted these reforms to protect the Society's reputation.[129] We see that after 1752, Transactions was explicitly defined as being run "for the sole use and benefit of this Society". However, the procedures put in place by this process bore few similarities with today's manifestations of peer review. The newly-formed Committee of Papers was more concerned with stopping the publication of clearly unsuitable papers than to do any real filtering for quality. The new rules did lend some legitimacy to the practices of *Transactions*, but another filtering mechanism was also in effect prior to the meetings: papers were only read at meetings if "communicated" by a fellow. The selection process for the meetings was still opaque, and decisions whether to present papers submitted were done by the president and officers, not through the peer review triad. Records indicate they often sought unofficial advice on this pre-screening selection, a process much less fixed and documented than the selection done at meetings. Because there was a certain unwillingness to reject papers that had been read at meetings, limiting the size of Transactions, with significant per article printing costs, was done through this prior selection by the president and officers.

Filtering or not, *Transactions* presented itself as unwilling to adjudicate on knowledge claims, another difference from today's peer review. The committee of papers was not comprised of experts. It was one thing to filter out anything that was trivial or clearly wrong (e.g. designs for perpetual motion machines), but submissions that did not fall into this category were given the benefit of the doubt, with disclaimers. The Society explicitly distanced itself from the *Transactions*, going as far as stating that they would not give their opinion as a body on any work published. Publication simply meant that the committee had found this interesting and significant, but not that the committee vouched for the "certainty of facts, or the propriety of the reasonings"[130] The Transactions were not intended as a repository of "officially sanctioned knowledge", but instead of interesting results. Since peer review is a part of a knowledge

---

[128] "Following Oldenburg's death in 1677, the *Transactions* was edited for seventy- five years by the secretaries to the Society. Few observers recognized that the editors were acting in a private capacity, not least because the content of the *Transactions* became increasingly identified with the activity of Society meetings. This left the Society vulnerable to the imputation of failing to enforce adequate standards in the *Transactions*, yet with no obvious means of exercising control, and little hope of being believed when it tried to deny responsibility" Moxham and Fyfe p. 870.

[129] A particularly persistent critic, John Hill, a botanist, but also actor and apothecary, who had failed in his candidacy for fellowship, entered the debate in the early 1750s. Hill went on the offensive against the Society. Because the Society was linked to the *Transactions*, albeit only loosely, Hill was able to criticize previously published papers that he considered particularly weak. His criticisms ranged from satirical to accusations of cronyism, as well a more scientific criticism. He published his critique as A review of the works of the Royal Society, doubling down on the link between the Royal Society and the *Transactions*. Moxham and Fyfe, p. 870.

[130] Moxham and Fyfe, p. 873.

production process, this is quite a philosophical difference from today's state of affairs.

The unwillingness to vouch for the validity of papers remained the official editorial stance and the function of *Transactions* until the early 19th century, when again criticism of the Society led to a change in protocol. In this case, this was the introduction of external refereeing. In November 1832 the duke of Sussex, by now president of the Society, announced a change of practice. From this point on, a paper would be approved for *Transactions* only if "a written report of its fitness shall have been previously made by one or more members of the Council, to whom it shall have been especially referred for examination."[131] The Royal Society now cites 1832 as the invention of refereeing. This development, which had actually started in 1831, was inspired by similar systems already in place at the French academies.[132] While the committee of papers had been explicitly permitted to solicit the advice of experts on matters on which the committee felt unqualified to rule, only five instances of such consultations are recorded in writing between 1780 and 1815. A number of more informal consultations could obviously have taken place, but nevertheless, the introduction of a systematic requirement of consulting external referees formalized the third part of the "Ziman triangle" as seen in chapter one, and is thus much closer to our present state of affairs.

The Royal Society is of course only one instance of the development of a system of evaluation. In the commercial press, publication by editorial fiat was certainly considered as valid for well into the 20th century. Customs and protocols about publication also varied between different countries, as we saw in the introduction with the discussion of Einstein's battle with the *Physical Review*. It is interesting to note that refereeing for the *Transactions* continued when commercial journals apparently spurned the practice entirely, for speed of publication.[133] In the early 20th century, refereeing was seen as an "obsolete holdover"[134] from an era dominated by amateur scientists rather than the newly emergent professional class. The cumbersome peer review practiced by societies was not a tempting strategy for commercial journals, who treasured their nimbleness and flexibility. For them, consultation could certainly take place, but it would be more informal, involving trusted acquaintances, hearkening back to the procedures of Oldenburg centuries earlier.[135] The history of the Transactions thus emerges more as an outlier than a

---

[131] Moxham and Fyfe, p. 874.

[132] Csiszar, 2016, p. 142.

[133] "Compared with the long, labour-intensive, and comparatively inaccessible publishing processes at learned societies, the swift editorial decision-making and more rapid publishing frequency of the independent journals made them attractive to authors looking to publish quickly, especially in fast-moving fields like physics." Moxham and Fyfe, p. 884, see also Melinda Baldwin. 2014. "'Keeping in the Race': Physics, Publication Speed and National Publishing Strategies in Nature, 1895–1939." *The British Journal for the History of Science* 47 (2): 257–79. https://doi.org/10.1017/S0007087413000381.

[134] Moxham and Fyfe, p. 884.

[135] "Independent journal editors could follow their own instincts and interests, with no need to represent or protect the corporate reputation of a sponsoring organization through mechanisms for collective responsibility. Their desire for speedy publication was better served by making executive decisions than by seeking referees' reports. Thus, in the early twentieth century, the practice of refereeing could be seen, in some quarters, as an obsolete holdover from an age of amateur dominance, out of touch with the needs of the new professional scientist – a remarkable transformation from the 1830s, when refereeing had been one of the chief demands of a reform

signpost of the evolution of peer review.

However, refereeing started to take hold in post WWII scholarly publication. There was not a uniform adoption of this practice. As Melinda Baldwin relates, the Editor of *Nature*, John Maddox, would publish some papers based on his personal evaluation, not involving others, as late as the early nineteen seventies. It was only in 1973 that his successor David Davies made external reviewing a prerequisite for publication in *Nature*. "When Davies arrived at *Nature*, one of his first goals was, as he put it, 'getting the refereeing system beyond reproach.'"[136] This change was in response to criticism that *Nature* was a journal of the British establishment, and as a result procedures that could raise the specter of cronyism and differentiation of treatment were eliminated.

Differently from the practices at *Nature*, a separate development from society journals and commercial journals was taking place in the funding agencies of the United States. Under pressure from increasing government oversight, agencies began to formalize procedures for peer review of grant applications after the second world war. Such practices propagated through to other publishers, society publishers as well as commercial ones, with a process that was employed for cause transforming into standardized practice[137]. The lack of comprehensive archives makes it unclear to what extent other publishers employed refereeing, but some correspondence has survived to indicate the practice was alive at the *Physical Review*[138] at least by the 1930s.

Arguments by scholars of peer review often declare that peer review is achieved or improved by some criterion being met. In order to understand what constitutes "proper" peer review, the scheme in question is held up to some standard, but this standard is often not well defined. From reading one gleans that the standards involve a) having clear criteria, b) being systematic, c) involving outside experts, d) having oversight whether by a board or a funding agency, and other markers that are put in place with the aim of increasing rigor. What we seem to have is a linear picture with complications, occasional flashpoints which lead to a revision of standards, but does not fundamentally change the idea that some kind of evaluation needs to happen by peers (editors, referees, committees…) ahead of publication.

If one goes deeper into the history of peer review, one notes that these moments: all-powerful editor, committee, referees, anonymity, etc. do not necessarily occur in a determined sequence and occasionally go back and forth. For instance, just as refereeing systems were on the rise and becoming increasingly established in the UK and the United States, academic reports on manuscripts diminished in significance and in frequency during this same period, as the unrefereed "Proceedings"-style journal *Comptes Rendus* came to be the most prominent

---

movement that championed the expansion of professional science and the imposition of more stringent qualifications upon men of science." Moxham and Fyfe, p. 884.

[136] Baldwin 2015, p. 346.

[137] For more on the evolution on review systems, see for instance Baldwin 2018, particularly 540-544.

[138] Lalli 2016

publishing output of the Académie des Sciences[139]. Various trends such as making authors also anonymous passed in and out of fashion with the changing of editors in chief at the *American Economic Review*, which made four procedural changes in the period between 1973 and 2011[140]. Other new initiatives occur all the time. For this reason an authoritative single "history of peer review" cannot really be written. The definitions of the terms in use are too variable, however, these variable terms can still be associated with philosophical ideals, such as objectivity.

As we have shown, to really understand the motivations for all these procedural changes described by historians, and to account for them in anything other than a contingent manner determined by cultural, political, and social circumstances, it's important to evaluate concepts that underlie the rationales for peer review and developments of peer review systems. At all times systems have been elaborated to further the credibility and legitimacy of the institution of peer review, often by claiming the mantle of objectivity, which we will explore more in the next chapter.

---

# Chapter Three: Objectivity

In order for science to work, it has to work everywhere. As commonly described, this is a requirement that science be objective, in the sense here of being the same no matter who encounters it, representing matters of fact about the world that are the same independently of us.[141] Objectivity can also mean that scientific results should not be affected by personal preferences, community attitudes, anything other than the plain facts. As a concept, objectivity is a value. As a society, we in the West approves of objective data, objective stances and objective processes. Objectivity is a thick ethical concept, which is a concept that has a descriptive value as well as a moral or normative value. This concept was coined by Bernard Williams in *Ethics and Limits of Philosophy* in 1985, referencing the already existing concept of a "thick description" as elaborated by Clifford Gertz, as discussed in Chapter One. Williams defines a thick concept with examples like coward, lie, brutality, gratitude, and claims that these are related to "reasons for action"[142] or at the very least "action-guiding". We do not feel neutral towards a thick ethical concept, whether good or bad, but rather make a moral judgment about the person to which it is applied. Thick ethical concepts exist between thin descriptive concepts like square or yellow, and thin normative concepts like good, bad, right, or wrong, which in themselves do not provoke a moral reaction in the absence of specifics. This designation of objectivity as a thick ethical concept does not tell the full story, for instance in that if we say that someone is a "good author" or a "good referee" we do make a normative judgment on that person in that role, but considering objectivity as a thick ethical concept starts us off on a path to understanding why it is so valorized.

Philosophically, there are several different ways in which objectivity is used to describe something. One way in which the word is used is to indicate that our theories and findings correspond to matters of fact about a shared external world.[143] The results theories and experiments produce can thus be said to be objective knowledge, which we can term *product objectivity*. Objectivity can also be thought of as a characteristic of beliefs, individuals, theories, observations, and methods.[144] It is perfectly possible to doubt (or leave as unknowable) that our theories have correspondence with reality and still believe that the scientific project is in some regard objective. In *Real Science*, John Ziman writes that "…the relative `objectivity' of physics is not due to the fact that it describes the world `as it really is'. It is because […] physics has evolved as an epistemic culture devoted to the measurable aspects of the world."[145] This could be referred to as *process objectivity*. We will return to this distinction later in this chapter.

---

[141] "According to this norm [of disinterest, see Chapter One]. it might seem that scientific knowledge should always be presented as cognitively *objective* – i.e. as referring to entities that exist quite independently of what we know individually about them." Ziman 2002, p. 155.

[142] Williams, Bernard. 2015. *Ethics and the Limits of Philosophy*. London: Routledge, p. 141.

[143] For examples of this sense of objectivity see for instance Nagel, Thomas. 1989. The view from nowhere. New York: Oxford Univ. Press., p. 14-17, and Williams 2015, p. 139.

It follows that if science is supposed to be objective, whether as product or as process, so should also the fruits of scientific efforts, i.e. results. The mechanism by which we bring accepted scientific results into the world is publication, so, to fulfill its function, that process would then also have to be objective, the argument goes. There is thus a prevalent idea that peer review should, like science, be "objective", and as such focus on the science in a submission rather than other criteria like relative importance and breadth of interest. These criteria are often described as being more subjective than ones that concern the validity of a submission, as we saw in the previous chapter on editorial criteria for publication. Other subjective assessments take the form of biases, taking into account factors such as the personalities involved, the reputation of the authors, and the individual tastes of editors and referees. As Ziman observes, the stance of objectivity is intended to protect us from subjective influence, writing about disinterestedness as a social norm that "functions primarily to protect the production of scientific knowledge from personal bias and other `subjective' influences."[146]

New approaches to peer review have emerged in order to address the concerns authors have with more conventional models of peer review. Examples need not necessarily entirely break the mold of "standard" peer review triad of author, editor, and referee. Rather, proposed changes to peer review often make tweaks to the existing system. New approaches span the whole spectrum from the extreme anonymity of triple-masked review, in which even the editors do not know the authors' identities when selecting referees, to fully open review in which all agents are known to each other, and the reports are published with the paper, as discussed in the previous chapter. But all suggested approaches have one thing in common, they are embarked upon in a desire to make the process more objective, with the paradoxical consequence that the same motivation applies whether the remedy is by making peer review more transparent as in open review, or more deliberately opaque for double- and triple-masked approaches.

Lorraine Daston and Peter Galison claim that objectivity as a positive value gained traction in the 19th century,[147] an attitude that Alex Csiszar, in 2015's "Objectivities in Print," links to emerging practices in academic peer review from the same period.[148] However, the history of

---

objectivity as a concept is not straight-forward. While the word objectivity is not used, Elizabeth Lloyd, in 1995's "Objectivity and the Double Standard for Feminist Epistemologies," draws an account dating back (at least) to Galileo, contrasting "primary" and "secondary" qualities. Primary qualities are described as absolute, constant, and mathematical, numbering extension, number, figure, magnitude, position, and motion. Secondary qualities, less valued, are described to arise from the senses, and are the effects of primary qualities. Even though the words objective and subjective are not used, this seems to encapsulate the dichotomy by another name.[149] Additionally, from a historical perspective, David Wootton's research reveals that the word "fact", in the sense of a state of affairs that can put an end to the most well established and revered theory that contradicts it, was something that emerged as a concept earlier than the 19th century, during the Renaissance.[150] Whether one calls it objectivity or something else, the division between what we today call objectivity and contrast with subjectivity has a long history. The above are just some accounts to indicate that many different accounts of objectivity exist. In a sense, the thing that brings them all together is a suspicion of the subjective. However, what Daston and Galison show on a move in the reach for objectivity as the negation of self. The worry about the subjective coincides with the increased use of the terms objective and subjective in the 19th century, as exemplified by their study of scientific image making.[151]

Daston and Galison charts the adoption of mechanical means for image production, replacing human illustrators who were suspected of inserting their own interpretation into images. Turning this analysis towards academic publication, Alex Csiszar draws an analogy with the implementation of systems such as the systematic consultation of external experts as well as the masking of the identities of referees as other moves to remove the subjective in favor of a more objective evaluation. This may be described well by making an analogy with mechanical objectivity, and as we recall, a system or process can be seen as a technology. But what is it that people mean when they ask for an objective peer review process? Is it increased mechanization, enhanced protocols, or something else? Since the participants in peer review are scientists, it is reasonable to assume that their notions of objectivity in peer review are similar to the notion of objectivity in science and other forms of research, which means that we have to take a hard look at what objectivity means.

There are several ways of looking at objectivity. Writing in 1992, Lorraine Daston notes that "We slide effortlessly from  statements about the 'objective truth' of a scientific claim, to those about the 'objective procedures' that guarantee a finding, to those about the 'objective manner' that qualifies a researcher."[152] As Daston goes on to enumerate, the word can be applied to signify the empirical or factual, the idea of some common, public repository of knowledge, the removal of self and its inconvenient emotions, and for the kind of rationality which demands the assent of all logical minds, and additionally to the idea of things in themselves independent of all minds. This long list corresponds to different conceptions of objectivity, but they are not all the same. I propose that objectivity as relevant to peer review can most usefully be thought of as corresponding to three types: objectivity as it relates to an external world of objects that can be investigated; objectivity as a description of the way in which we as individuals or as communities go about performing these investigations; and objectivity in the sense of our own personal manner of engaging with the world, for instance with our biases and prejudices.

## Objectivity as realism

The first conception concerns a correspondence to an external reality. Much has been written about this kind of objectivity, about to what extent we can gain knowledge of the real world, to what extent our theories can be thought of as "true", and to what extent entities in the world are "real". Per John Ziman, scientific realism signifies a belief in public invariants, being the implication that "some of the features on the mental `maps' on which we individually base our actions are cognitively objective."[153] This is what scientific realists refer to as an external world.

The problem with making these kinds of bold pronouncements is skepticism, which is another scientific value that we have previously discussed. There is a division between a pragmatic instrumentalism, which takes frequently and reliably observed events such as the predictable falling of bodies as evidence for the functioning of science, and constructivism, which holds that since all scientific knowledge is a human construct, and therefore a fabrication, scientific results are as much 'discovered' as 'made'.[154] But regardless of which school of thought one belongs to, there is an argument that this debate can be bracketed for the purpose of this study of peer review. While these are interesting questions, once we establish that we observe that there is a thing we call science, and scientific practice, then we can argue that the practitioners of this science do bracket any concerns they might have about realism while they are in the moment of doing science. As an example, in *Representing and Intervening*, Ian Hacking remarks that scientists don't doubt the objective existence of electrons when they can use them to reliably produce images of entirely different things with an electron scanning microscope. As he puts it, "Experimenting on an entity does not commit you to believing it exists. Only manipulating an entity, in order to experiment on something else, need do that."[155] This bears some resemblance to the treatment by Bruno Latour in 1987's *Science in Action*. For Latour, once a concept becomes established theoretically, Latour envisages it as a ``black box'' -- a plug-and-play concept ready to be slotted into its place in the scientist's conceptual machine. Taking as examples the discovery of the structure of DNA and the development of the Eagle microcomputer, Latour draws a distinction between knowledge as objects of study and as tools. "Learning to use the double helix and the Eagle to write programs reveals none of the bizarre mixtures they are composed of; studying these in 1952 or 1980 reveals it all."[156] Likewise, John Ziman argues that scientific knowledge is very intermingled with what he calls "common sense" in that scientific knowledge "cannot be supposed to be `unreal' without supposing the same of the whole life-world."[157] If scientific knowledge were unreal, then we cannot take for granted that other people experience the same world as us. A generally realist approach is thus necessary in order for science to be a rational pursuit. "In effect, the norms of academic science require scientists to behave as if they believed in a shared external world which is sufficiently uniform that they can usefully exchange information with one another about it. One might even say that they do this as a matter of policy, but it becomes so ingrained by training and practice that it is completely `taken for granted.'"[158] The same, for Ziman, applies to scholars in the humanities and social sciences, "even when they are remarking on the influence of ideological constructs, reified ideals, and other unreal entities in social life!"[159]

---

[153] Ziman 2002, p. 316.

[154] *Ibid.*, p. 318.

[155] Hacking, Ian. 2010. *Representing and Intervening: Introductory Topics in the Philosophy of Natural Science,* p. 263.

This notion of the limitations of objectivity regarding realism also carries over to peer review. It is not rational to be writing a paper detailing discoveries if one is not taking for granted that the reader is also in the world being described. It is not rational to review a paper if one doubts the underlying existence of a shared world that can be described, in some manner, through the procedures described in a paper. The very existence of scientific publication seems to be evidence to show that, for operational purposes, scholars believe in an external world about which things can be learned.

## Process objectivity

Even if the question of realism can thus be sidestepped, there still is the question about process objectivity. Even if we take an external world and the scientific project for granted, as Ziman advocates, our processes are still up for debate. Helen Longino[160] posits a kind of contextual empiricism to describe process objectivity, which is based on experience, but at the same time she makes clear that the relevance of particular experiences to hypotheses is mediated by background assumptions. These assumptions can play a role in reasoning, the plausibility of a hypothesis, in the choice of methods, what data is/is not relevant, and more. The background assumptions are generated and controlled by the interaction amongst scientists through what

---

[156] Latour, Bruno. 1987. *Science in Action: How to Follow Scientists and Engineers through Society*. Cambridge, Mass.: Harvard University Press, p. 7.

[157] Ziman 2002, p. 318.

[158] *Ibid.*, p. 319.

[159] *Ibid.*, p. 320.

[160] Longino 1990; Longino, Helen E. 1995. "Gender, Politics, and the Theoretical Virtues." *Synthese* 104 (3): 383–97.

Longino calls transformative criticism. For Longino, successful criticism must come from a variety of perspectives, ideally as many as possible.[161]

The background assumptions should not be taken to be all bad, since they are part of what makes a scientist. One of the consequences of Longino's picture is that the same mechanism (criticism) "accounts for both the suppression and the expression of social values, interests and ideology in the sciences. Idiosyncratic values are suppressed, while values held by all members are invisible (as values, interests, or ideology)."[162] The second consequence is that the producer of knowledge is the community, and not the individual scientist, through mechanisms like peer review, which Longino holds happens in places far beyond the consideration of the paper in a journal, but also affects the choice of project, hypothesis, method, system, as well as the post-publication reception of the work, and what research builds on it. These are all part of community evaluation of research, which Longino claims must satisfy four criteria: "(1) there must be recognized avenues for the criticism of evidence, of methods, and of assumptions and reasoning; (2) there must exist shared standards that critics can invoke; (3) the community as a whole must be responsive to such criticism; (4) intellectual authority must be shared equally among qualified practitioners."[163]

These are the standards of judgment involved in peer review. Referees are consulted because of their standings as qualified practitioners who share epistemological standards with the editors and journals for which they review. While authors may object to individual items of referee criticism, they never claim that a referee does not have the overall standing to offer criticism. The referee may be inappropriate for the subfield, but usually this is not such an inappropriate choice that epistemological standards are not shared. The referee does not arrive with an underprivileged perspective, but rather very much as part of the "in group." An issue with referee selection is including people from unrepresented groups, but this problem has its origins in decisions far before the peer review process starts. Decision are made about who gets to be in the in group a long time before a paper is ever sent for review. We will revisit this question later in this chapter when we discuss bias.

Heather Douglas[164] notes that invocations of objectivity are not just endorsements that contains an imperative of agreement, but carries meaning with it. To unpick those meanings she lays out a mapping of senses of objectivity. She identifies eight senses of objectivity that contain a way of telling whether the term objectivity really applies, a distinction she terms "operationally accessible." Operationally accessible conceptions of objectivity stand in contrast to, for instance, a claim of knowledge that is outside of all human experience and that is independent of human thought, such as religious experience. She divides up her senses of objectivity into three classes: objectivity-1: processes where humans interact with the world; objectivity-1: focused on human reasoning, and the role of values; objectivity-3: focused on the social processes that structure epistemic procedures.

---

[161] "What controls the role of background assumptions is interaction among scientists, interaction consisting in criticism of assumptions involved in observation, of assumptions involved in reasoning, of assumptions involved in thinking a given hypothesis plausible, of assumptions involved in the application of particular methods to the solution of particular problems. To be successful in uncovering such assumptions, criticism must proceed from a variety of points of view, ideally as many as are available." Longino 1995, p. 384.

[162] Longino 1995, p. 384.

[163] For to every one who has will more be given, and he will have abundance; but from him who has not, even what he has will be taken away. — Matthew 25:29, RSV.

[164] Douglas, Heather. 2004. "The Irreducible Complexity of Objectivity." *Synthese* 138 (3): 453–73. https://doi.org/10.1023/B:SYNT.0000016451.18182.91.

Inside of the first class, objectivity-1, Douglas distinguishes between manipulative objectivity, in which we interact with the world in a reliable fashion,[165] similar to Hacking's and Latour's examples mentioned above, and convergent objectivity, for which evidence from disparate approaches yield the same or coherent results, each boosting[166] the strength of the claim. Considering either of these approaches calls for an examination of the process by which one reaches one's conclusions, which informs the level of trust one can have in the objectivity of the result.

Objectivity-2 by contrast concerns individual thought processes, and here Douglas identifies three subcategories, The first, a) is a prohibition of using values in place of evidence. Just wanting something to be true does not make it so, and some distance (detachment) from the object of study is required. Douglas calls this Detached objectivity. A further subcategory b) is called Value-free objectivity, and is really expanded from type a, but represents a much broader sense of objectivity in which all subjectivities are banished. In this conception, values are held as inherently subjective things, which means that their role in a process contaminates it by definition. Here we see a difference between a and b in that b contains no differentiation between the values. In Value-free objectivity any value is a contaminant and should be treated all the same. Douglas identifies a problem with b. While it is irrational to ignore evidence, differentiation between errors in degree of severity is not irrational, and neither is choosing a particular field of study because of one's personal interests. Values can have a positive effect on the course of research, at least as articulated by Longino.

The third subcategory of objectivity-2 is c) Value-neutral objectivity, which acknowledges values but holds that we can overcome the effects of our values to attain objectivity by assuming a "balanced" position. This approach is particularly important for areas of study where values play important roles in making decisions, but where there is no clearly better value, this stance can be productive. Many value-neutral conflicts constitute current legitimate debates. However, Douglas highlights that a pose of value neutrality can come with the danger of reflective centrism, valorizing the mid-point between extremes, which would be inadvisable if e.g. racist and sexist values were at one extreme of the spectrum. Here advocating some middle ground comes in conflict with our other values, such as the wrongness of racism and sexism.

The third form of objectivity Douglas discusses is objectivity-3, which concerns the case when social processes are involved in knowledge production. Unlike objectivity-2, objectivity-3 always involves working with other people, and not just with one's individual thoughts. Douglas divides this category up in three parts as well, the first being a) Procedural objectivity. In this mode, social processes are said to be objective if the same outcome of a process is always produced regardless of the operator of the process. This sense of objectivity allows society to impose uniformity of processes. The identity of the individuals are interchangeable, in that the process should have the same outcome regardless of who performs it. This has the consequence of removing individual judgment. Douglas observes that a focus on this kind of rule-based objectivity lends credence to a field, but notes that "the elimination of idiosyncrasies, while raising public trust, does not ensure the elimination of values." Instead of individual values, the values are encoded in the process itself, a critique that can also be leveled at Daston and Galison's Mechanical objectivity.[167]

---

[165] "[Manipulative] objectivity is important outside of the laboratory as well as in it. When we can use objects around us, we trust our accounts of their existence and properties as reliable. If I can reach out and drink from the glass of water, and it quenches my thirst, and I can fill it back up again, repeating the whole process reliably, I have good reason to trust the reliability of relevant beliefs about the glass" *Ibid.* p. 457.

[166] Once weaknesses of this particular approach that Douglas highlights is that the independence of the approaches has not necessarily been established.

The second subcategory of objectivity-3 Douglas calls Concordant objectivity, which is often related to Interactive objectivity. Douglas describes that these two are often subsumed under the category of intersubjectivity. The first replaces individual judgment with a check on whether individual judgments of a group of people agrees, whereas the second requires a discussion process rather than the more detached polling for the former approach. Both of these approaches are vulnerable to the limitations imposed by shared values, with its attendant accusations of groupthink, or orthodoxy. Particular to interactive objectivity, a question concerns the choice of the participants in this discussion. For peer review, it is eloquently summed up by the question, who counts as a peer?

These approaches as described by Douglas do not always have sharp boundaries, but there are striking differences. Convergent objectivity-1 and concordant objectivity-3 differ in that one person can achieve convergent objectivity-1 by for instance using more than one experimental technique, but more than one person is required for concordant objectivity-3. Nothing about convergent objectivity-1 requires group activity, but it is essential to concordant and interactive objectivity-3. At the same time, Douglas finds it useful to invoke more than one sense of objectivity when used in practice. In an example that is particularly relevant for peer review, Douglas suggests that "A review report might be objective because the reviewer took pains to consider all the disparate research and perspectives on a contentious topic (value-neutral objectivity-2), and in a debate over the topic that followed, a group of experts came to similar conclusions and found, when they met as a review panel, they had nothing to add (interactive objectivity-3)."[168]

That being said, the outcome of the peer review process, the published article, can be viewed as being objectively true by the standards of the peer review process. Both senses of objectivity, product and process, are thus relevant here, but the product objectivity *follows* from the process objectivity, where an article is found valid and appropriate for publication through the peer review process. The peer review process starts with an assumption of a shared world about which factual statements can be made, and yields published articles that are a product intended to be objective by the standards of its process.

---

## Persona

The third part of the objectivity triptych is the stance of objective manner. Part of this question has already been covered in the discussion of the referee persona in the introduction. Robert Merton had previously declared that scientists are constrained to be disinterested,[169] in which he was referring to their professional behavior. The neutral, passive voice in which research is communicated is an affectation that implies the absence of a subjective experimenter and the presence of objective facts. Writing about this voice, John Ziman posits that "It is adopted to conceal the personal and the social factors that may have originally motivated the research and might influence it towards a particular outcome. The author presents himself as a mere name, a disembodied instrument of factual observation or logical inference, morally detached from the events or arguments reported."[170] In addition, a norm of humility exists. No research paper is published without thorough references and acknowledgment of what preceded it.[171] As Ziman points out though, in reality this fools no one, scientists are anything but disinterested in getting their work published. However the public pose that is adopted is in accordance with a social norm that the researcher is a humble, disinterested, detached seeker of knowledge. This is not necessarily a contradiction in terms. How an investigation is conducted is not a comment on why it should be conducted, as David J. Gray writes.[172] Scientists can argue very well for the relevance of their discipline while conducting it in a disinterested fashion.

Another interesting take on the creation of scientists arrives via Hubert Dreyfus. In his account of expertise, set out at various places in his work, sometimes with his brother Stuart Dreyfus, the emphasis is on how skills are acquired, whether mechanical or intellectual skills. Dreyfus and Dreyfus present a five-stage model of skill acquisition through verbal or written instruction.[173] The Dreyfus's model is a continuum model, going from Novice to Advanced Beginner, moving then through "Competent" and "Proficient" phases before ending at the "Expert" stage.[174] The continuum from the first to the final stage is a transition from "knowing that" to "knowing how", where the first stages are characterized as an adherence to rules that might be given without context and the final stage does not rely on conscious adherence to rules. This expert is totally engaged in skillful performance in the normal course of practice. "When things are proceeding normally, experts don't solve problems and don't make decisions, they do

---

[169] Merton, Robert K, and Norman William Storer. 1998. *The Sociology of Science: Theoretical and Empirical Investigations*. Chicago: University of Chicago Press.

[170] Ziman 2002, p. 38.

[171] "By systematically citing formal scientific sources for everything that is not their own work, researchers limit immodest claims to personal originality." *Ibid.*, p. 39.

[172] Gray, David J. "Value-Relevant Sociology: The Analysis of Subjects of Social Consequence, Including Implications for Human Well-Being." *The American Journal of Economics and Sociology* 42, no. 4 (1983): 405–16., p. 409.

[173] Dreyfus and Dreyfus 1986, p. 19.

[174] *Ibid.* p. 20-22.

what they normally do."[175]

One can say that researchers are not born but made. Helen Longino has observed that the process by which one becomes a researcher is one of induction. "One does not simply declare oneself a biologist but learns the traditions, questions, mathematical and observational techniques, `the sense of what to do next,' from someone who has herself or himself been through a comparable initiation and then practiced." This training brings with it a sense of social objectivity of shared norms. Science (or any research) cannot be done in isolation, so it makes sense that norms and values would influence procedural objectivity. The communal norms are reinforced by the social structures that police research, since research is evaluated by peer review, it is held up to the standards of the community of peers. Having been trained to value this social objectivity over subjective expression, and to adopt this persona, why would researchers when in the role as authors not expect the same professional behavior from their refereeing and editing counterparts?

The manner of a scientist or researcher who values objectivity as an approach is to eschew partiality in examining natural phenomena, and also in evaluating a research paper. Such partiality, when localized in the individual, is often also called bias. Thus, one definition of objectivity of manner is freedom from personal bias. Different from freedom from professionally and collectively held values, bias is seen as an individual failing rather than a capitulation to orthodoxy and groupthink. Of a fashion, this form of objectivity is related to objectivity$_2$ as described in the previous section, but a crucial difference worth highlighting is when the influence of values is not a conscious one.

Carole Lee and coauthors made a study of bias in peer review in 2013, "Bias in Peer Review," in which they describe various studies of bias as they may have an effect on peer review. They defines bias as "the violation of impartiality in the evaluation of a submission."[176] What does this mean? Ideally, in the upholding of criteria for a submissions, impartial reviewers should arrive at the same conclusion. This should seem to be the result of an objective evaluation in which the operators of the objective machinery are interchangeable, per Douglas's Objectivity-3. However, the situation is often more complicated than that. It is questionable if the criteria for publication are simple enough for this requirement to apply as a measure of freedom from bias. Michèle Lamont observes, from her study of grant applications to funding agencies, that "between proposals the criteria for comparison and evaluation are continually changing, as different proposals are regrouped based on different principles and compared."[177] In addition, criteria change not only between submission as time passes and new items are added to the literature, but different standards also obtain between subfields. "Reviewers evaluate fairly when they use standards that are most appropriate to the object of evaluation. Rather than applying a single universal criterion indiscriminately, they specify which criteria, or lenses, are most appropriate to assess the strengths and weaknesses of the object under evaluation."[178] So, if the criteria are in flux, could an editorial procedure apply consistent standards other than in the broadest sense, leaving a lot up to the individual interpretation of the reviewer?

The idea of reviewer agreement as a measure of impartiality is also complicated by a need to qualify disagreement. Reviewers might differ on the importance of including a certain section in a paper, or on matters of interpretation, while at the same time recommending eventual publication. In addition, for interdisciplinary papers, reviewers are often recruited from the disparate fields the submission brings together, and these reviewers would not only have different perspectives but would in fact be explicitly consulted for this reason. This is appropriate not only for interdisciplinary research, but whenever more than one reviewer is consulted to attain a multiplicity of perspectives in order to obtain better criticism (per Longino). A more qualitative evaluation of reviewer reports reveals that disagreement between reviewers is an intended effect in many cases. In Lee's 2012 monograph paper, "A Kuhnian critique of Psychometric Research on Peer Review", it is noted that "When we shift focus from the numerical representation of a reviewer's assessment to the content on which such assessments are grounded, we can identify cases in which interrater disagreement reflects normatively appropriate differences in subspecialization, as well as normatively appropriate differences in the interpretation and application of evaluative criteria."[179]

But even if reviewer disagreement is admissible, some attitudes are viewed as being less than objective. When it it no longer a matter of professional difference from a variation of perspectives, the variance is known as bias. When the bias is applied to characteristics of an authors, concentrating not on the ideas and results in the paper but on who the author is, this is known as author characteristics bias. According to Carole Lee[180] and coauthors in the Peer Review Bias study of studies, such bias challenges the ideal of impartiality, suggesting that reviewers are not evaluating submission in terms of their content and ideas independently from the author's identity. Notably this bias exists whether the identity of the author is known or not, or if it is just surmised. Assumptions along the lines of "this is a low-quality paper, I bet it was written by [a member of some disfavored group] also falls within the category of this bias, even when subconscious, and when the bias is implicit. The forms of bias relevant to peer review include bias informed by researcher prestige, affiliation, nationality, language, gender. Additional bias less focused on actual authors include confirmation bias, conservatism as a reaction for unorthodox results, opposition to interdisciplinary research, and a preference for positive rather than negative results and replication studies. The underlying assumption here is that research from men, women, prestigious institutions or not prestigious ones is roughly equivalent in quality in the aggregate, however, for established reasons of social and financial inequality this may not be the case, an effect that will apply well before any paper is submitted to any journal.

---

[178] Mallard, Grégoire, Michèle Lamont, and Joshua Guetzkow. 2009. "Fairness as Appropriateness Negotiating Epistemological Differences in Peer Review." *Science, Technology & Human Values* 34 (5): 573–606. https://doi.org/10.1177/0162243908329381 p. 578.

[179] Lee, Carole J. 2012. "A Kuhnian Critique of Psychometric Research on Peer Review." *Philosophy of Science* 79 (5): 859–70. https://doi.org/10.1086/667841, p. 863.

[180] Lee et al. 2013, p. 6.

The relative prevalence of all of these have been studied in the various studies discussed by Lee *et al*. Whatever the results of these studies, what seems more important is how and why this criticism of peer review is raised. The very fact that there is talk about bias of this form tells us more about the regard in which peer review is held, and its importance in research, than it does about the fallibility of individuals and systems.

We recall that it was a fear of individual frailties and lack of rigor that lay behind a move towards objectivity in the 19th century in the form of mechanical objectivity, as described by Daston and Galison. However, it was not just a desire for accuracy that informed this practice, but a moral dimension. "Although mechanical objectivity was nominally in the service of truth to nature, its primary allegiance was to a morality of self-restraint."[181] Their narrative of the emergence of objectivity describes a nigh-on fanatical denial of the self, in which subjectivity "is not weakness of the self to be corrected or controlled, like bad eyesight or a florid imagination. It is the self."[182] While no one seems to be asking for mechanical objectivity in peer review, this is exactly the kind of thinking that leads to an overly simplistic way to think about personal bias.[183] To many disappointed authors, it seems that bias simply emerges when a reviewer disagrees with the authors on a journal's more qualitative criteria. The author then concludes that the reviewer has let subjective criteria color the judgment, a judgment which is (it is assumed) meant to be objective. The base expectation is an objective evaluation.

However, given that reviewers are approached because of who they are, and are often asked to give a judgment on behalf of a community on factors beyond mere validity, the clearness of division between subjectivity and objectivity becomes blurred. Is it reasonable to expect reviewers to carry out this degree of compartmentalization a total separation of self from the task of evaluation, to which the reviewer has been set because of who they are in the review process?

## Whither objectivity?

Subjectivity thus seems inescapable, so does this mean that we just throw up our hands in despair and surrender to subjectivity, including bias? Not at all. In considering criticism, which is one of the functions of peer review, Helen Longino (1990) identifies three types of criticism, evidentiary, which raises questions on theoretical concerns, and conceptual, which comes in two parts and can can firstly critique the soundness of the initial hypothesis, the coherency of the hypothesis with established theory, or secondly the relevance of the evidence to the hypothesis, criticizing the assumptions underlying the study itself. As Longino points out, these three types of criticism are part and parcel of normal critique, into which novices are initiated as practitioners of the discipline. Her final notion of conceptual critique, regarding relevant evidence, is however different from the other types of criticism. This last criticism questions the background beliefs or assumptions through which some state of affairs is counted as evidence. If one is to be objective in the face of concerns about choices of relevant evidence, we "require a way to block the influence of subjective preference at the level of background beliefs."[184] that is, to compensate for bias, in this case from the author. The very act of criticism however can (and should) challenge these background beliefs. The challenge should not necessarily overturn them, especially when they are allowing for evaluation of subjective criteria like importance or interest of a result in a manuscript. Responding to criticism or providing further support for the original proposal strengthens the argument or serves to check the background beliefs. This means that criticism is transformative, and, what more, that it exists on a level beyond mere evidential reasoning. Since it concerns background beliefs of both author and reviewer, it is not possible to discount the subjective perspectives of all who are involved in the peer review process. There has to be a spectrum of subjectivity that ranges from legitimate expressions of disciplinarity up to outright bigotry, and the legitimacy of the subjectivity should be subjected to Longino's transformative criticism. Longino holds that as long as the background beliefs can be articulated and criticized, these can be "defended, modified, or abandoned in response to such criticism."[185] Clearly unfounded background beliefs, if put through the same process of criticism, should also then be abandoned. This would mean the background beliefs, including biases, become not only a subject of the evaluative criticism, but also part of the criticism. From this, Longino concludes that objectivity is a community characteristic born out of practice, not something that can be reduced to the individual,[186] unlike Douglas's account.

---

[181] Daston and Galison 1992, p. 117.

[182] Daston and Galison. 2007, p. 374.

[183] As an example of this attitude, "However, disclosure of [conflicts of interest] may not exclude indoctrinated or subconscious biases that may affect the review process. [...] Ideally, the review process relies on the personal responsibility of the reviewer to assess and critique a work based solely on its merit, free from personal bias. [...] This type of bias, either personal or institutional, can sway the literature through promoting or suppressing work." Adler, Adam C., and Stephen A. Stayer. 2017. "Bias Among Peer Reviewers." *JAMA* 318 (8): 755–755. https://doi.org/10.1001/jama.2017.9186.

From Longino's account of transformative criticism, and the accounts of objectivity as it applied to peer review, we conclude that the focus on bias as an individual failing that exists as a separable part of the peer review process is a misunderstanding of the nature of peer review critique. This is not to deny bias in the sense of racism, gender bias, institutional bias et cetera exist, but rather that they do not exist independently of the self that is either performing the work or criticizing it, and thus cannot be removed in a modular fashion. This means that the usual division of so-called objective versus subjective criticism is unhelpful in considering the function of peer review. What is needed is a more human account of peer review, which cannot be reduced to a procedure existing independently of its participants, and cannot be rescued by "mechanical objectivity."

The objectivity at play in peer review is one of judgment, and it is not empirical. The peer has learned in their training (that we covered earlier in this chapter) in the norms of the community and holds up the work to be evaluated against this standard. The norms have an empirical underpinning, in that they are connected to empirical science, but in the moment of peer review judgment questions and not empirical ones are asked and answered.

---

[184] Longino 1990, p.73.

[185] *Ibid.*, p. 73-74.

[186] "Objectivity, then, is a characteristic of a community's practice of science rather than an individual's and the practice of science is understood in a much broader sense than most discussions of the scientific method suggest." *Ibid.*, p. 74.

# Chapter Four: Anonymity

> "Peer review may be necessary, but is open to bias and abuse, especially when referees are anonymous. To shelter under a cloak of anonymity is cowardly, and it is surprising that the editors of many eminent journals still permit this practice. Anonymous letters should be ignored and placed in the waste paper basket where they belong."

Harry Morrow Brown in a 2003 letter to the BMJ in response to the article "Little evidence for effectiveness of scientific peer review" by Caroline White.[187]

Few aspects of peer review generate as much heated discussion as anonymity. The principle of anonymity appears in most forms of peer review, where referees, authors, and even editors can be subject to not knowing the identity of one's interlocutor. While the style of review of journals can vary significantly, most journals have anonymous referees, and many also have anonymous authors. For some journals, the identity of authors is also not known to editors.

The most common form of anonymity in peer review, certainly in the natural sciences, is the so-called single-blind peer review, in which the referee is known to the editors, but not to the authors. The authors are known to the referees as well as the editor. The pragmatic reasons given for this include that the referee is, when anonymous, able to offer candid comments without fear of repercussions. This is again motivated by the desire of achieving an ever more objective method of evaluation of science. As we have seen, critics of peer review, particularly scientists rather than students of peer review *qua* peer review, usually critique individual systems for being insufficiently objective, as if these failures were aberrations rather than systemic problems built into the machinery. The way anonymous peer review is spoken of, as single- or double-blind review, hints at the illusion of objectivity in the method. The term blinding as a verb describing the concealing of identities of referees or authors encourages comparisons with e.g. double-blind studies for medications, and carries with it the image of peer review as a scientifically controlled experiment where this parameter can be controlled exactly. In addition, in common parlance, being blind is a binary condition; you are either blind or seeing, and no nuance is permitted. However, because referees and authors are embedded in a community, their identities cannot be switched on and off at will, and referees are consulted specifically because of who they are. In

---

fact, the expectation of the authors is that the referee is drawn from a finite list of relevant experts. In some communities, this list could be quite short. A frequent contention of a referee report that the authors think is insufficiently accurate is that the referee is obviously not sufficiently expert in the field, and therefore inappropriate.

It is therefore misleading to talk about peer review as blind or not, as if the identity of the reviewer were irrelevant. In some ways peer review is more like a parlor game than a scientific experiment. For this reason, we will use "anonymous referee" and "mutually anonymous" as synonyms for single- and double-blind review, not only to avoid ableist language, but also to give the practice no more epistemological weight than it really possesses.

It is also important to realize that author anonymity is not the reciprocal of referee anonymity. The author arrives at the review process with something to prove, to show or demonstrate results. The referee is the gatekeeper who will evaluate the paper. The paper itself is an expression of the author's will, and as such the author is a different participant in the discourse. In some academic fields the anonymity of the author in the review process is the standard, however, even then the roles are different, and the author and referee assume anonymity for different reasons. In addition, many famous authors are best known by their pseudonyms for a variety of reasons, but these reasons had little to do with peer review. We will return to anonymous authors and mutually anonymous review later in this chapter.

## What is anonymity?

Anonymity has historically had a bad reputation. Recently, it seems we are living through a plague of online trolls that are either fully anonymous or hide behind pseudonyms. This is certainly the case for controversies such as GamerGate[188], and the astounding amount of abuse experienced on online places such as Twitter.[189] A frequent advice to those who would get their news and other information online is "don't read the comments!" which are frequently anonymous or signed by soubriquet. Hubert Dreyfus has developed Søren Kierkegaard's writings on what he called "The Press" into a critique of the Internet, and the anonymity on there. To recap Kierkegaard's argument, when he attacks the Press he accuses it of creating what he calls the Public. The Press speaks for the public but no one stands behind public opinion[190], which to

---

[188] Hathaway, Jay, "What Is Gamergate, and Why? An Explainer for Non-Geeks." Gawker. Accessed December 10, 2018. http://gawker.com/what-is-gamergate-and-why-an-explainer-for-non-geeks-1642909080.

[189] There is a tremendous amount of cases of abuse on Twitter, for examples, see "Twitter (finally!) takes aim at anonymous egg accounts", https://www.wired.com/2017/03/twitter-abuse-tools/, "Two cases of Twitter abuse highlight the obscure nature of suspensions", https://www.theguardian.com/technology/2017/jan/09/twitter-abuse-martin-shkreli-lauren-duca-alexandra-brodsky, "Twitter announces new measures to tackle abuse and harassment", https://www.theguardian.com/uk-news/2014/jan/24/two-jailed-twitter-abuse-feminist-campaigner (all retrieved 11 April 2017).

[190] "As Kierkegaard puts it: `A public is neither a nation, nor a generation, nor a community, nor a society, nor these particular men, for all these are only what they are through the concrete; no single person who belong to the public makes a real commitment.'" Dreyfus, Hubert L. 1999. "Anonymity versus Commitment: The Dangers of Education on the Internet." *Ethics and Information Technology* 1 (1): 15–20. https://doi.org/10.1023/

Kierkegaard is an appalling lack of responsibility. In extending this analysis to the Internet, Dreyfus interprets Kierkegaard to mean that the only alternative to "this anonymity and lack of commitment was to plunge into some kind of activity – any activity – as long as one threw oneself into it with passionate involvement.[191]" Passionate involvement is thus the only cure to this dangerous form of anonymity.

However, the anonymity of referees in peer review is not akin to the anonymity of an undifferentiated mass, like the online trolls on various discussion boards, in the comment threads on youtube.com, or the anonymity described by Dreyfus/Kierkegaard. What Kierkegaard and later Dreyfus railed against was the lack of accountability that comes with complete anonymity, but the anonymity in peer review is at best partial. For one thing, we know several things about the referee already. Not just anyone is asked to review a paper. In fact, referees are chosen specifically because of who they are. Addressing Kierkegaard's critique directly, they could be said to already be passionately involved in their field, certainly sufficiently involved to give up their time for unpaid refereeing. In addition, they would be accountable at least to the editor who chose them. So what is this kind of anonymity that still has so much identity in it?

Kathleen Wallace[192] analyzes anonymity a bit differently from the completely unknown public mass discussed by Dreyfus. To her, anonymity is a form of nonidentifiability, which is about more than simple naming. In spite of the etymology, the point of anonymity is that the person who is anonymous is not identifiable, not that they do not have a name. In the same vein, a pseudonym can do the same work as being simply anonymous. Particularly relevant to peer review, Wallace's account of anonymity describes a relation between an anonymous person and others, in that the anonymous person is known "only through a trait or traits which are not coordinatable with other traits such as to enable identification of the person as a whole."[193] As a particularly relevant example Wallace asks us to consider the author of a book. It is perfectly possible not to reveal the true name of the author of a book by use of a pseudonym such as "Elena Ferrante", and still have "Elena Ferrante" be known as the author of the book. In peer review, for anonymous review, a label such as Referee A allows for pseudonymous identification of the anonymous referee as the author of the report in question. In future rounds of review, a second report might arrive from Referee A, who is still the same A as was consulted in the first round of review. However, the fact that Referee A can be identified as the writer of report A does not allow for coordinating that information with such things as name, institutional affiliation, address etc. Other traits may of course be revealed through the report that will allow the author of the paper under review to guess the identity of the referee, which makes the referee non-anonymous even though anonymity has not been intentionally breached. Breaching anonymity does not require complete knowledge of a person, only enough to coordinate relevant traits such as a report and a name and affiliation.

Other traits that makes the anonymous person different from the absence of any identity is the ability to act. I may never meet a certain person from the other side of the world, and be completely unaffected by their existence, but that does not make them *anonymous* to me. Anonymity is not equivalent to being socially unknown. As Wallace points out, isolated hermits are not known by a large number of people, but we do not refer to these as being anonymous.[194] The anonymous referee is however affecting the author and the editor by their action of providing a report.

The author of an article under review will know a number of other things about an anonymous referee. They will know that the referee is selected and consulted by the editor. They will know that the referee is an expert in this field, though they will here have to trust the editor's judgment in selecting the proper expert. They will also know some more trivial things such as that the referee is human, speaks or at least reads the language in which the article is written, and was alive at the time the report was submitted. Most importantly they will know that the referee is a *peer*, though the referee might not be of the same institutional rank as the authors. Post-docs can review full professors, and vice versa.

Given however that anonymity has such a bad reputation, why is peer review using anonymity not just incidentally but to safeguard the integrity of the process? To explain this we have to unpack what this kind of anonymity means. Wallace points out that there are three general purposes or goals of anonymity:

    1.   anonymity for the sake of furthering action by the anonymous person, or agent anonymity

    2.   anonymity for the sake of preventing or protecting the anonymous person from actions by others, or recipient anonymity

    3.   anonymity for the sake of a process, or process anonymity.[195]

For peer review, all three apply. The first is the case of the anonymous author who believes they will benefit from remaining anonymous. It also serves the referee who can thus better act as a representative of their field, in that the personal preferences of an identified referee will not be a topic for debate. The second is to shield referees from repercussions of a negative report, since authors may be in positions of power to hurt a critical referee's career. The third is tied up with integrity of the review process and is related to the use of anonymity in other arenas too, such as the law or test-taking.[196] Even in ethics, anonymity in the form of noncoordinatablity can play a role, for instance "The famous Rawlsian 'veil of ignorance' is a form of anonymity (participants being rendered unable to coordinate themselves with their particular social traits, i.e., locations in

---

[194] "A hermit may be 'nameless' or unknown but is not typically referred to as 'anonymous'; rather a hermit is an unrelated, socially disconnected agent whose life does not and whose actions do not, for the most part affect or are not affected by the lives of others in a social environment." *Ibid.,* p. 24-25.

[195] *Ibid.,* p. 29.

[196] "For example, test-taking, research studies, peer review of work and the judicial arena are all areas in which anonymity has been thought to serve the purposes of neutrality and impartiality." *Ibid.,* p. 30.

society) for the sake of arriving at principles of justice."[197]

## Rise of the anonymous referee

The first instances of anonymity in peer review involved referees whose identity was unknown to the authors. The step of making referees, or even authors, anonymous for the purpose of improving peer review could only make sense once pre-publication peer review had been instigated. As we saw in Chapter Two on the history of peer review, communications surrounding manuscripts under consideration started off even more opaque than this. In many cases of early peer review, the procedure was that a referee would review a manuscript on assignment from the editor , but no report was sent to the author apart from an editorial decision letter, which might or might not summarize or paraphrase a referee's impressions. This may in principle count as anonymous reviewing, since clearly the action of providing the report has an effect on the author, even if the author is not engaging with the referee directly. For the purposes of considering the anonymous referee as a participant in the review process, however, it is more useful to look at the procedures where reports are shared with the authors. Without this, the author cannot engage with the referee as a pseudonymous person, and the "Ziman triangle" is missing one of its legs.

It is however certain that when the shift was made to sending reports by anonymous referees to authors, the resulting lack of transparency was almost immediately singled out for criticism. Alex Csiszar relates what happened when *Philosophical Transactions* moved to using anonymous referees to help decide on publication of submissions. He cites a commentary piece published in 1845 that "painted a picture of referees as scheming judges quite possibly `full of envy, hatred, malice, and all uncharitableness'. Hidden away in some secret chamber, this scientific judiciary, the article implied, used the cover of anonymity to advance their personal interests — perhaps through undetectable acts of piracy — at the expense of helpless authors."[198] Clearly anonymity stirred strong feelings, even though the practice also had its defenders, "Through anonymity, as one uncredited editor argued in 1833, `the individual is merged in the court which he represents, and he speaks not in his own name, but ex cathedra (with full authority)'"[199]

Of course, criticism of anonymity in the peer review process has manifested itself frequently and more recently. A more modern but no less forceful criticism of the practice on anonymous referees came in 1974 in the New Scientist, in which Robert Jones, a biochemist, wrote, "On occasion, the act of submission of a paper can place the author at the mercy of the malignant jealousy of an anonymous rival. Manifestations of antipathy can take many forms, which range from contemptuous mockery […] to outright theft. To have suffered such an experience generally induces sympathy from colleagues, but condolences are not enough. What is needed is justice;

---

and justice can be secured only by altering the system."[200] Jones was not implying that referees are universally bad actors, but rather that the nature of the peer review system left it open to abuse. In response to concern like this one, Nature published an editorial, "In defense of the anonymous referee"[201] in which they offered three defenses of anonymity: 1. To protect referees from irate authors, 2. To allow for genuine criticism, 3. That the anonymous referee is also a future reader.

In parallel, and from the world of medical research, in 1974 David F. Horrobin published a debate piece for the British Medical Journal (BMJ) in which he railed against a referee of a grant application of his, in which he included his point-by-point rebuttal of the referee report. He declared that the problem with the existing peer review system is that it is a closed one, and he argues for more openness. "The closed system is a real bar to rapid progress. There is a common belief that a benevolent dictatorship is the best method for the achievement of efficient advance. Karl Popper's *The Open Society and its Enemies* should have destroyed that fallacy forever but alas his works go largely unread."[202] Sixteen years later, Horrobin would go on to declare that it is the duty of editors not just to exercise quality control of the journals and the fields they represent, but to prevent referees from stifling innovation. "Editors must be conscious that, despite public protestations to the contrary, many scientist reviewers are against innovation unless it is their innovation."[203]

The BMJ published a rebuttal editorial to Horrobin's 1974 piece, in which they claimed (without citing evidence) that events such as the Horrobin case were very rare, and took the opportunity to outline the mechanics of the BMJ peer review system, arguing for the existing anonymous review system for practical reasons.

> Should referees be made to sign the reports sent to the author […]? The answer must surely be no, since few referees would be prepared to wound with the unvarnished truth, and the practice would almost certainly lead to personal conflicts between them and the authors, or the heads of their departments. Should papers be sent to referees shorn of their list of authors? This, again, is scarcely practicable. The referee may spend much of his time trying to decide on the authorship, which anyway is often evident since earlier work from the team is cited. To some extent the credence placed on results depends on the scientific reputation of the research team responsible for it. It is naive and unrealistic to expect work to be assessed solely at its face value.[204]

---

[200] Jones, Robert. "Rights, wrongs and referees." *New Scientist* **61**, no. 890 (1974): 758-759.

[201] Nature, "In defence of the anonymous referee", *Nature* **249** (5458), 601., (1974).

[202] Horrobin, D. F. "Referees and Research Administrators: Barriers to Scientific Research?" *British Medical Journal* 2, no. 5912 (April 27, 1974): 216, p. 218.

[203] *Ibid.,* p. 1441.

[204] BMJ, 1974. Editorial: Both sides of the fence. *BMJ* 2, 5912, 185-186.

Several more critiques of the practice of anonymous review would follow.[205] However, arguments for anonymity, at least pragmatic ones, would largely remain unchanged, even if referees can sometimes choose to sign their reports.[206] While BMJ have since adopted open peer review,[207] at the time, both BMJ and Nature grounded their decisions to embrace anonymity for referees from Wallace's second purpose (recipient anonymity). The Nature editorial went one step further, expanding on the referee's role in its third point raised, stating in that, "The referee serves as more than just the expert vouching for technical plausibility. He is also the potential reader vouching for relevance, wide appeal, and readability."[208] This sentiment exemplifies the "peer" role of the reviewer. As a representative of the field the individual referee stands in for the whole of the subdiscipline. As such their criticisms cannot be dismissed as merely a personal opinion. While arguments for anonymity of referees can be pragmatic, they also stem from a desire to increase objectivity. If criticism is done from behind a veil of anonymity, it should allow one to obtain more frank and candid criticism. It also appears, per Wallace's recipient anonymity, that researchers are less likely to want to review if more persons than the authors and editors will read their reviews. Subsequent accounts have backed up this view, and there have been attempts to qualify and quantify this. Jerry Suls and René Martin write, on peer review quality in different modes, of trials that assessed review quality as a function of reviewer identifiability. While review quality seemed unchanged, they highlight downsides, specifically that "[u]nder identifiable conditions, potential referees were more likely to decline invitations to review, and those who did review were more likely to recommend manuscript acceptance."[209]

## Why is anonymity accepted?

Even if the journals' position is clear, one might wonder why authors accept critique from anonymous referees. Surely any criticism garnered in the peer review process would be more convincing if it came from a known authority, a person whose expertise and reputation were beyond reproach? It may be said that anonymity arose as a form of protection from repercussions of critiquing one's peers in the open, but the move towards anonymity can also be put down to a desire to make the witnessing more objective, to let the scientific facts "speak for themselves". If a referee can adopt the scientific position of a view from nowhere, then their criticism takes on a more objective air. The changing identity of the referee coincided with the idea of a common knowledge base that we call objective, which is not a product of the individual scientist's impartiality, but rather of a system set up to guarantee this impartiality.[210] It is a movement away from appeals to an authority, such as to God or the king, and towards convincing by the force of the arguments themselves. The rise of empiricism allowed for experiments to produce new facts[211], and now peer review would fill the same role.

---

[205] For a few examples, see Bastian, Hilda. 2015. "Weighing Up Anonymity and Openness in Publication Peer Review." Absolutely Maybe (blog). May 13, 2015. http://blogs.plos.org/absolutely-maybe/2015/05/13/weighing-up-anonymity-and-openness-in-publication-peer-review/; Gaudet, Joanne J. 2014. "An End to 'God-like' Scientific Knowledge? How Non-Anonymous Referees and Open Review Alter Meanings for Scientific Knowledge." Working Paper. http://ruor.uottawa.ca/handle/10393/31415; Moriarty, Philip, and Paul Benneworth. 2015. "Should Post-Publication Peer Review Be Anonymous?" *Times Higher Education (THE)*. December 10, 2015. https://www.timeshighereducation.com/features/should-post-publication-peer-review-be-anonymous; Polka, Jessica K., Robert Kiley, Boyana Konforti, Bodo Stern, and Ronald D. Vale. 2018. "Publish Peer Reviews." *Nature* 560 (7720): 545. https://doi.org/10.1038/d41586-018-06032-w.

[206] https://www.nature.com/authors/policies/peer_review.html; https://journals.aps.org/prl/authors/editorial-policies-practices#anonymity

[207] https://openscience.bmj.com/pages/policies/#OMaterials

[208] *Nature* 1974., p. 601.

[209] Suls, Jerry, and René Martin. "The Air We Breathe: A Critical Look at Practices and Alternatives in the Peer-Review Process." *Perspectives on Psychological Science* 4, no. 1 (January 1, 2009): 40–50., p. 43.

The scientific persona of the referee turned into a gatekeeper[212]. It should be recalled that one reason for the rise in the use of referees at all was the increase in the amount of science being created and the increasing desire for publication entities to be more selective. For a long time, the problem journals had was to generate enough material to justify the subscription charges, not to keep substandard research out. The number and scope of scientific journals grew rapidly in the 18th and 19th centuries, but as we saw in the history chapter, "peer review remained very much the exception rather than the rule. One reason for this was purely economic. Journals were typically sold by subscription and editors were under tremendous pressure to simply fill the requisite number of pages by a deadline."[213]

Alex Csiszar details a particular case study of the emergence of the expert anonymous referee at the Royal Society. In 1831, William Whewell brought in the system of sending all paper submitted to Philosophical Transactions to external reviewers. Whewell's idea was to publish the reports as interesting commentary on results, an idea borrowed from l'Academie Française, where such reports were written for the sake of the King. However, when the idea was put into practice by Whewell and astronomer John William Lubbock, jointly reviewing a submission by George Airy, the proposed scheme changed rather dramatically. Lubbock wrote a completely different report than Whewell, a very critical report from the vantage point of the expert, and his report was never published.

Csiszar's claim is that this incident highlighted two radically different view on the person of the referee. Whewell was a generalist where Lubbock was the expert. Whewell's contention was that the commentary itself was of interest, possibly even more than the actual paper, but Lubbock viewed the peer review dynamic as an arena for critique. "Whewell was the authoritative generalist, glancing down on the landscape of knowledge. He was unconcerned with — and probably not in a position to critique — the details. Such referees were, according to the Royal

---

[210] Csiszar 2015, p. 148-149.

[211] "Before discovery history was assumed to repeat itself and tradition to provide a reliable guide to the future, and the greatest achievements of civilization were believe to lie not in the present or the future but in the past, in ancient Greece and classical Rome." Wootton, p. 61.

[212] Csiszar 2016, p. 308.

[213] Chapelle, Francis H. "The History and Practice of Peer Review." *Groundwater* 52, no. 1 (January 1, 2014): 1–1. doi:10.1111/gwat.12139.

Society's president, `elevated by their character and reputation above the influence of personal feelings of rivalry or petty jealousy'.[214] Lubbock was a younger specialist, Airy's equal. This allowed him to take a fine-tooth comb to Airy's arguments; it also put him in the position of reviewing a direct competitor."[215] Whewell crucially did not see the peer review process as a filter, and initially his vision of the proper role of peer review won out, but eventually that of Lubbock became the common practice, and soon reports became for the eyes of the editor only. Whewell himself changed his tune; writings of his from 1836 show that he embraced the role of the referee as a gatekeeper, excluding articles that should not be published. This new role for referees was probably one motivator for the referees being anonymous, for their own safety. According to Csiszar, there is plenty of precedent for this stance in England, where "signing one's name to explicit criticism of a colleague would have been ungentlemanly."[216] Csiszar describes a preference for criticism from an anonymous piece that took the stance of speaking for the public. Today, as he notes, The Economist maintains this practice. "Through anonymity, as one uncredited editor argued in 1833, `the individual is merged in the court which he represents, and he speaks not in his own name, but ex cathedra (with full authority)'.[217] Justifications of the anonymity of the scientific referee took a similar view."[218] This persona took a little while to catch on. Csiszar writes that in the beginning the idea of anonymous referees was viewed with suspicion and fear of scientific misconduct by unaccountable anonymous referees was high. However, eventually this became accepted practice, and eventually review became seen as a guarantor of quality.[219]

But what does it mean to speak on behalf of the community? The referee is not just asked to be a future reader, but to make recommendations on behalf of an entire community. We can only arrive at the idea of these public expressions making sense if we separate what is written from the writer, and admit that ideas have valence in themselves regardless of who expresses them. This move requires a sense of knowledge as a shared body of information, rather than appeal to an authority figure. A useful example of the previous kind of thinking is provided by David Wootton, who describes pedagogical dissections in the Middle Ages before empirical thinking was accepted. Dissections would take place to illustrate the reading of a text of received wisdom, but not for the dissection to provide information on its own.[220]

---

[214] *Proc. R. Soc. Lond.* **3,** 140–155 (1832), cited in Csiszar 2016, p. 308.

[215] Csiszar 2016, p. 308.

[216] *Ibid.*, p. 308.

[217] *New Monthly Magazine* **39**, 2–6 (1833).

[218] Csiszar 2016, p. 308.

[219] "It was only near the turn of the twentieth century that the idea began to take hold that editors and referees, taken as one large machinery of judgement, ought to ensure the integrity of the scientific literature as a whole." *Ibid.*, p. 308.

[220] "Medieval anatomists had frequently lectured by reading Galen aloud and commenting on his text, while assistants opened up the body: the body was intended to illustrate what Galen said, not to correct him when he was wrong. But, even when medieval anatomists had performed their own dissections, what they found (or thought that had found) was what Galen had told them to find. Mondino de Liuzzi (1270-1326), for example, the author of the

The idea that evidence of one's senses had to take precedence over received wisdom took a long time to gain foothold. This would seem strange to us in the current age. Science is of course based on logic, which is intended to be universal. However, in a social process like peer review, separating knowledge derived from logical arguments from that which is received wisdom is not always easy. As we saw in Chapter One from Latour and Longino, one has to have an idea of a common body of knowledge in order to build what we today call science. In a similar fashion, Wootton describes the rise of a scientific community citing the example of William Gilbert (1544-1603), the magnetician, who, in his seminal work *De Magnete* (1600), "acknowledges a little community of experts, many of whom are known to him personally… Where all previous experimenters, from Galen to Garzoni, appear to have worked in isolation, we have here for the first time a functioning scientific community."[221] Wootton later describes that by some decades later, the circumstances under which Evangelista Torricelli (1608-1647) worked in during the invention of the barometer constituted a scientific network, as the start of the institutionalization of science. According to Wootton this led to "a new commitment to the idea of scientific progress. In a draft preface to an unpublished book on the vacuum (c. 1651), Pascal distinguished between forms of knowledge that were historical in character and depended on the authority of the sources on which they relied (theology was the key example), and forms of knowledge that depended on experience."[222] It is this foundation of empirical knowledge that is now shared in a network of scientists which constitutes the basis of science and scientists both, and gives rise to the idea of the expert.

However, all of these accounts problematize the narrative of ideas standing for themselves, whether they rely on Latour's black boxes or on Longino's socially trained scientists who live in their own special world. Reviewers may be anonymous, and the identity of the reviewed may be irrelevant, but they still have to be experts drawn from the relevant community, otherwise how would peer review be different from soliciting random opinions in an online poll? As we have shown, the anonymity of the participants in peer review is different from an *ad hoc* Internet audience, with one difference being that the peer review system is just that, systematic, with the assumption of participants in good faith. The anonymity of peer review is not the chaotic atmosphere of the Internet in general against which we sometimes rail, but a guided process in which anonymity is applied judiciously for the good of the process.

In this way, it seems clear that the desire for anonymity stems in part from an attitude that ideas should stand for themselves regardless of who expresses them, but clearly this is not how peer review works. The paper is not in standard peer review open to be reviewed anonymously by the public. Instead specific experts are consulted, matched through expertise to specific manuscripts. Anonymous they may be, but we take it on trust that they are relevant experts in the field in question. If referees are anonymous, how can we know that? Authors have to trust that the journal's editors will consult appropriate referees, and that, if inappropriate referees have first medieval textbook on how to perform a dissection, had plenty of hands-on experience, but he still found at the base of the human brain the rete mirabile (miraculous network) of blood vessels that Galen claimed was there, despite the fact that it is not there at all – it is only present in ungulates." Wootton, p. 183.

[221] *Ibid.,* p. 331.

[222] *Ibid.,* p. 341.

been consulted, that this can be remedied by the editors or by editorial procedures such as an appeal to an editorial board.

This trust is not automatic. The approach of consulting peers, who provide anonymous reports, is not without its risks. Active researchers in the field will be the most suitable in terms of expertise, but these are also likely to be a direct competitor. Indeed, in smaller subfields, the referee most likely knows the authors well and may in addition be working on rival projects. A competitor, reviewing a manuscript, might stand to gain by delaying its publication, so a balance has to be struck between the benefits of reviewer expertise with the possibility that a direct competitor is consulted. It is the job of editors to know sufficiently about the field to avoid these kinds of conflicts of interest, whether intentional or unintentional. Referees have absorbed this ethic and frequently alert editors to conflicts of interest. Nevertheless, the worry about competition and referees taking advantage of their privileged position as early readers is one thing that has led to the call, and for some subject fields, adoption of, mutually anonymous peer review.

## Anonymous authors

It might naïvely be argued that there is a symmetry to mutually anonymous peer review. When both authors and referees are anonymous, this seems fair and equitable. This, one might think, would end referee bias against less renowned authors in favor of more famous authors, help less historically successful authors, and aid authors outside of the orthodoxy of academia. As a defense of this latter group, the fact that Einstein did some of his most significant work as a patent clerk in Bern is frequently cited, with the intimation that a scientist "operating outside the system" would not find publication space for their ideas today.[223] However, the exchange inside of academic peer review is not a symmetrical process. The author is not, unlike the referee, acting as a representative of the community, but rather as someone trying to convince said community of their findings. The burden of proof in the writing of the paper is on the author.

Mutually anonymous peer review is *de rigeur* in some academic fields, but other fields stubbornly resist its call. Most natural science subject area journals do not practice mutually anonymous peer review, relying instead on modes where the referees know the authors' identities. Editors of natural science journals were adamant that mutually anonymous peer review would not suit their journals, or their communities. In a letter to Physics Today, the then-Managing Editor of the American Physical Society, Samuel A. Goudsmit, who would later go on to found Physical Review Letters, wrote that making authors anonymous was "impossible" because reviewers would easily guess the identity of the author or authors. In addition one would have to remove "all references to previous work by the same author, all descriptions of special equipment and other significant parts of the paper."[224] It may be that for experimental papers making authors anonymous is difficult for these reasons (how many large particle accelerators exist in the Geneva area of Switzerland?) but one would like to separate these practical objections and separate them from the desire to maintain control over publication capital by relying on name recognition.

---

[223] See for example "The Patent Clerk's Legacy", Gary Stix, Scientific American, September 2004, http://www.scientificamerican.com/article/the-patent-clerks-legacy/

[224] Ward, W. Dixon, and S. A. Goudsmit. "Reviewer and Author Anonymity." *Physics Today* 20, no. 1 (January 1, 1967): 12–12. doi:10.1063/1.3034118.

One of the most influential pieces on mutually anonymous peer review was an experiment on peer review itself, presented in 1982, by Douglas P. Peters and Stephen J. Ceci.[225] Peters and Ceci selected twelve already-published research articles from twelve highly regarded journals in psychology, from renowned authors at top-level institutions. They altered the names of the authors of the articles and sent them back to the journals that had originally published them with new names and fictitious institutional affiliations. Thirty-eight referees and editors formed part of the evaluators, out of which only 8% detected that these were resubmissions of previously published articles. Nine of the twelve articles were sent for anonymous peer review and eight out of the nine were rejected. Sixteen of eighteen referees consulted recommended against publication, and the relevant editors agreed with this recommendation.

These were not obscure pieces – far from it, they were highly cited contributions from leading journals. Peters and Ceci present a detailed analysis of the factors that might have led to this stunning result but come up with only two possible explanations: that the referees consulted were unusually incompetent (which is statistically unlikely) or that the referees and editors were negatively predisposed to articles coming from unknown authors from obscure (actually fictitious) institutions. This study is, to Peters and Ceci, and many others, a strong argument in favor of changing the peer review system to a fairer one, with their primary suggestion being mutually anonymous peer review.

But would such a system be practicable? The article garnered many responses published in the same issue as the Peters and Ceci study. These responses ranged from objections about the methodology to attempts at claiming that this said more about the state of psychology than about peer review as such. For instance, in one response to Peters and Ceci, a survey by an editor of *Physical Review Letters*, Robert K. Adair, found that in an experiment run for the physics journals he was familiar with, referees could correctly identify 80% of the submitting authors, despite efforts to mask author identity.[226] The concern that referees could easily identify authors even if their names were concealed is not restricted to physics, in fact four randomized trials in biomedicine with concealed author identities found that referees could identify the authors 23%–42% of the time.[227] In addition, even if referees do not identify the authors, they will make assumptions, and act upon those assumptions in ways that may benefit or hamper the chances of the authors.[228]

Another question that arises about this study is, did Peters and Ceci break the rule of

---

operating in good faith? Editors and referees assume that an author submits to the journal in good faith and not in some kind of "gotcha" exercise. It might be argued that it is an unreasonable epistemological burden on editors and referees to also suspect the motives of the authors. Often these kinds of studies start out with the designers having a clear idea of the conclusion they would like to prove, as in the case of Bohannon's "sting" against predatory open access journals,[229] but depending on the methodology the successes of deliberate stings are more or less successful.[230] For cases of actual scientific misconduct, it is instructive to note how long it took for Jan-Hendrick Schön's data fabrication to come to light, largely because the idea of outright data fabrication was so outlandish.[231]

Another problem is that Peters and Ceci did not have access to a lot of the relevant material. They did not uniformly have access to the original reports for the papers as first published, the original decision letters, or the identities of the reviewers either for the papers in their original form or for their experiment.In the part about accountability in their paper,[232] this is most apparent. Peters and Ceci do not take into account that journals may have databases on referees where they build up a track record, letting editors know what referees can and cannot be trusted. Even if anonymous, reports are written for the editor and the authors, and studies have shown that having to justify a recommendation to the entire readership often leads to unwarrantedly positive and less detailed reports.[233] Other trials have reported no improvement from either concealing or unmasking referee identities, or signed reports.[234]

Ultimately, it seems there is a logical flaw in Peters and Ceci's study, in that one does not see how mutually anonymous peer review would have helped this case. If Peters and Ceci are challenging the original decision to publish the twelve articles they chose, that is one thing, but they seem to be invoking a concept that "publishability" is a definable quality that transcends authorship, never mind time, place, and any subsequent developments in a field. As a result, Peters and Ceci probably ask too much of peer review as a method, surprisingly not recognizing

---

that journal publication is a social system that is highly dependent on and calibrated to the needs of another social system, the scientific discipline the journal serves.

With all of these conflicting accounts, it is hard to make sense of the question of mutually anonymous peer review, and it might be tempting to put the whole question down to individual journals' and disciplines' preferences. These are not necessarily even static, however. An interesting case study of mutually anonymous peer review is recounted by David Pontille and Didier Torny, concerning the journal American Economic Review (AER). This journal made four procedural changes on how it ran its peer review system between 1973 and 2011. In 1973, inspired by a motion at an American Economic Association (AEA) congress in 1972, which called for mutually anonymous review for the better protection of young researchers, AER initiated an experiment on mutually anonymous review. This experiment concluded that mutually anonymous review was fair and practicable, and came down enthusiastically in favor of mutually anonymous review. Seven years later, the new editor-in-chief reversed the policy, stating no motivation other than personal taste. Following concerns about gender bias, a later editor-in-chief revived the practice of mutually anonymous peer review even though a survey showed that half of the referees had correctly identified the authors whose work they were reviewing.[235] The survey had however garnered opinions from reviewers implying that the imposition of mutually anonymous peer review had not hindered the review process. Thus, in spite of no statistically significant proof for its efficacy, mutually anonymous review was reinstated. In 2011, the journal again reverted to referee anonymity only, because it had become too easy to unmask authors through Internet searches. Acting on the belief that true anonymity was unachievable in the Internet age, AER decided to do away with what was by then a popular practice. In spite of a petition signed by five hundred scholars demanding its return, which they saw as necessary due to the still manifest effects of reviewer bias, AER did not alter this decision, and was not swayed by the vow of the signatories that they would not to use search engines to try to unmask authors they were reviewing.[236]

Technological obstacles offers a material argument against author anonymity, but it is not a useful argument against the very real effects of bias against women, authors from less renowned institutions, authors from the Global South, and independent scholars. It also seems a spurious argument that a referee could identify an author. Peer review is not a game of Clue where successfully unmasking the murderer, murder weapon, and location, brings the entire game to a screeching halt. What are the real consequences of referees successfully guessing the identity of authors? The obvious benefit to having author identities known to the referees would be to more easily identify if a contribution is new and a substantial enough advance to merit publication, but one might respond that a competent referee should be on top of the literature anyway, and that the authors' bibliography should be sufficiently comprehensive so as to put the research into context.

As long as peer review maintains the idea of the referee persona as a "modest witness" that is allowed to be a disembodied and objective entity of dispassionate judgment, then questions of bias cannot be thoroughly settled, because our referee cannot both be such a modest witness and have to be shielded from the authors' identity lest biases, implicit and explicit, take over. This would seem to doom the project of peer review by any shared standards. However, as Helen Longino points out, our idea of objectivity is a property of groups, not individuals. We cannot be objective on our own. "[T]he objectivity of scientific inquiry is a consequence of this inquiry's being a social, and not an individual enterprise."[237] If it is a social enterprise, then there can be no view from nowhere, and no objective evaluation divorced from an evaluator, who themselves is trained into sharing the epistemological standards of the relevant scientific community.

---

[235] Blank, Rebecca M. "The Effects of Double-Blind versus Single-Blind Reviewing: Experimental Evidence from The American Economic Review." *The American Economic Review* 81, no. 5 (1991): 1041–67.

[236] Pontille and Torny, 2014.

Those who would wish to dismiss mutually anonymous peer review and other approaches to eliminate the effects of bias will inexorably cite practical impediments to justify not maintaining the practice, just like AER did. The wealth of literature about implicit bias does suggest that implementation of mutually anonymous peer review would at the very least be a step in the right direction. In order not to let the perfect be the enemy of the good, the arguments for anonymity of authors are hard to resist. However, when we return to Blank, we have to ask ourselves how effective mutually anonymous peer review (double blind review) truly is, since her observation is not only that it does not favor women but also that anonymous authors have a rougher time in the review process[238]. As later work by Carole Lee and coauthors points out, while the gender imbalance in STEM would lead one to believe that gender bias is rife in peer review, actual studies of gender bias in peer review are inconclusive at best, and in many cases show no such bias.[239] More recently, the Nature journals have been experimenting with allowing author anonymity, but are not imposing it across the board[240], with only 12% of authors choosing to be anonymous for all the Nature journals (*Nature*, Nature research journals, *Nature Communications*). While it was noted that the acceptance rates for mutually anonymous papers that were sent to referees were lower (25%) than those where the author was not masked (44%). It is uncertain if one can learn very much from a voluntary program like this, in contrast to the AER experiment, since the anonymous authors will necessarily be a self selected group. Considering the outcomes of the experiment, it was noticed that the take-up was much higher from "disadvantaged regions" than from the West, with India (32%) and China (22%) the most frequent users of mutually anonymous reviews, with by contrast only 8% of French and 7% of American authors choosing this option. The authors' identities were further not concealed from the editors, which is a significant factor since the Nature journals are known to reject a high fraction of submissions without external review, with only 8% of anonymous manuscripts being sent for review compared to 23% for the non-anonymous group.[241]

---

[237] Longino 1990, p. 67.

[238] The primary conclusion of this study is that there are significant differences in acceptance rates and referee ratings between single-blind and double-blind papers. Most strikingly, double-blind papers have a lower acceptance rate and lower referee evaluations. In addition, double-blind reviewing results in different patterns of acceptance rates and referee ratings by institutional rank of author. While the data are consistent with an argument that women fare better under a double-blind reviewing system, the estimated effects are small and show no statistical significance." Blank, p. 1042.

[239] "For example, if there is gender bias in review, we would expect double-blind conditions to increase acceptance rates for female authors. However, this is not the case ([…]. Nor are manuscripts by female authors disproportionately rejected at single-blind review journals […] Even when the quality of submissions is controlled for, manuscripts authored by women do not appear to be rejected at a higher rate than those authored by men." Lee et al. 2013, p. 8 and references therein.

[240] https://www.nature.com/authors/policies/peer_review.html

The fear of bias is however real enough to motivate journals to offer anonymity for authors, either across the board or as a choice. Anonymity for referees and anonymity for authors are both motivated by a reach for a more objective means of scientific evaluation. Anonymizing the author is however a prioritization of the value of accountability over confidentiality as a means of improving objectivity. The move to anonymous referees seems more successful and less controversial than its counterpart. This may be because the goal of making referees anonymous (as a representative voice of the community) is more clearly defined than the reasons given for mutually anonymous peer review. For the latter, concerns about bias as well as practical considerations come into play when considering implementing such a system, resulting in a muddled motivation with a diverse set of stakeholders. In many cases, in the Internet age anonymous authors run against the increasing climate of openness, making participation in preprint services problematic, since referees could easily search online for the relevant preprint and thus find out the identity of the authors. If the motivation for anonymity of authors is a more equitable distribution of ideas, then blocking communication of ideas through these services until peer review has had its say – a process that can last months or even a year or more – runs against current developments in opening up scientific communication.

## The tension between anonymity and open science

The practice of anonymity of referees, authors, and sometime having the authors anonymous even to editors, runs in stark contrast to the movement for increasingly open science and by extension open peer review. Anonymity is applied to increase objectivity of a process, however the drive for publishing reports, sometimes signed, together with other open science projects, is justified in the name of transparency, also in the name of objectivity. The negotiation between drives for openness, particularly with the advent of the Internet, and the drive for objectivity through confidentiality continues. We saw in Chapter One some of the open peer review schema that have been put into place, but this drive is also related to the Open Science movement, which will be dealt with in a later chapter. It is hard to imagine that there is one optimal peer review system that will satisfy all stakeholders, but one thing they do have in common is that they rely on trust in the process, and in all the agents involved in the process, as we shall see in the next chapter on Trust.

---

# Chapter Five: Trust

Like any social process, peer review requires trust. Trust in peer review brings up questions of expertise, since one of the things authors have to trust in is the good behavior and participation of appropriate experts in vetting their results. Philosophical issues brought up by trust include the reasons for trusting, and their nature. They can be cognitive, affective, or a combination of the two. Power relations can also affect the nature of a trusting relationship. In addition, it is an open question whether trust is a thing like trustworthiness or if it is localized in the relationships in question. In the following I will first sketch out the trusting relationships in peer review and go into more detail on the above conceptions of trust to see how they map onto the system we are studying.

Trust is important not just in peer review but in science in general. As John Ziman writes, "The scientific culture depends fundamentally on personal honesty and mutual trust."[242] We trust published results because they are the outcome of a trusted process, peer review. In addition to a journal's ranking by impact factor, to the academic community, certain journals are seen as more trustworthy than others.[243] Following from the account of witnessing science in Chapter Three, we see in peer review again the influence of witnesses, a reach for an objective evaluation obtained by the multiplication of perspectives, as seen in Chapter One. In science in general we need to trust that results are correct. One way we do this is by the practice of publication, through the mechanism of witnessing science discussed previously, where first referees, then readers, successively become witnesses. Not only does the public need to trust science (and scientists) but scientists need to trust fellow scientists and scientific institutions like journals. As described previously, in Chapter Two, peer review emerged to lend legitimacy to the selection of articles for publication, and hence has a need for trust in its basis.

Peer review is a system of interactions between peers, in which authors, referees, and editors are all counted as peers. We recall Ziman's description of peer review as being an activity that balances "three distinct interests – those of the author, of the editor, and of the referee", and his description of these as peers, "any professional scientist of some standing may be called on from time to time to play any of these roles."[244] With this description, Ziman does not seem to account for professional editors, but since many of these come out of the research community before they end up serving as editors, they are at the very least close to the subject and are often considered as peers. If they have also been through the same formation process described by Longino,[245] then they will share the epistemological values of the authors and referees. However, in peer review, the trust mechanism is mutual but uneven. The editors and the referees both trust that the authors are in earnest when they write of the research they have performed, and also trust that the authors are not fabricating data, or engaging in other scientific misconduct[246]. The authors trust that the editors will handle their papers according to established procedures, otherwise they would not submit to that particular journal.[247] The editors also have to trust that the referees will not evaluate a paper maliciously, and will declare any conflicts of interest, whether real or perceived. While the rewards of publication are great and necessary for professional advancement, trust does a lot of work in peer review. To examine the mechanism of this trust, it is necessary to take another look at the actors.

---

[242] Ziman 2002, p. 267.

[243] T. Wilkie, "Sources in Science: Who Can We Trust?," *The Lancet* 347, no. 9011 (May 1996): 1308–11, https://doi.org/10.1016/S0140-6736(96)90947-2.

[244] Ziman 1984, p. 64.

Why do authors submit to the judgment of journals when they could create a system amongst researchers themselves, or just publish everything online? Rationales for some kind of differentiation between articles through publication has been touched on previously in Chapter One, but if we are to examine trust, why do authors subject their work voluntarily to the gauntlet of peer review? If peer review is purely transactional and only the outcome is important, perhaps we can arrive at an account of trust from a purely cognitive perspective?

## A cognitive account of trust

One way of looking at a motivation for trust in peer review is that it makes sense, logically, to have trust between peers. But what is this nature of peer to peer trust? David Hume contented that human behavior is as predictable as any other natural phenomenon, so one should be able to arrive at an account of trust from this analysis. In *A Treatise on Human Nature,* he considers the nature of mutually beneficial arrangements. People do not naturally gravitate to such arrangements in Hume's model, indeed, for him, any sense of justice we have is not natural. However, when they realize the benefits of cooperation, they take to it quickly by Hume's account. "When this common sense of interest is mutually expressed, and is known to both, it

---

produces a suitable resolution and behavior."[248] To take advantage of cooperation it is necessary that people are aware of the advantages of cooperation, which, Hume holds, is not possible prior to a society.[249] The qualities of the mind, Hume says, are selfishness and confine generosity to a few individuals (friends and family). It is this, together with scarcity in nature of resources, that compels us to construct a system of justice. A sense of public interest is not the prior motivation for justice, and reason alone would not be enough to inspire a system of justice. The impressions that give rise to a sense of justice are not natural to the mind, "but arise from artifice and human conventions."[250] Hume's argument why public interest is not pursued naturally relies on the idea that a society that had public interest as its primary, or at least foundational basis would have no need for laws or rules. In such a society, individuals "would never have dreamed of restraining each other by these rules; and if they pursued their own interest, without any precaution, they would run head-long into every kind of injustice and violence. These rules, therefore, are artificial, and seek their end in an oblique and indirect manner; nor is the interest, which gives rise to them, of a kind that could be pursued by the natural and inartificial passions of men."[251]

When only selfishness and limited generosity operate, cooperation is impossible, even if, rationally, it would make a lot of sense. Hume illustrates this with an analogy, "Your corn is ripe today; mine will be so tomorrow. 'Tis profitable for us both that I shou'd labour with you today, and that you shou'd aid me tomorrow. I have no kindness for you, and know that you have as little for me. I will not, therefore, take any pains on your account; and should I labour with you on my account, I know I shou'd be disappointed, and that I shou'd in vain depend upon your gratitude. Here then I leave you to labour alone: You treat me in the same manner. The seasons change; and both of us lose our harvests for want of mutual confidence and security.[252]" As far as Hume is concerned, we have to train ourselves to enter into these mutually beneficial relationships, in the expectation of a favor returned. This makes a consequentialist case of the goodness of trust. But trust cannot be solely transactional, as evidenced by instances of trust where there are no material benefits at question. To solve this problem, Hume makes an exception for acts of kindness between friends, making thus a distinction between *interested* and *disinterested* actions. *Interested* actions take the form of promises. With promises a moral obligation to carry them out emerges, which Hume localizes in the same mechanism that makes respect others' property. A promise only makes sense inside of a human society with conventions,[253] and Hume also described the convention of "abstaining from the property of others." "It is only a general sense of common interest; which sense all the members of the

---

society express to one another, and which induces them to regulate their conduct by certain rules. I observe, that it will be for my interest to leave another in possession of his goods, provided he will act in the same manner with regard to me."[254] Extrapolating from these interactions, Hume finds that we develop conventions that are the basis of a functioning society. "…the rule concerning the stability of possession… arises gradually, and acquires force by a slow progression, and by our repeated experience of the inconveniences of transgressing it." What's more, from this convention arises "ideas of justice and injustice, as also those of property, right, and obligation. The latter are altogether unintelligible without first understanding the former."[255] Thus, Hume bases his entire idea of a just society on respect for private property. In this vein, when we enter into a situation that demands trust, based on past events, experience, we can be *reasonably certain* things will happen in an expected way. Experience of similar things happening causes us to expect certain events. Thus, trusting starts as belief which turns into experience, which leads to a projected trust onto future encounters.

An account that build a picture of trust on respect for private property seems inadequate, however. Research is not property in anything but the intellectual sense, and the way that scientists build off each others' work seems not quite captured by Hume's model. Scientists can, after all, already be said to be engaged in a common project, that of the advancement of their field, and must form affective bonds in order to collaborate in this fashion. Performing experiments motivated by previously published research is not viewed as theft, but is rather expected behavior for scientists. Therefore private property seems inadequate as a basis for an account of trust in peer review. What if the trust dynamic is instead centered on the roles that people play, as if peer review is a game of sorts?

## The Hardwig Affair

A fascinating debate took place in the pages of *The Journal of Philosophy* some decades ago, offering insight about trust in peer review, if only as a cautionary tale. It started innocuously enough with an article by John Hardwig called "Epistemic Dependence" in 1985.[256] In a discussion of the foundations of belief, considering the intellectual authority of experts, Hardwig asks how we gain knowledge of things that are beyond our epistemic grasp. Even in a scientific collaboration, no one member of the collaboration could run the experiment by themselves, however, collectively they can be said know the entire project through their relationships with other members of the collaboration. The question arises then as to the origin of this knowledge. Hardwig offers two solutions to the dilemma: either vicarious knowledge is possible, or the knowledge is in fact located in the community. The first is problematic because it relies on a form of knowledge where we do not have the evidence for an item of knowledge ourselves, but would know it because another person is known to know it. The second alternative is that knowledge is held collectively, such that, while an individual could not say "I know *p*", they could say "we know *p*," where "we" stands for a community that is not reducible to a class of individuals. Both of these explanations move away from epistemic individualism. Hardwig concluded that he found both alternatives disturbing since this move "reveals the extent to which even our rationality rests on trust and because it threatens some of our most cherished values — individual autonomy and responsibility, equality and democracy."[257]

---

[254] *Ibid.*, Book 3, Section 2, p. 206

[255] *Ibid.,* Book 3, Section 2, p. 207.

[256] Hardwig, John. "Epistemic Dependence." *The Journal of Philosophy* 82, no. 7 (July 1985): 335. https://doi.org/10.2307/2026523.

Two years later, Michael J. Blais wrote a response to Hardwig's article, "Epistemic Tit For Tat", in which he declared both of Hardwig's alternatives unacceptable.[258] Blais utilizes an iterative version of the classic game theory exercise the Prisoner's Dilemma[259] to reframe Hardwig's account of trust in a strategic, rather than moral, sense. The version of the Prisoner's Dilemma articulated by Blais involves an imagined exchange in which you agree to a transaction with a supplier that the goods and the money for the goods be left in two prearranged places, one for the money and the other for the goods. The question to be answered is whether it is better to keep your word (cooperate) or not (defect). Blais's example involves recurring transactions and utilizes game theory to try to elucidate the best strategy. He uses this analysis to try to examine how trust works in science, and what the reasons for trust are, and sets out trusting strategies that, in his analysis, further the progress of science. Identifying that the reward in science, if treated like a Prisoner's Dilemma problem, is reputation, Blais makes the connection between reputation and publication in journals, noting that cooperation is the best strategy here, and that the short-term benefits of defection are outweighed by the consequences of exclusion, making cooperation strategically favorable and promoting a climate of trust. "During apprenticeship, one soon learns that short-term gains of defection are far outweighed by the long-term rewards of cooperation, whether one is morally trustworthy or not. To be allowed to stay in the game requires cooperation, but the punishment is permanent exclusion."[260]

The next year, John Woods published a rebuttal in the same journal. Woods argues, also from game theory, that it is misleading to assume that cooperative dynamics in science follow a Prisoner's Dilemma model. Taking as an example the submission of an article to a journal for publication, Woods claims that "The conservatism of scientific acceptance (tough peer review, high rejection rates by the good journals, etc.) is important, but it is overdescribed in the minimax strategy[261] of the classical game in which the best course for a author in a one-shot submission to a scientific journal is to send off a cheating piece of work, and the best course for the journal is to turn it down sight unseen."[262] Continuing in this framework, Woods describes the journal's decision of rejection as "defection" in game theory parlance.[263]

---

[257] *Ibid.,* p. 349.

[258] Blais, Michel J. "Epistemic Tit for Tat." *The Journal of Philosophy* 84, no. 7 (July 1987): 363. https://doi.org/10.2307/2026823.

[259] *Ibid.,* p. 365-368.

[260] *Ibid.,* p. 372.

[261] A game theory decision rule to minimize the possible loss for a worst-case scenario.

This provoked a response from Blais in 1990 in which, in addition to disputing Woods' analysis, he  also made the case the Woods had mischaracterized the function of a journal in his game theory characterization.

> When two players (an author and a journal) play the knowledge game within the scientific framework, each cooperates by doing the job well and defects by not doing it well. For the author, doing his job well (cooperation) consists in submitting only correct results that have not been plagiarized, to the best of his knowledge; not doing his job well (defection) consists in cheating. On this, Woods and I are in good agreement. For the journal, cooperation consists in correctly reviewing submitted papers, publishing good papers as space permits, rejecting bad ones, refraining from nepotism, and, in general, being as objective as possible; defection, on the other hand, consists in haphazard reviewing, nepotism, favoritism, nonobjectivity, refusing to publish good papers for bad reasons, or publishing shoddy material. A journal has not defected just because it refuses to publish a paper; on the contrary, the journal is doing its job—is cooperating—when it rejects a bad paper. The journal defects when it refuses to publish a good paper, not when it simply refuses to publish.[264]

Blais concludes by reiterating his conclusion that "moral trust is not a necessary foundation for the reliability of the accumulating knowledge afforded by the methods of cooperative scientific investigation, that, as Axelrod[265] states, `[T]he foundation of cooperation is not really trust, but the durability of the relationship.'"[266]

However, this was not the last word in this exchange. The next year, John Hardwig published a new paper in *The Journal of Philosophy*, "The Role of Trust in Knowledge", which addressed all of the previous exchange by Blais and Woods. Against the objections of epistemologists, who overwhelmingly claim that knowledge rests on evidence, not trust, Hardwig states that "Modern

---

knowers cannot be independent and self-reliant, not even in their own fields of specialization."[267] Writing more specifically about peer review, and its role in detecting scientific misconduct, Hardwig added that "The number of really well-qualified referees for peer reviews is often inadequate, given the quantity of articles submitted and the complexity and multiplicity of techniques involved in research."[268] Hardwig here is concentrating on what role peer review can play in ferreting out scientific misconduct,[269] but peer review is notoriously poor at this. John Ziman agrees, "And yet, in spite of peer review and other safeguards, it is relatively easy to get fraudulent results into the literature, and to profit from them careerwise for a while."[270] So if peer review cannot spot these mistakes, what is it good for?

## Peer review and trust

Ironically, the reason why peer review is poor at detecting scientific misconduct is just because of this climate of trust. What the peer review process is actually for is critique inside the culture of trust. To work, peer review assumes that the actors are acting in good faith. Overall, all actors trust the other participants to adhere to some kind of code of conduct. It is expected that authors do not fabricate results, plagiarize, or leave colleagues who did significant work off the author list[271]. It is expected that referees declare any conflicts of interest to the editor, do not seek to torpedo papers with fallacious arguments, and do not attempt to use their power as referees to lobby for citations of their own work[272]. It is expected that the editor does not set out deliberately to compromise the procedure by fabricating reports, colluding with referees, and ignore journal policy in order to get the result they want[273].

A conspicuous case where trust broke down in peer review was the Sokal affair. A physicist, Alan Sokal, submitted a hoax article called "Transgressing the Boundaries: Towards a

---

[267] Hardwig, John. 1991. "The Role of Trust in Knowledge." The Journal of Philosophy 88 (12): 693–708. https://doi.org/10.2307/2027007 p. 693.

[268] *Ibid.,* p. 703.

[269] "The phenomenon of scientific misconduct reveals that a more thorough-going trust than mere strategic trust is involved in science. The consensus within the biomedical sciences is that neither peer review nor replication is likely to detect careless, sloppy, or even fraudulent research." *Ibid.,* p. 703.

[270] Ziman 2002, p. 267.

[271] See for instance https://www.nature.com/authors/editorial_policies/authorship.html; https://www.sciencemag.org/authors/science-journals-editorial-policies; https://publicationethics.org/authorship

[272] See for instance https://www.nature.com/nature-research/for-referees; https://www.sciencemag.org/authors/peer-review-science-publications

[273] See for instance https://www.springernature.com/gp/editors/code-of-conduct-journals; https://publicationethics.org/files/Code_of_conduct_for_journal_editors_Mar11.pdf

Transformative Hermeneutics of Quantum Gravity[274]" to a journal outside of his professional field: *Social Text*.[275] The paper contained blatantly erroneous statements but was dressed up in what Sokal referred to as "fashionable nonsense", terms that he had gleaned from the field of postmodern cultural studies. Sokal unmasked himself in the pages of the now-defunct literary magazine *Lingua Franca* revealing his intentions. "While my method was satirical, my motivation is utterly serious. What concerns me is the proliferation, not just of nonsense and sloppy thinking *per se*, but of a particular kind of nonsense and sloppy thinking: one that denies the existence of objective realities, or (when challenged) admits their existence but downplays their practical relevance."[276]

The reactions to the hoax sparked a number of exchanges[277] and eventually an expanded account of Sokal's analysis of the state of postmodernism in a collaboration with Jean Bricmont.[278] But ultimately, for the status of peer review, a case can be made that the hoax proved nothing but naivety and sloppiness on the parts of the editors.[279] This and other less-publicized cases of misconduct do not seem to undermine the reverence in which peer review is held. When Jan Hendrik Schön was found to have fabricated data in several publications in 2002,[280] the response was not to put the institution of peer review under greater scrutiny. On the contrary, one of the sanctions passed down against Schön in 2004 by the DFG, the Deutsche Forschungsgemeinschaft (German Research Foundation), was an interdiction forbidding him to serve as a reviewer himself for eight years.[281] In fact, when scandals in peer review do break they are usually explained away as being due to individual bad actors.[282] It should be noted that it is hard to track scientific misconduct since many cases of fraudulent submissions are simply not published. The site Retraction Watch (https://retractionwatch.com), which covers retractions of papers published in scientific journals, maintains a database.[283] Searching this reveals that 1,113 retractions occurred in 2015.[284] In that year, according to Web of Science, 2,906,121 articles were published in the journals they index. Of course, some of these retractions are due to honest mistakes, not scientific misconduct, but even casting the net this wide does not give the impression scientific misconduct is a substantial problem in academic publishing.

## Power relations and trust

The consequences of scientific or publishing misconduct are serious, and indicate the high stakes at play in peer review. In terms of trust, this raises the issue of power differentials in peer review as a philosophical concern. Sissela Bok, in her book *Lying*, examined the question of trust from the perspective of both liars and the deceived, and developed and account centered on an atmosphere of trust.[285] Her account led Annette Baier to present a treatment of trust and relative power. Baier starts from an analysis of trust from purely cognitive reasons, but observes, from previous work done on trust, that an excessive focus on contractarian relationships leads to an impoverished picture of trust "The domination of contemporary moral philosophy by the so-called Prisoner's Dilemma problem displays most clearly this obsession with moral relations between minimally trusting, minimally trustworthy adults who are equally powerful.[286] [...] For the more we ignore dependency relations between those grossly unequal in power and ignore what cannot be spelled out in an explicit acknowledgment, the more readily will we assume that everything that needs to be understood about trust and trustworthiness can be grasped by looking at the morality of contract."[287] This brings up an important point that the agents are not entirely free to take or leave publishing in journals. For authors, it is a necessary process that is part of one's identity as a researcher, as we have seen. Likewise referees feel pressure to participate in peer review out of an obligation towards their communities.

As Baier puts it, when we have no choice, trust becomes less an expression of autonomous attitudes and more about reliance. "We may have no choice but to continue to rely on the local shop for food, even after some of the food on its shelves has been found to have been poisoned with intent. We can still rely where we no longer trust."[288] How then do we avoid falling into the

---

trap of reducing trust in peer review to reliance? What is the difference between trusting others and merely relying on them? It seems to be reliance on their good will toward one, as distinct from their dependable habits, or only on their dependably exhibited fear, anger, or other motives compatible with ill will toward one, or on motives not directed on one at all.[289] Of course there are dependable habits of authors, referees, and editors, but the motivation for all of these actors is one of service to a scientific community. As we highlighted earlier in this chapter, the consequences of misconduct can be very serious, but fear of consequences is not the only thing that keeps agents in peer review from behaving badly. Authors submit to journals because they have to publish somewhere for their work to count as science, and referees are also authors who depend on the smooth functioning of the system. The stakes are simply too high to risk not publishing research.

It would be a mistake, however, to take from Baier's analysis that trust between anything but "articulate adults, in a position to judge one another's performance"[290] is simple reliance. Baier as an example treats infant trust, where some trust is needed even as the infant realizes they are completely dependent on the parent. Total dependence does not necessarily equal trust, infants can make gestures of rejection of the mother and the mother's breast, "but surviving infants will usually have shown some trust, enough to accept offered nourishment, enough not to attempt to prevent such close approach."[291] There is still a trust relation here even if it is not contractual. For relations of less drastic inequality, Baier also treats the subject of traditional marital relations, which is traditionally not an arena of relations between parties of equal power.

This is not to say that the peer review trust relations are intimate as described by Baier, nevertheless, the relations inside of peer review are neither straight-forward nor symmetrical. The authors fears ill-will from referees and, to a lesser extent, editors. The editors fear fraudulent results and dishonest authors who would try to game the peer review system to publish inferior manuscripts. The referees fear being taken advantage of and overburdened by editors. The authors may be the supplicants in peer review, wanting something (publication) out of the process, but the whole process relies on everyone doing their job. Certainly authors submit manuscripts to journals in order to have them published, but that publication is meaningless unless they obtain the four things that journals can give, registration, archiving, dissemination, and validation through peer review, so they have to trust the journal to do its job, and trust that editors will pick referees who will evaluate the manuscript honestly. This is not a contractual obligation, authors can submit elsewhere, referees can decline to review, (editors are the only exception here), but the expectations are set by the community. As Baier puts it, "Trust of any particular form is made more likely, in adults, if there is a climate of trust of that sort. Awareness of what is customary, as well as past experience of one's own, affects one's ability to trust. We take it for granted that people will perform their role-related duties and trust any individual worker to look after whatever her job requires her to. The very existence of that job, as a standard occupation, creates a climate of some trust in those with that job."[292] What the agents in peer review trust the others to do is not something specific like accept the paper but play the role they are supposed to play. Be a critical referee, Be an honest author. Be a balanced editor. The mechanism of trust in peer review is a mechanism of *trust between peers*. The validity of the judgment is guaranteed by the process by which one is evaluated by ones peers, with shared norms and notions of objectivity, as we have already seen described by Longino in Chapter Three. As being part of a community is an affective relationship, how does this view on trust play into peer review?

---

[289] *Ibid*, p. 234.

[290] *Ibid*, p. 240.

[291] *Ibid*, p. 241.

## An affective account of trust

A more personal model of trust that bases itself on affective attitude is offered by, amongst others, Karen Jones, who has a model of trust that may be instructive to look at for peer review. Hers is a two-part model, part cognitive, part affective. She characterizes trust as an attitude of optimism about *goodwill* and *competence*.[293] By these terms she means goodwill about you, and competence to fulfill your request. If A trusts B, the idea is that B will be moved to fulfill A's request because B knows that A is *counting* on B to do so. A is also optimistic that B has the requisite competence to perform the task at hand. Jones regards such a person as *trustworthy*, though "one is not trustworthy unless one is willing to give significant weight to the fact that the other is counting on one, and so will not let that consideration be overruled by just any other concern one has."[294]

Optimism does not mean here "to look on the bright side". This is not to say that the optimism in Jones's mechanism is a qualified or restricted optimism in terms of its magnitude, rather that the optimism only extends to the domain of the task. "For example, the optimism we have about the goodwill and competence of strangers does not extend very far. We expect their goodwill to extend to not harming us as we go about our business and their competence to consist in an understanding of the norms for interaction between strangers."[295]

The attitude so described seems affective, but it has cognition at its core. The optimism is grounded on experience about the trusted, or about people like them, in similar positions. By contrast, *distrust* is then pessimism about goodwill and competence. It would seem that there only has to be distrust about one of these two for the trusting relationship to fall apart. In addition to both goodwill and competence having to be in place, an *expectation*, cognitively derived, has to be added to the affective optimistic attitude, in order to have an adequate account. "The affective element of trust needs to be supplemented with an expectation, namely, the expectation that the one trusted will be directly and favorably moved by the thought that someone is counting on her."[296] Jones illustrates this by considering two examples, those of unwelcome trust and the case where a generally benevolent person falls short of fulfilling their task.

---

[292] *Ibid*, p. 245.

[293] Jones, Karen. 1996. "Trust as an Affective Attitude." *Ethics* 107 (1): 4–25.

[294] *Ibid.*, p. 8.

[295] *Ibid.*, p. 6.

For the case of unwelcome trust, the trust someone has in us may feel coercive, in that we do not welcome the trust placed in us "[W]hat we object to when we do not welcome someone's trust is that, in giving it, she expects that we will be directly moved by the thought that she is counting on us and, for one reason or another, we do not want to have to take such expectations into account, across the range of interactions the truster wants."[297] This seems understandable as a complaint, but is scarcely relevant to peer review, where the agents cannot really withdraw their labor from the peer review system *as a whole*, even if authors can choose different journals and referees can decline to review papers for a host of reasons.

The second case, of a generally benevolent person who falls short, is illustrated by its obverse, the person who is generally benevolent towards you but is not particularly moved by you trusting them. In Jones's example, a doctor, but who will treat your condition whether they like you or not. "Suppose that the only operative motive in your interaction with me is concern about my well-being. Regardless of what I count on you to do, you do it only if it maximizes my well- being, and if it does that, you would do it anyway, whether or not I counted on you to do so. I would be justified in having an attitude of optimism about your goodwill while refraining from seeing you as trustworthy."[298]

But both of these cases seem to fall outside of the peer review dynamic. Much like the purely cognitive model, it is questionable if the affective model is really an adequate description of peer review. In peer review expectations are of *fairness* and competence, not goodwill and competence.[299] Furthermore, what motivates referees to review a paper is not a function of goodwill towards the authors, rather a feeling of obligation and duty towards the community.[300] A certain amount of *quid pro quo* is also present; it is unreasonable to expect good and prompt reviews on your own papers if you yourself are unwilling to do your part as a referee. If there is any expectation of goodwill, it will be the minimal one that Jones describes when discussing trust between strangers, which makes the account of the affective model of trust less suitable for discussing trust in peer review. Unlike most trusting relationships, there can be no expectation of the outcome of review process, however, one can have an expectation of the nature of the relationship with the other agents. The expectation is of the nature of fulfillment, authors expect editors to pick competent and prompt referees, heed advice, and not be arbitrary. Referees and editors trust that the authors have in fact done what they write about. Additionally, editors trust that referees will not use the reviewer privilege, consisting of access to pre-published information, to their own advantage and the authors' detriment. Finally, all agents carry expectations from the journal that they will abide by the policies and practices in line with a

---

[296] *Ibid.*, p. 8.

[297] *Ibid.*, p. 9.

[298] *Ibid.*, p. 10.

[299] *Ibid.*, p. 8.

[300] Ziman 2002, p. 42.

journal's perceived identity.[301]

The requirement of goodwill in many trust models is complicated by another aspect of the role of peers, the community norm of skepticism. It has been recognized for a long time that science thrives on healthy skepticism, indeed, about a thousand years ago the Arab scholar Alhazen wrote, of evaluating scientific work, that "The duty of the man who investigates the writings of scientists, if learning the truth is his goal, is to make himself an enemy of all that he reads, and, applying his mind to the core and margins of its content, attack it from every side. He should also suspect himself as he performs his critical examination of it, so that he may avoid falling into either prejudice or leniency."[302]

This skepticism need not be hostile, but the stance is significantly less positive than the optimism of goodwill. From the vantage point of authors, seeking publication of results and thus subjecting one's work to the sometimes harsh judgment of peer review may be viewed as a necessary evil, or a requirement of the academic system, but the relationship can certainly not be characterized as coming from anything other than professional goodwill. Scientists take care to avoid conflicts of interest and routinely decline to review the works of personal friends, or, if editors, refuse to handle submissions of people if they would not be able to have the work evaluated without a conflict of interest. The authors may well have expectations of speedy handling and a smooth process, but these are localized in the importance of the results and not in some feeling that the editors have goodwill towards authors on a personal level. In fact, a journal that gets a reputation of editors being motivated more by goodwill towards (some) authors than by commitment to rigorous peer review soon becomes *distrusted*. Of course, editors will use their experience ("so-and-so usually does interesting work" in the pursuit of the evaluation), but in a rigorous review process, the input of referees is required for an "objective" evaluation, rather than publication decisions being made by editorial fiat.

This is not to say that good intentions are not important. The dynamic of peer review is susceptible to the same issues as Hume's free rider problem, in that the burden of work will not be even. In this objection which Hume himself raised to his account, as one scales up the number of participants, the accountability of individuals is decreased.

"Two neighbours may agree to drain a meadow, which they possess in common; because 'tis easy for them to know each other's mind; and each must perceive, that the immediate consequence of his failing in his part, is, the abandoning the whole project. But 'tis very difficult, and indeed impossible, that a thousand persons shou'd agree in any such action; it being difficult for them to concert so complicated a design, and still more difficult for them to execute it; while each seeks a pretext to free himself of the trouble and expence, and wou'd lay the whole burden on others."[303]

In peer review, some agents will end up being disproportionally more active referees than

---

others. If a journal seems to utilize only a select coterie of reviewers, accusations of cliquishness may result, which erodes trust in a journal from those who see themselves as outside this clique. In spite of this inequality, participation in peer review, whether as an author or a referee, is viewed as being part of the scientific persona.

## Expertise and trust

It is worth reminding ourselves of what is expected on peers inside the peer review dynamic. Peer review is in essence the interaction of experts. Expertise in general brings forth the following philosophical issues: who gets to call themselves an expert,[304] the demarcations between an in-group and out-group,[305] and, for peer review in particular, the idea that the institution of peer review is underpinned by expertise and hence derives its legitimacy from it.[306] We have seen the effects of this institution of peer review previously in the introduction, and note that the term "peer review" was coined as a means of drawing boundaries about who could and could not evaluate science.[307] Expertise is a central plank of peer review. The conversation and interactions happen between experts. There is no entry into this from non-experts.

Much writing on expertise concentrates on experts as sources of information, and this is indeed how the general public often experience experts, in the form of, for instance, expert witnesses in social and political contexts[308]. Evan Selinger and Robert Crease contrast traditional philosophy of science with Science and Technology Studies (STS) in terms of how they deal with expertise and find them both deficient. The former, traditional philosophy of science, focuses on scientific creativity but does not examine where that creativity comes from, whereas STS "treats expertise as "distributed," externalized into particular settings such as laboratory and social networks, and standardized in technologies, criteria of scientificity, protocols for evaluating proof, and the rhetorical means of recruiting allies."[309] Selinger and Crease by contrast want to introduce a phenomenological analysis of expertise, drawing on Hubert Dreyfus's model of expertise, that we saw in Chapter Three. We recall his five-stage process, and that the transition at play here is the one from "knowing that" to "knowing how", with the expert being completely engaged in skillful performance by the end of the transition.[310] Such an expert is the peer in peer review when dealing with what Fleck calls vademecum science, but not, crucially, when dealing with journal science, a place where things do not "proceed normally." Nevertheless, in the peer review mode, the expert referee needs to reconcile journal science both with other journal science and with vademecum science, which can only be done if the engagement with vademecum science is done in the effortless manner described by the Dreyfus brothers' final stage of expertise.

---

[304] "Scientific recognition takes a variety of forms, graded to the various stages of a successful career." Ziman lists as examples of these publications, attributions in the form of having a discovered effect, theory, or method named after one, honorific awards, etc. Ziman 1984, p. 70-71.

[305] "What it means to be a `specialist' is to be in there with all these goings on, twenty-four hours a day. To be a non-specialist is not to be in there. I you are outside, things inevitably become simplified. " Harry Collins, *Are We All Scientific Experts Now?*, New Human Frontiers Series (Cambridge: Polity, 2014), p. 85.

[306] "The referee is the lynchpin about which the whole business of Science is pivoted. His job is simply to report, as an expert, on the value of a paper submitted to a journal. " John M Ziman, Public Knowledge: An Essay Concerning the Social Dimension of Science (Cambridge: Univ. Pr., 1974), p. 111.

[307] "'Peer review' was a term borrowed from the procedures that government agencies used to decide who would receive financial support for scientific and medical research. When 'referee systems' turned into 'peer review', the process became a mighty public symbol of the claim that these powerful and expensive investigators of the natural world had procedures for regulating themselves and for producing consensus, even though some observers quietly wondered whether scientific referees were up to this grand calling" Csiszar 2016, p. 308.

[308] Roger A. Pielke, *The Honest Broker: Making Sense of Science in Policy and Politics* (Cambridge ; New York: Cambridge University Press, 2007).

This model of expertise bears upon peer review because the way in which expertise operates inside of peer review is in practice. It is also compatible with the induction of peers as described by Longino in Chapter Three.[311] Referees are asked to evaluate the process of how results were obtained, not to repeat experiments or theoretical work themselves. For this reason the expertise that is called on for reviewing is the expertise of doing, not a mastery of assembled facts. Since what we are talking about here is journal science, in Fleck's parlance, evaluation cannot happen solely by consultation with vademecum science, which might be done by any dedicated fact checker, but instead requires witnessing a process and judging it appropriate and valid.

In other words, what is being trusted is not so much the goodwill of actors but their expertise, their "professional behavior". This is more the trust towards a mechanic in some sense, but it is different in that the trust is inside of a community. It's like a mechanic who knows all the tricks unscrupulous mechanics might play.

There is another dimension of the role that expertise plays in the peer review process, being that of the editor. As we saw in Chapter One there is a great variety of editors and the degree of their involvement in a field can vary sharply from journal to journal. There is nothing particularly unusual about the expertise of referees as compared to the authors they review. Referees are however selected not only on their expertise but more on their ability to parse and uphold the standards of a journal. In a way the expertise of editors in handling peer review is actually more interesting since it is a skill that depends on the social relationships and not so much on knowledge of science. Harry Collins has expounded on the idea of interactional expertise, particularly from the perspectives of scholars of Science and Technology Studies, in how they relate to the fields and communities they study. Put simply, if one studies research in (for example) gravitational waves one will necessarily absorb enough knowledge about gravitational waves to qualify for a degree of expertise. Collins, writing with Robert Evans, makes a distinction between three levels of expertise:

1) No Expertise: That is the degree of expertise with which the

fieldworker sets out; it is insufficient to conduct a sociological analysis or do quasi-participatory fieldwork.

2) Interactional Expertise: This means enough expertise to interact interestingly with participants and carry out a sociological analysis.

3) Contributory Expertise: This means enough expertise to contribute to the science of the field being analysed.[312]

Interactional expertise is not restricted to STS researchers, sometimes this is the best way to describe lab directors, project managers, and other people who are adjacent to whatever scientific work is occurring. Collins and Evans calls such expertise "referred expertise", which is in general not contributory expertise, but is different to pure interactional expertise. It is tempting to conclude that inasmuch as the manager contributes anything, it is management, not science. However, one needs to be conversant with the science in order to fill this role. As Collins and Evans write:

The resolution seems to be that to manage a scientific project at a technical level requires, not contributory expertise in the sciences in question, but experience of contributory expertise in some related science. In other words, the managers must know, from their own experience, what it is to have contributory expertise; this puts them in a position to understand what is involved in making a contribution to the fields of the scientists they are leading at one remove.[313]

This also seems to describe well the expertise of editors. Editors are often not direct experts in the fields of the papers that they handle, and at any rate, if they are professional editors, they will not primarily be active researchers.[314] However, a large part of managing peer review consists in discerning between experts. Elsewhere, Collins writes of "meta-expertise", which is the ability to discriminate between experts. This discrimination takes several forms, and is helped by a general understanding of the related science, but also a knowledge of the individuals involved and their context, backgrounds, and idiosyncrasies. In addition technical knowledge, the assessment of the skills of someone in a related field, and the above-mentioned referred expertise,[315] are all tools in an editor's toolkit.[316]

---

[312] Collins, H. M., and Robert Evans, "The Third Wave of Science Studies: Studies of Expertise and Experience," Social Studies of Science 32, no. 2 (April 1, 2002): 235–96, https://doi.org/10.1177/0306312702032002003, p. 254.

[313] *Ibid.*, 2002, p. 257.

[314] Hirschauer, p. 84.

[315] Collins 2014, p. 74-79.

[316] J.G. Ray, "Judging the Judges: The Role of Journal Editors," *QJM* 95, no. 12 (December 1, 2002): 769–74, https://doi.org/10.1093/qjmed/95.12.769; J Smith, "Journalology--or What Editors Do.," *BMJ* 301, no. 6754 (October 3, 1990): 756–59, https://doi.org/10.1136/bmj.301.6754.756; David Card and Stefano DellaVigna, "What Do Editors Maximize? Evidence from Four Leading Economics Journals" (Cambridge, MA: National Bureau of Economic Research, March 2017), https://doi.org/10.3386/w23282.

## Trust and distrust

A way of looking at trust that draws on Dreyfus's conception of expertise comes from Robert C. Solomon. In discussing trust, Solomon creates a dynamic picture of trusting as a shared practice,[317] which has trust as a continuous social practice rather than a specific state of mind. "Trust is invisible (or `transparent') in many (or most) trusting relationships insofar as it tends to provide the background rather than the focus of our activities."[318] It's not quite as simple as trust being an atmosphere or medium however, except for inasmuch as we exist inside of an atmosphere without paying particular attention to the atmosphere as such. But to take trust for granted is dangerous, "it ignores trust just when we need to be most concerned about it, when trust is flagging or betrayed, when trust is gone and needs to be replaced."[319] Trust is invisible only in a sense, in the sense that "we are not paying attention to it and, perhaps, are so skillfully engaged in it that we could not describe what we are doing even if asked."[320] This seems very reminiscent of the "expert" mode of operation of expertise as described by Dreyfus, with the expert totally engaged in the skillful perfomance of practice.

In a system of peers the expectation is that the agents can be trusted, though one cannot rely on the outcome of a peer review process. This account seems to apply to the case of peer review very well, since each interaction in the peer review process can be viewed as its own trust relationship. Solomon outlines how in every trusting relationship individual acts and their reception forms the relationship, "[T]rust in relationships is built out of such "trivia," such routine everyday making and keeping of promises and committments. I want tu suggest that trust in any relationship is built out of (and destroyed in) such routine  frustrations, promises, and committments."[321] Indeed, there is nothing to say an author should put the same trust in referees A and B both for the same submission, especially if their critique diverges. The fact that these are peers is something that is guaranteed by the editor, or rather by the journal's policies. The trust might in fact then be placed in the journal as an institution, that such and such a journal is known for its rigorous peer review.

What is most interesting about Solomon's account from the point of view of peer review is his take on *distrust*, which bears considering in the context of the skepticism that is part of the reach towards objectivity of the scientific method, as seen in Chapter Three. Distrust is not the opposite of trust but "an essential aspect of trust itself."[322] that against which trust is defined. Neither is distrust a failure of trust or a preliminary to trust, in the vein of "we will wait to see if we can trust you", but rather a tempered trust. To Solomon, distrust is an essential aspect of trust itself, going as far as to say that the dialectic between trust and distrust is "the most exciting" part of trust.[323] To illustrate, if one has no experience with someone, one does not by default trust them, but it would be wrong to say this automatically means one *distrusts* them.

---

[317] "Trusting", Solomon, Robert C. in Dreyfus, Hubert L., Mark A. Wrathall, and Jeff Malpas, eds. 2000. *Essays in Honor of Hubert L. Dreyfus*. Cambridge, Mass: MIT Press, vol 2., 229-244; p. 233.

[318] *Ibid.,* p. 236.

[319] *Ibid.,* p. 236.

[320] *Ibid.,* p. 237.

[321] *Ibid.,* p. 239.

Solomon outlines three types of trust that he categorizes by their relation to distrust, being simple trust, which is childlike and devoid of mistrust; blind trust which is absolute and denies even the possibility of distrust; authentic trust, which recognizes the possibility of distrust and betrayal, but has resolved on side of trust. "I distinguish between *simple trust*, naive trust, trusting as yet unchallenged, unquestioned […], *blind trust*, which is not actually naive but stubborn, obstinate, possibly even self-deluding, and *authentic trust.*"[324] This latter one is the trust that is of interest in considering the mechanics of peer review. Authentic trust exists in a dialectic with distrust, it "embraces distrust and involves the willful overcoming of it."[325]

Peer review seems to exist in this strange balance of trust and distrust. Pure distrust would not work in peer review, as attempts to simplify peer to a manifestation of game theory makes clear, as seen earlier in this chapter.[326] This conclusion can also be drawn from the treatments of Hume, Baier, Jones and indeed Solomon. Trust means we sometimes have to trust in outcomes that are not certain. At the same time, excessive blind or simple trust would, in the case of peer review, just lead to everything being published, regardless of apparent validity, which would go against the culture of skepticism.[327] The consequences of publishing everything, even if one does not filter on criteria as importance and interest, would undermine the scientific project as an epistemological activity. Science could survive a series of trivial publications such as investigations of the boiling point of water at sea level, but if papers emerge where this boiling point is claimed to be 100, 157, or 32 degrees Celcius and if there is no way to differentiate between these accounts, then no knowledge can be gained from science.

---

[322] *Ibid.,* p. 241.

[323] *Ibid.,* p. 241.

[324] *Ibid.,* p. 241.

[325] *Ibid.,* p. 242.

[326] "Pure distrust, as Kant and almost anyone else has speculated, is no basis for a coherent society." Solomon, p. 243; "The ultra-Hobbist child who fears or rejects the mother's breast, as if fearing poison from that source, can be taken as displaying innate distrust, and such newborns must be the exception in a surviving species." Baier, p. 241.

[327] "Scientists take nothing on trust. That is to say, scientific knowledge , whether new or old, should be continually scrutinized for possible errors of fact, or inconsistencies of argument." Ziman 1984, p. 85.

### Conclusion on trust in peer review

To sum up, out of the accounts considered, the nature of trust in peer review seems to cleave relatively close to Solomon's account of *authentic trust*, existing as it does in constant dialogue with distrust. When considering trust in peer review, the question of objectivity looms large.  In keeping with the interpretation that peer review has some kinship with the kind of account of science that relies on the scientific model, the trust seems to be a hope that agents will dispassionately view and evaluate the science.  The motivations of different agents vary, but the hope is that the outcome of the process will be beneficial to all parties.  Authors will be published, or at the very least receive constructive feedback that they can use to improve the work, if the matching to the journal was not as complete as had been hoped.  Referees will be awarded a chance to participate as representatives of their community, influencing the work of their peers. Editors will for their part hope to uphold the standards of their journals and publish only work that qualifies, while improving submitted papers along the way as a second-order benefit.

Science as well as peer review requires trust. Empiricism requires skepticism, which is very close to distrust. To obtain an objective process, we try to eliminate subjectivity. But trust is subjective, and so is distrust. The trust is in people, not a reliance on senses or on instruments. Instead of reliance, peer review relies on trusting the actors to uphold a form of objective judgment and to not be damagingly subjective. This trust stands in contrast to the fact that science is built on skepticism, even distrust. However, by shared epistemological standards, it can be said that everyone has the same level of distrust, and share a scientific community's definition of objectivity in coming to the judgment in the dynamics of peer review. What happens, however, when we consider these mechanisms of trust outside of the immediate scientific community?

# Chapter Six: Opening up the black box of peer review

In the previous chapter, we considered the trust mechanisms inside of peer review, but peer review is also something that is trusted by the exoteric circle, the general public. Recently there has been a drive towards opening up science.[328] Calls for opening up science manifests themselves in two main ways, opening up the assessment protocols that control not only what science is published[329] but what science is attempted at all, [330]and opening up access to published research beyond a small specialized community.[331] Both of these movements have philosophical implications. The first extends from the trust discussion in the last chapter, as the public takes more of an interest in the production of science, but also complicates the earlier discussions on objectivity and anonymity. The second is intimately related to the first, and is also motivated by a sense of the appropriate public ownership of science, especially that carried out using public funds. The question of Open Access, especially concerning some models with mandates by funders on acceptable forms of publication and venues for scientific results, affects the role of the author and complicates the dynamics of the self-regulating peer community.

---

[328] Benedikt Fecher and Sascha Friesike, "Open Science: One Term, Five Schools of Thought," in *Opening Science: The Evolving Guide on How the Internet Is Changing Research, Collaboration and Scholarly Publishing*, ed. Sönke Bartling and Sascha Friesike (Cham: Springer International Publishing, 2014), 17–47, https://doi.org/10.1007/978-3-319-00026-8_2.

[329] Elizabeth Walsh et al., "Open Peer Review: A Randomised Controlled Trial," *British Journal of Psychiatry* 176, no. 1 (January 2000): 47–51, https://doi.org/10.1192/bjp.176.1.47; Emily Ford, "Defining and Characterizing Open Peer Review: A Review of the Literature," *Journal of Scholarly Publishing* 44, no. 4 (July 2013): 311–26, https://doi.org/10.3138/jsp.44-4-001.

[330] Tony Ross-Hellauer, "What Is Open Peer Review? A Systematic Review," *F1000Research* 6 (August 31, 2017): 588, https://doi.org/10.12688/f1000research.11369.2.

[331] Peter Suber, *Open Access* (Cambridge, Mass.; London: MIT Press, 2012).

## Opening peer review

Regarding the question of opening peer review, as we have already seen, there are many different versions of peer review that have been suggested or implemented with the motivation of providing a better process, and some of these operate through increased openness. These have consequences for the reach towards objectivity as discussed in previous chapters. For instance, a scheme of peer review that jettisons anonymity in favor of open review, with published reports, and naming referees, is banking on the idea that the benefits of accountability outweigh the downsides of fear of repercussions for a negative reviewer.[332] Beyond this matter-of-fact statement, what this does is prioritize subjectivity in one domain (personal accountability) and does away with the idea of objectivity being best attained in a closed and confidential space. It is interesting to note that at the same time that these approaches of opening up peer review are suggested, there is also a strong push for making the review process mutually anonymous or "double blind" in fields where this is not already standard operating procedure, fields where anonymity of authors has not been the norm, as seen in the chapter on anonymity. Making more agents inside the peer review process anonymous is seemingly motivated by a desire to increase the objectivity of the process, by controlling for bias from the other direction, that is, bias from referees onto authors. Unlike the open review system, this approach does not seek to increase accountability, but rather to protect the referee from themselves.

However, mutual anonymity is not as much as a departure from the norm as some of the other ideas for opening up peer review. A particularly open idea of peer review is crowdsourced review, in which any community member can contribute to article review. The referees are not selected by the editors and reviewing is open to all, without any limit on the number of reviews. There can be some editor mediation, but it is not a requirement.[333] Reconnecting with Helen Longino, crowdsourced peer review seems appropriate for her model of criticism, since she values a multiplicity of perspectives.[334] However, Longino also talks about criticism stemming from the social values of a community,[335] so how does this desire for a greater range of perspectives square with scientific communities, which, due to the epistemic burden of attaining enough expertise, is necessarily limited? As we recall, Longino has four criteria for community evaluation of research, "(1) there must be recognized avenues for the criticism of evidence, of methods, and of assumptions and reasoning; (2) there must exist shared standards that critics can invoke; (3) the community as a whole must be responsive to such criticism; (4) intellectual authority must be shared equally among qualified practitioners."[336] Evaluation by nonexperts seems to fall foul of point four at the very least, and possibly point three as well, if the community resents the oversight of outsiders. This is only one example where notions of expertise seem to come in conflict with open science, and we will attempt to reconcile this conflict later in this chapter.

---

[332] For an example of the debate, see this pair of articles: T. Groves, "Is Open Peer Review the Fairest System? Yes," *BMJ* 341, no. nov16 2 (November 16, 2010): c6424–c6424, https://doi.org/10.1136/bmj.c6424 and K. Khan, "Is Open Peer Review the Fairest System? No," *BMJ* 341, no. nov16 2 (November 16, 2010): c6425–c6425, https://doi.org/10.1136/bmj.c6425.

[333] Examples of journals using crowdsourced peer review in one fashion or another include Journal ofInteractive Media in Education (JIME), Shakespeare Quarterly, Atmospheric Chemistry and Physics (ACP). See Emily Ford, "Defining and Characterizing Open Peer Review: A Review of the Literature," Journal of Scholarly Publishing 44, no. 4 (July 2013): 311–26, https://doi.org/10.3138/jsp.44-4-001.

[334] See the Objectivity Chapter of this dissertation, and also Longino 1995, p. 384.

[335] Longino 1995, p. 384.

Less radical forms of open science do not seek the public input explicitly in the review process, coming down on the side of expertise in the dichotomy above, but publishes reports together with the published paper, and sometimes names the referees. The philosophical consequences of these kinds of approaches do not come in conflict with Longino's scheme in the same way that sidestepping expertise does, but they do downplay or deemphasize the role of referees as standing in for a community, as described in the chapter on anonymity, and on the nature of reports being confidential advice. One thing that is lost in a review process when the referee is identified is what Kathleen Wallace referred to as "recipient anonymity", "anonymity for the sake of preventing or protecting the anonymous person from actions by others,"[337] Practically, the consequence could be that the referee's scope for action is curtailed because of a fear of retaliation, but what is more likely to happen is that the fraction of referees who decline to review will rise.[338] On a deeper, more philosophical level, what this does is to change the role of the referee from a vicarious representative of the relevant scientific community to an explicitly identified peer. In a sense, this is *embracing* subjectivity, even though the motivation is an increasingly objective process. The calculation here is that hidden bias, facilitated by anonymity, is a bigger threat to objectivity than any potential loss of candor from the referees as a result of open review. The loss of any claim to the "view from nowhere" objectivity that comes with a nebulous referee who stands in for an entire field, replacing them with a named colleague or even a competitor changes the dynamic from one of arbitration to one of scientific debate. We recall Daston and Galison's account of mechanical objectivity, in which, "[w]here human self-discipline flagged, the machine would take over."[339] While the approaches for open peer review are not suggesting replacing referees with machines, what is interesting here is that this approach of mechanical objectivity represented a restraint of the self. Opening up review and embracing the subjective is a trade-off, the opposite direction to what Daston and Galison described, "Although mechanical objectivity effaces some features of the scientist, it demands other traits; it has a positive as well as a negative sense. In its negative sense, this ideal of objectivity attempts to eliminate the mediating presence of the observer: some versions of this ideal rein in the judgments that select the phenomena, while others disparage the senses that register the phenomena, and still others ward off the theories and hypotheses that distort the phenomena. In its positive sense, mechanical objectivity requires painstaking care and exactitude, infinite patience, unflagging perseverance, preternatural sensory acuity, and an insatiable appetite for work."[340] Do signed reviews reverse this process, which Daston and Galison claims "glorifies the plodding reliability of the bourgeois rather than the moody brilliance of the genius."?[341] Since mechanical objectivity moralizes against the professional sins of embellishment and unwittingly "seeing *as* rather than seeing *that*", will the form of the evaluation change its character to be more a debate between author and referee as mediated by the editor? Put differently, will the vicarious nature of the anonymous referee endure when reports are signed? We previously made the connection that the institution of the anonymous referee has some resonance with the work of John Rawls on the Veil of Ignorance[342], highlighted by Wallace, as a form of anonymity when the referee (or author) cannot be coordinated with respect to their social traits, beyond the most obvious that the referee is an expert and (should) have expertise that is relevant to the paper at hand.[343]

---

[336] Longino 1990, p. 76.

[337] Wallace, p. 29.

[338] "Reviewers who were to be identified to authors produced similar quality reviews and spent similar time on their reviews as did anonymous reviewers, but they were slightly more likely to recommend publication (after revision) and were significantly more likely to decline to review." S. van Rooyen et al., "Effect of Open Peer Review on Quality of Reviews and on Reviewers' Recommendations: A Randomised Trial," *BMJ* 318, no. 7175 (January 2, 1999): 23–27, https://doi.org/10.1136/bmj.318.7175.23

[339] Daston and Galison 1992, p. 81.

This expertise is the relevant determinant of who is permitted to review papers and who counts as a peer. If peer review is opened up, then the expert becomes a public figue, with certain consequences for how they act. A famous case where the expertise of scientists and that of "lay people" who nevertheless held relevant expertise occurred in the United Kingdom as a consequence of the Chernobyl disaster. The parameters of the case: scientists from the British Ministry of Agriculture, Fisheries, and Food (MAFF), alarmed by high levels of radioactivity in sheep, imposed blanket bans on the sale of lamb and sheep in Cumbria, but the scientific methods they used to estimate radioactivity levels and the amount of time the radioactivity from cesium poisoning would endure was calculated badly. The local farmers had expertise about the soil, rain patterns, and the digestive habits of sheep, none of which the scientists had. The case, first chronicled by Brian Wynne,[344] has been discussed and is described in greater detail by Harry

---

[340] *Ibid.*, p. 82-83.

[341] *Ibid.,* p. 83.

[342] Rawls, John.. *A theory of justice.* Cambridge, MA: Belknap Press of Harvard University Press. (1971).

[343] Wallace, p. 29.

[344] Brian Wynne, 'Sheep Farming after Chernobyl: A Case Study in Communicating Scientific Information', *Environment*, Vol. 31, No. 2 (1989).

Collins, alone,[345] together with others,[346] by Christopher Hamlin,[347] and by Robert Crease with Kyle Powys White.[348] The case has bearing on how to reconcile expertise by experts and by non-experts, or as Dreyfuss would call it, beginners and experts, with the intermediate stages we saw in the last chapter. If reports are made public, it will affect the nature of the report, particularly when non-experts read the reports. As Hamlin puts it, "For the scientists, to be publicly scientific is to be definitive. Doubts are only for the ears of other scientists."[349] A public expert has a different role than one consulted in private, because they need to also be credible to the reader, which may be a larger circle than the immediate esoteric one. John Hardwig argues that even inside the esoteric circle, an untrusting attitude of suspicion would impede scientific process. The extra trust required in open peer review should however not be framed as a necessary evil. On the contrary, "It is a positive value for any community of finite minds, provided only that this trust is not too often abused."[350] However, as Fleck points out in his distinction between journal and vademecum science, and also popular science, journal science is written in a much more modest way, bearing as Fleck puts it "the imprint of the provisional and the personal."[351] Scientists writing to other scientists, whether in paper form or in report form, adopt a different tone of voice than they do when speaking to the general public. Hence, the means of achieving trust in the esoteric circle and the exoteric circle can be very different. In the esoteric circle, expertise does a lot of the work of establishing trust, but outside, in dealings with the public, further steps have to be taken to achieve credibility. There are plenty of cases of public mistrust in science, so-called "poisoned well scenarios", where the public mistrust science and scientists due to social factors.[352] What is the guarantee that greater insight in the mechanics of science will breed greater trust? "Climategate", the leak of emails from the Climatic Research Unit (CRU) at the University of East Anglia (UEA) in 2009, caused a loss of trust in science *because* of greater insight into the mechanics of climate research, and the "warts and all" picture was interpreted, in

---

bad faith as well as good, as reducing the credibility of climate science.[353] For any open peer review where the referees share their reports, either signed on unsigned, they suddenly now need to be credible to readers. If the articles in question are published in journals with a wide readership such as *Nature* or *Science,* the circle of potential readers, and hence readers one needs to have credibility with, is widened. This can be quite consequential in terms of the credibility of peer review. As Whyte and Crease point out, "The public benefits of scientific research activities are often compromised when ordinary citizens distrust scientists."[354]

Involving the public in assessment need not necessarily be in the same role as expert referees. For certain studies, there are relevant stakeholders to the study baked in, and not only ethics but also proper epistemology demands they have a voice. Recalling the Cumbrian case, Collins and Evans set out a model which we discussed in Chapter Five, dividing expertise up into "no expertise", "interactional expertise", and "contributory expertise." Ultimately their model turns into a schema of contributory expertise for the core set, supplemented by "lay expertise" from appropriate bystanders.[355] However, Hamlin considers this division arbitrary and points out that it does not deal well with boundary conditions. Considering Collins' case study of gravitational waves (pre-LIGO), Hamlin makes the point that for this field, the core set (esoteric circle) and the controversy mutually define each other, in that the controversy is what the gravitational wave community are engaged in debating. Any lay people, even scientists in different fields, are likely to have no more than an academic interest in the debate. For the Cumbrian case, however, the controversy does not map onto the core set and different stakeholders can indeed have very different motivations and can experience controversies differently.[356] For MAFF, the controversy is about health and safety, while for the farmers the ability to sell meat is intimately tied to livelihood, though of course it would be a mistake to ascribe only one aspect of the problem as being the sole preoccupation of any group.

But talking about stakeholders in this fashion risks instrumentalizing them, as if their only concerns would be concrete ones. For more abstract science, there should regardless be a way of bringing a broader set of perspectives into the evaluation process. Such an ambition calls to mind Jürgen Habermas's description of expert cultures in his work on communicative rationality. These are the specialized spheres of art, science, and morality (by which Habermas means the legal and ethical), populated by experts in each domain, artists and critics for art, scientists and engineers for science, and  theologians, lawyers, ethicists etc. for the morality sphere. Habermas warns against the consequences of expert cultures becoming too isolated,  "In its cognitive, moral, and evaluative components the cultural tradition must permit a feedback connection with the specialized forms of argumentation to such an extent that the corresponding learning processes can be socially institutionalized.  In this way cultural subsystems can arise – for science, law and morality, music, art, and literature  -- in which traditions take shape…"[357] In Habermas's analysis, the expert cultures, of art, science and technology, and moral/legal communities, have become separated from the lifeworld. Habermas later repeated this assertion his piece "Modernity: An Unfinished Project", in which he writes (in this case principally about art),  "Like art, science and technology on the one hand and moral and legal theory on the other have become autonomous. […] institutionalized science and scholarship and the moral-practical discussions that have been separated off into the legal system have become so distant from everyday life that the program of the Enlightenment could be transformed into a false sublation in these spheres as well."[358]

What concerns Habermas here is that science (for example) has become an instrumental exercise in the domination of nature.[359] To gather in the expert cultures into communicative rationality, Habermas recommends a tripartite solution: break the fixation with cognitive-instrumental rationality, encourage communication between expert cultures, and direct the cognitive potential of the expert cultures back into the lifeworld.[360] Differentiation between subsystems and value spheres is an aspect of rationalization of society.  Through instrumental/ purposive rationality, each value sphere (science, value, art) constructs its stock of knowledge that is accumulated by a nascent expert culture.  A divide subsequently opens up between the value spheres and lifeworld that leads to cultural impoverishment.

> The three cultural value spheres have to be connected with corresponding action systems in such a way that the production and transmission of knowledge that is specialized according to validity claims is secured; the cognitive potential of developed by expert cultures has, in turn, to be passed on to the communicative practice of everyday life and to be made fruitful for social action systems, finally, the cultural value spheres have to be institutionalized in such a balanced way that the life-orders corresponding to them are sufficiently autonomous to avoid being subordinated to laws intrinsic to heterogeneous orders of life. A selective pattern of rationalization occurs when (at least) one of the three constitutive components of the cultural tradition is not systematically worked up, or when (at least) one cultural value sphere is insufficiently institutionalized, that is, is without any structure-forming effect on society

---

[357] Habermas, Jürgen, trans. Thomas McCarthy, *The Theory of Communicative Action, Volume 1: Reason and the Rationalization of Society*, Boston, MA: Beacon Press (1984), 71-72.

[358] Habermas, Jürgen, *Modernity: An Unfinished Project*, in *Critical Theory Essential Read*, 1st ed. St. Paul, MN: Paragon House, (1998), ed. Ingram, David and Julia Simon-Ingram, 351.

[359] Friedel Weinert criticizes Habermas in his essay "Habermas, Science and Modernity," where amongst other things he accuses Habermas of holding that science to have lost its connection to analyzing nature.  Weinert claims that Habermas, in his characterization of modernity, "views science predominantly as an instrumental enterprise, which produces means for the domination of nature." Weinert, Friedel. "*Habermas, Science and Modernity*," *Royal Institute of Philosophy Supplements* **44** (1999), p. 340.

[360] Habermas vol 1 1984, p. 240-241.

as a whole, or when (at least) one sphere predominates to such an extent
that it subject life-orders to a form of rationality that is alien to them.[361]

Should this third part, the communication back into the lifeworld, be one that breaks down
the separation between popular science and journal science, as articulated by Fleck? We will
consider this further when we discuss another movement to open up science: Open Access.

## Opening Access

Open Access describes a movement to make research papers available for general use at the
point of publication. This can be either by depositing preprints or even the latest draft of an
article in a public repository such as arXiv (arXiv.org) for physics (green OA), or that the author
or funder pays an article processing charge that defrays all the costs and overhead of the
publisher (gold OA). Open Access often coming with very liberal copyright agreements that
allows for reuse as long as credit is given to the original author or authors. Such rights beyond
just reading is called *libre* Open Access, in contrast to read-only, which is *gratis*.

On one level, when considering the readers, Open Access problematizes the neat division of
potential readers into experts and non-experts. Returning to Fleck, with his division of types of
scientific publications, this would mean that members of the exoteric circle should no longer
content themselves with the output of popular science,[362] and should instead have free access to
the output of journal science, regardless of any expertise barriers to accessibility.[363] Another
argument made for Open Access involves the role of publicly funded research. A frequently-
stated picture of academic publishing is that journals receive publicly-funded research from
authors, which is then packaged and sold back to the public, bypassing the taxpaying public's
interest in reading about current research. As Peter Suber puts it, "Conventional publishers
acquire their key assets from academics without charge. Authors donate the texts of new articles
and the rights to publish them. Editors and referees donate[364] the peer-review judgments to
improve and validate their quality. But then conventional publishers charge for access to the
resulting articles, with no exception for authors, editors, referees, or their institutions."[365] In an
article on Open Access and Ethics, Michael Parker writes that this "is an argument about the
nature of the researcher's social licence to practice and the responsibilities of academics and those who spend public money to contribute to the dissemination of knowledge."[366] While this picture might be criticized for ignoring the value added by peer review, and the work of professional editors, it has gathered increased traction as the Open Access debate rumbles on, particularly in the medical sciences where the results on diseases and their treatment garner more public interest.[367]

---

[361] *Ibid.,* p. 240.

[362] Fleck, p. 173.

[363] Writing about the notional *archive* of primary literature of science, Ziman notes that, "in principle, this is
open to anyone who can get to an academic library and make sense of the specialized technical language in which
most of it is written." Ziman 2002, p. 34.

[364] Suber here does not consider professional editors, who are paid to manage peer review for submissions,
whether part-time of full-time.

[365] Suber, p. 37.

The arguments for Open Access could use, as a justification, utilitarian principles to claim that increased readership may be good for science. Proponents often claim that giving access to more people in itself will result in better science, as Michael Parker articulates, "By far the most common argument made for open-access by scientists and science funders is one that springs from a widely-held belief that the sharing of data will lead to more rapid scientific progress and a reduction in unnecessary duplication of scientific effort."[368] As Parker points out, this is of course an empirical claim and, because of the great variance that exists inside academic publishing, comparing like for like situations is complicated. The metrics used are either problematic, such as the Impact Factor, or can be easily gamed, such as download statistics and page views. Some studies have indicated a definitive citation advantage[369] but other studies claim the opposite.[370] However, this argument only addresses one aspect of Open Access and not the

---

rights of reuse that come with Open Access.

One aspect of Open Access that does have some philosophical consequences is the right of reuse. In order to properly explore this issue we have to look more deeply at the author and their relationship to the work in question. A central point about intellectual property is that it is about more than the ownership of a physical book, or of all book copies that exist of a certain work. The ownership is localized in the underlying work itself. Immanuel Kant, writing about plagiarism, makes this distinction, "This right of the author is, however, not a right to the object, that is, to the copy (for its owner is certainly entitled to, say, burn it in front of the author); rather, it is an innate right, invested in his own person, entitling him to prevent anyone else from presenting him as speaking to the public without his consent – a consent which cannot be taken for granted by any means, since he has already conceded it to someone [to his publisher]."[371] The important thing to focus on here is the thoughts, the words, that come from a person and are thus an extension of that person's personality, which no one else can own, unlike a physical manifestation of the book. In his essay 'Answering the Question: What is Enlightenment?' Kant also expresses this distinction between the copy and the thoughts behind it. To create content is using one's reason in public: "By the public use of one's reason I understand the use which a person makes of it as a scholar before the reading public."[372] This is also the way that academic credit works, no one but the discoverer can claim credit for the discovery, even as people can use previous discoveries to build work of their own.

Inasmuch as one owns one's intellectual work, another discussion of the role of the author is possible through considering the work of Hegel. His description of the role of the author comes out of his rationale for property in general, "In relation to needs […] the possession of property appears as a *means*; but the true position is that, from the point of view of freedom, property, as the first *existence* of freedom, is an essential end for itself."[373] In Hegel's system property gives existence to freedom, and as such is a matter of a right.[374] When it comes to "intellectual production" as property, Hegel like Kant makes a distinction between ownership of the work in the abstract and manifestations of the work in e.g. book form "The distinctive quality of intellectual production may, by virtue of the way in which it is expressed, be immediately transferred into the external quality of a thing, which may then in turn be produced by others."[375] However, just because one owns a copy of something does not mean one owns the work and the rights to it. "Since the person who acquires such a product possesses its entire use and value if he holds a *single* copy of it, he is the complete owner of it as an individual item. But the author of the book or the inventor of the technical device remains the owner of the *universal* ways of reproducing such products and things, for he has not immediately alienated these universal ways and means as such but may reserve them for himself as his distinctive mode of expression"[376] Hegel justifies the notion of the author's intellectual property not by "arbitrarily imposing the condition" that the rights to the original remain with the author even as it is copied and passed on because it is an individual expression. However, readers can use the information in the work as inspiration for their own research, and to Hegel this is in fact the point of intellectual production. "…the destiny of a product of the intellect is to be apprehended by other individuals and appropriated by their representational thinking, memory, thought etc. Hence the model of expression whereby these individuals in turn make *what they have learned* […] into an *alienable thing* will always have some distinctive *form*, so that they can regard the resources which flow from it as their property, and may assert their right to reproduce it."[377] These readers thus becomes authors of work in their own right. Since science, as Hegel puts it, consists of the repetition of established thoughts, having an established means of honoring the rights of the originators of those thoughts is essential to avoid the damaging effects of plagiarism, which to Hegel removes the incentive to create intellectual works.[378] If the possibility of dissemination of ideas were not present then it would not be possible to protect intellectual property other than as trade secrets about which some or all details are hidden for fear of plagiarism.

---

[371] Kant: On the Unlawfulness of Reprinting, Berlin (1785), Primary Sources on Copyright (1450-1900), eds L. Bently & M. Kretschmer, www.copyrighthistory.org

[372] Immanuel Kant. Translator: Daniel Fidel Ferrer. 2013, Answer the Question: What Is Enlightenment? http://archive.org/details/AnswerTheQuestionWhatIsEnlightenment.

[373] Hegel, Georg Wilhelm Friedrich, trans. Allen W Wood, and Hugh Barr Nisbet. *Elements of the Philosophy of Right*. Cambridge [England]; New York: Cambridge University Press, 1991. §45.

[374] "*Right* is any existence in general which is the *existence* of the *free will*." *Ibid.,* §29.

[375] *Ibid.,* §68.

Hegel's insistence on preventing plagiarism is very important for academic writing, since research carries with it the notion of discovery. For him, the notion of individual property is centered on the idea of it being an expression of personality, not on monetary rewards. However, in the online world, for subscription journals, readers of articles do not have full and free use of the article in the sense envisaged by Hegel,[379] in that they are not the complete owner of the object. They are not allowed to pass it on like one can pass on a book, for instance. Indeed, the switch to digital media has allowed for publishers to erect new barriers to sharing, as illustrated by Michael Falgoust, who makes the interesting point how digital rights management has undermined first sale principles by restricting the ability to pass things on, going against utilitarianism. "In the current intellectual property system, the doctrine of first sale facilitates a wide distribution of creative works. First sale allows the resale, loan, or gift of a copy of a work, a novel for example, after the original sale. The existence of used book stores and lending libraries rely on first sale. Second-hand retailers and libraries both serve to distribute creative works to the economically less well-off, secondhand retailers by offering their wares for a lower price and libraries by lending works at no charge. Without first sale, these institutions might be eliminated, limiting the audience's access to creative works. Even current practices in releasing electronic versions of books has undermined first sale through technology. Electronic books ('ebooks') are sold with Digital Rights Management software, often preventing users from lending their ebooks as they do with physical books."[380]

---

[376] *Ibid.,* §69.

[377] *Ibid.,* §69.

[378] "[Plagiarism] can easily have the effect that the profit which the author or inventive entrepreneur expected from his work or new idea is eliminated, reduced for both parties, or ruined for everyone." *Ibid.,* §69.

[379] "New restrictions on electronic journals add a permissions crisis on top of the pricing crisis. For publishers of online toll-access journals, there are business reasons to limit the freedom of users to copy and redistribute texts, even if that leaves users with fewer rights than they had with print journals. But these business reasons create pernicious consequences for libraries and their patrons." Suber, Open Access, p. 34.

Hegel's definition of authorship is of interest in the context of *libre* open access, because it appears to come in conflict with it. For Hegel, as we showed, the interesting moment for intellectual property inasmuch as it relates to the externalization of it is the transformation of an idea into a physical object, e.g. a book, or in this case, a journal article. The use of an intellectual work by a reader involves taking "what they have learned"[381] and turning it into another product (article), which in turn becomes the intellectual property of the reader turned author. Intellectual property is maintained in the academic journal world in that the author is identified as the owner of the work, in terms of credit if nothing else. But Hegel's notion of the author of the work remaining, "the owner of the *universal* ways of reproducing such products and things," is complicated by *libre* open access. As J. Britt Holbrook points out, the requirement to permit derivative works has different consequences in different parts of academia. "I can imagine how my individual reputation could be severely compromised by derivative works — say, a bad translation of my work — for which I had been forced to "grant" permission under a mandatory [*libre*] license."[382]

To sum up, the simple access part of Open Access, access alone, whether green or gold, is less philosophically problematic than the reuse part. The exact way in which access is achieved does not seem to be philosophically fraught. The question of how many people have access to a work is of philosophical concern only inasmuch as it touches on what the public, having funded research with taxes, is due. So from the author's perspective they should have no concerns about Open Access and plenty to cheer about it, since the more people read their research, the more people can follow up on it. As Parker writes, "By far the most common argument made for open-access by scientists and science funders is one that springs from a widely-held belief that the sharing of data will lead to more rapid scientific progress and a reduction in unnecessary duplication of scientific effort."[383] This is a utilitarian justification, and indeed Robert Frodeman and Adam Briggle point to the long history of science funding, and the emergence of an attitude that "what is good for science is inherently good for society. Any investment in science will contribute to a reservoir of knowledge that society can then draw from in order to solve its problems."[384] It remains to be seen how liberal rights of reuse can be, and how this will play out in different disciplines.

---

[380] Falgoust, Michael. "The Incentives Argument Revisited: A Millean Account of Copyright." *The Southern Journal of Philosophy* **52**, no. 2 (n.d.): 163–83. https://doi.org/10.1111/sjp.12059. p. 171.

[381] Hegel, *Right*, §69.

[382] https://jbrittholbrook.com/2019/02/08/feedback-on-guidance-on-implementation-of-plan-s/, accessed August 4th 2019.

[383] Parker, Michael. "The Ethics of Open Access Publishing." *BMC Medical Ethics* **14** (March 22, 2013): 16. https://doi.org/10.1186/1472-6939-14-16., p. 16.

[384] Frodeman and Briggle, p. 6.

## Conclusion on openness

Alex Csziszar has made the claim that changes to peer review occurs at times of societal upheaval correlating with crises in the public standing of science. At this time, journal publishing and peer review is upheld as a gold standard for the proper practice of research. As Alex Csiszar concludes in his work on the Scientific Journal, "Many of the challenges of the dominance of the big academic publishers have focused on revolutionizing the economics of producing and of accessing the literature, but have left the definition of the literature—peer reviewed research papers collected in periodicals, written for an audience of other specialists—relatively untouched. Even preprint databases, which dispense with periodicity entirely, have maintained the genre of the scientific paper as the preeminent form of knowledge expression in science."[385] However, its future is uncertain. As a recent paper on alternative schemes of peer review states, "'Classical peer review' has been subject to intense criticism for slowing down the publication process, bias against specific categories of paper and author, unreliability, inability to detect errors and fraud, unethical practices, and the lack of recognition for unpaid reviewers."[386] In addition to criticism of just peer review, there are concerns about the legitimacy of scientific results, and an exaggerated expectation from the scientific literature, significant enough that some suggests that journals as we know them will soon cease to exist.[387] Such a change would doubtless again be motivated by an appeal to increasing the objectivity of science and the scientific project.

The calls for opening up peer review and opening up access to the products of peer review could be said to occur inside of a narrative of greater openness, spurred at least in part by the arrival of the Internet. Several journals have announced plans to publish reports with published papers.[388] The calls for increased openness in peer review have consequences that put these calls in direct opposition to previously-adopted strategies of anonymity as a means of bolstering the objectivity of the process, embracing subjectivity in the service of accountability. Yet at the same time calls for a wider role for anonymity is entering peer review in the natural sciences, indicating that the link between anonymity and objectivity is still viewed as a viable strategy. Calls for openness can also involve an increased role for public participation in peer review, in ways discussed previously in other aspects of the practice of science.

---

[385] Csiszar 2018, p. 282.

[386] Richard Walker and Pascal Rocha da Silva, "Emerging Trends in Peer Review—a Survey," *Brain Imaging Methods* 9 (2015): 169, https://doi.org/10.3389/fnins.2015.00169.

[387] [C]ontemporary worries that the scientific enterprise itself might be breaking down, exemplified by the replication crisis, are closely bound up with a mismatch between expectations that the scientific literature ought to be a repositor of carefully vetted claims and a rather less tidy reality. Some believe that the collapse of the journal system and of prepublication referee systems is a foregone conclusion, soon to be replaced by a variety of Internet platforms and data analysis that will make science increasingly open, efficient, and secure, from knowledge claims to lab notebooks." Csiszar 2018 p. 283.

[388] "*Nature* will publish peer review reports as a trial", *Nature* **578**, 8 (2020), "Bringing transparent peer review to the physical sciences and beyond", https://ioppublishing.org/news/bringing-transparent-peer-review-to-the-physical-sciences-and-beyond/ (2019), "Registered Reports are coming to *PLOS One*", https://blogs.plos.org/everyone/2020/01/14/registered-reports-are-coming-to-plos-one/

In the discussion of openness trust plays a big part, and it is worth remembering the need for a willingness for vulnerability that trust entails, as discussed in the last chapter. But trust is also required from the public in science and the means of knowledge production, and a request for more insight into this from the point of view of the public is understandable. Expertise, and the perceived legitimacy of that expertise, is also an important aspect of the increased demand for openness, and is consequential for trust. The balance between transparency and objectivity speaks directly to the role of the referee, whether they are to be considered as a peer who is in open dialogue with the authors, or a vicarious representative with a perspective that purports to speak for an entire community. The desire to reintegrate expert cultures into the entire lifeworld also plays a role in this demand for openness and transparency.

When it comes to open access to publications the demand for reuse, not just access, complicates the rights of the authors of the integrity of their own work. Scholarship always involved building on previous knowledge but the *libre* Open Access extends to reuse of the work itself for derivative works, which has consequences for the author's ownership of their work. The request for access for a wider audience seems like a simple demand, but blurs ideas of different types of scientific publications, and who they are for.

# Conclusion

The objectivity at play in peer review has certain characteristics that makes it different from other accounts of objectivity in science. It does not engage with the questions about the reality of the objects under study, instead actors take the reality of scientific concepts for granted while they are engaged in the moment of peer review. The referees may question the applications or implementation of theory to explain the results, but in the moment of peer review they do not question whether theories in general can explain observations, for instance, or whether electromagnetism is real. It would be incoherent to engage in peer review of a scientific paper while simultaneously doubting the ability of science to make statements about the world and states of affairs in it. This bracketing of realism is not only an epistemological stance, but also a social and political one. While engaged in peer review, no agent is seriously doubting the value and appropriateness of the peer review process *at that moment*. If peer review is to be viewed as a technology in a phenomenological sense, this is it fading into the background, only becoming conspicuous when it breaks down, when misconduct such as bias, data fabrication, or other unprofessional conduct, occurs. Having a paper rejected does *not* constitute breakdown (or "defection" in game theory parlance) rather it is part and parcel of normal, functioning, peer review.

Two relevant forms of objectivity are at play in academic publication evaluation: *product* objectivity and *process* objectivity. The product objectivity is in the final published article, found valid and appropriate for publication through evaluation in the peer review process. The process objectivity is how this evaluation happens. Editors and referees hold up an article for comparison with epistemological standards of a community, and authors will have prepared an article that they believe will meet these standards. All actors are intimately familiar with these standards as members of a peer group. In describing these standards, we drew from Helen Longino who, in elaborating her contextual empiricism to describe criticism, bases her scheme on experience that is mediated by background assumptions, assumptions which reflect group values.[389] The background assumptions stem from the interaction amongst scientists in transformative criticism, which ideally involves as many perspectives as possible. The outcome of this interaction is product objectivity in the form of scientific knowledge, formulated through the commonly-agreed-upon process. which is generated by the community that created these background assumptions.[390] Similarly, in peer review, the referees and authors are all peers working on the same problem. The editor is also a peer, albeit not necessarily as deeply involved in the particular topic of the article being reviewed. Nevertheless, the editor is an expert in the process, and as such shares background assumptions with authors and referees.

---

[389] Longino 1995, p. 384.

[390] Longino 1990, p. 74.

Peer review as a process references Longino's socially informed objectivity, which has objectivity as "a characteristic of a community's practice of science rather than an individual's,"[391] checking claims against this objectivity. Longino's model is however concerned with scientific praxis as a whole, and we are interested in a small but crucial part of it. The difference becomes very clear when she in fact discusses peer review. Longino claims that concerns over the breakdown of peer review are overblown because they reflect "an individualist conception of knowledge construction. Peer review prior to publication is not the only filter to which results are subjected. The critical treatment *after* publication is crucial to the refining of new ideas and techniques."[392] This is a valuable insight of social processes in knowledge production, but it has a drawback, it downplays the difference between peer review at a journal and scientists interacting in general, which is too broad a definition of peer review for our purposes. While it is worthwhile to observe that scientists interact significantly outside of the manuscript submission and evaluation process, peer review at a journal is a very special moment in scientific knowledge production and should be considered separately from other interactions. Notably, Longino is here also talking about knowledge construction, which falls into product objectivity and not the process objectivity we are concentrating on.

Considering the roots of modern peer review, the requirement to witnessing which we saw articulated in the account of Shapin and Shaeffer[393] is indeed also an attempt to evaluate scientific claims through the use of a form of process objectivity. However, once *virtual witnessing* through properly written accounts of results in publications became acceptable as a means of validating claims, the evaluation moves from sense experience—I know this is true because I *saw* it—to judgment—I know this is true because I am *convinced* by the arguments, that this is a *faithful* account of what occurred, and I *trust* that the authors did what they said they did. The virtual witnessing model of objectivity turns to consider the reader, the consumer of the published article, which means again the focus is on product objectivity and not on the evaluative mechanisms that form the process objectivity we are considering.

Daston and Galison also describe a form of process objectivity in their study of scientific image-making, however, the process objectivity at play in peer review is not an objectivity of representation.[394] When Daston and Galison step through their moments of true-to-nature, mechanical objectivity, and trained judgment, they are at all times talking of a representation of something empirical, whether it is a drawing of a flower, an x-ray diffraction pattern, or some kind of curve fitting or false-color image generated through the melding of mechanically-obtained data and expertise. However, the objectivity in peer review is not empirical, nor sensorially or instrumentally mediated. The requirement placed upon referees is not for them to repeat experiments, observations, or derivations, but instead they are asked for their judgment. This judgment is informed by community standards that originate in the social. Criteria beyond mere validity such as importance and interest come into play as well in many instances of peer review, and while these are often derided as "subjective", they are also evaluated from the perspective of someone embedded in a community with its socially informed mode of objectivity. Having identified the social as a necessary component of peer review, one comes to the conclusion that it is misplaced to ask for peer review to "focus on the objective science," because peer review can only do its work by being thoroughly informed by subjectivity through the relevant community and its epistemic standards. Subjectivity cannot be eliminated in peer review because there has to be a knower that knows, both in the case of authors, referees, and editors. One cannot have a scientist without subjectivity, or a scientist without a community of other subjective beings. At best, we can manage or account for subjectivity. As Daston and Galison observed, "Objectivity fears subjectivity, the core self […] But there is no getting rid of, no counterbalancing post-Kantian subjectivity. Subjectivity is the precondition for knowledge: the self who knows. This is the reason for the ferociously reflexive character of objectivity, the will pitted against the will, the self against the self This explains the power of objectivity, an epistemological therapy more radical than any other because the malady it treats is literally radical, the root of both knowledge and error."[395]  This is why accounts of subjectivity creeping into peer review are received with such consternation. So, if subjectivity cannot be eliminated, it must be managed.

---

[391] Longino 1990, p. 74

[392] *Ibid*. p. 69.

[393] Shapin and Schaffer, p. 60.

[394] Daston and Galison 2007, p. 17.

Efforts to manage subjectivity in peer review have traditionally been through anonymity. Anonymous referees stand in for a community in a way somewhat reminiscent of Rawls's original position, except here it is the authors who do not know who the referee is, and not the other way around. If the author is also anonymous, the reverse applies, with referees (and sometimes even editors) not knowing who the author is, only knowing that they are a peer.

Two competing trends are trying to modify anonymity in order to achieve better evaluation. The first is a move towards increased transparency, with proposals to publish referee reports with the paper, albeit often unsigned. The second is a push towards *further* anonymity, by also making authors anonymous in fields where this is not already the normal practice.[396] This creates a conflict between accountability and confidentiality. These efforts might superficially seem to be opposed, but the intention in both cases is to bolster the objectivity of the process, taking aim at different parts. There is a tension at the heart of objectivity in peer review, because issues with objectivity are not summed up by one root value, like anonymity or transparency. Increased transparency aims to boost objectivity through increased accountability and public confidence in the process, by emphasizing the value of anonymity over confidentiality. Increased anonymity on the other hand aims at increased objectivity by eliminating or reducing bias and its effects, not just protecting the anonymous referee from the author, but also protecting the referee from themselves and their biases. Both approaches come from the fear of the self that cannot be

trusted.[397]

What about increasing objectivity by even more radical transparency? When the term "peer review" was coined, it made very clear that the reviewing was going to take part internally in a community of trusted experts.[398] Even as the push for opening science up continues, and debates over Open Access rage on, there is not yet a way of replacing that expertise and those trusting relationships, as well-meaning as any democratizing effort of evaluating science could be. This is not to say that peer review could not be improved by citizen oversight, but there is no replacing expertise, especially as science becomes more esoteric. To borrow from Fleck's terminology, the esoteric circles are multiplying and simultaneously shrinking. The elaboration of such mechanisms for oversight while maintaining the epistemological standards is a philosophical challenge facing peer review going forward. Mechanisms for oversight also require trust in both directions between scientists and the public, asking for a trust between peers as well as a trust from the public to accept the scientific products of peer review. The balance of trust with the skepticism required by science is only possible if there are shared epistemological standards that provide legitimate avenues for criticism. As a consequence, the model of the dialectic of trust and distrust described in Chapter Five leads to a problem with increased transparency, since the distrust required in skepticism is hard to translate when communicating outside the peer group. Historically, scientists have expressed science to the public in very confident tones, with no trace of the tentative modesty found in journal science publications.[399] How can one open up a scientific community for public participation while maintaining legitimacy, balancing the epistemological advantage that expertise confers while simultaneously allowing for a more diverse range of perspectives in the peer review process?

In conclusion, science depends on peer review for an account of moving beyond result to actual science. Science is empirical, but peer review depends on judgment, and not one that is empirical. It is not the instrumental or empirical objectivity that is invoked in the gathering of data and the performing of experiments. Peer review provides an opportunity to discuss a manifestation of objectivity in scientific practice from a new perspective: making judgments about empirical matters without invoking empiricism in the act of judging. In the attempt to bolster this objectivity, several approaches are tried, but the difficult emerges because objectivity is not defined by one core value, but a balance of transparency, confidentiality, trust, representation, and living up to community standards. Moreover, the objectivity in peer review is a community characteristic born out of practice, and is not reducible to the individual.

What philosophy in general and this dissertation in particular changes about peer review is that we need to embrace the consequences of this different mode of objectivity. We have to acknowledge that our judgments are based on community standards and are only objective in this narrow sense. Indeed, aspects we have highlighted about anonymity, trust, and expertise all reference a community and its standards. In order to keep what is good about peer review we have to stop overstating our case on what peer review can and cannot do. If we, and here I speak as a journal editor, constantly exaggerate what peer review does, scholars as well as the public will stop holding it up as the mark of quality it should be.

---

[397] "All epistemology begins in fear — fear that the world is too labyrinthine to be threaded by reason; fear that the senses are too feeble and the intellect too frail; fear that memory fades, even between adjacent steps of a mathematical demonstration; fear that authority and convention blind; fear that God may keep secrets or demons deceive." Daston and Galison 2007, p. 372.

[398] Baldwin 2018, p. 540.

[399] Fleck p. 174-175.